\documentclass[reprint,amsmath,amssymb,aps,prb,longbibliography]{revtex4-2}

\usepackage{graphicx}   
\usepackage{dcolumn}    
\usepackage{bm}         
\usepackage{siunitx}    
\usepackage{booktabs}   
\usepackage{mhchem}     
\usepackage{epstopdf}   
\usepackage[caption=false]{subfig}
\usepackage{xcolor}
\usepackage{hyperref}
\hypersetup{breaklinks=true}
\hypersetup{
    colorlinks=true,
    linkcolor=blue,
    filecolor=magenta,      
    urlcolor=blue,
    pdftitle={},
    pdfpagemode=FullScreen,
}
\usepackage{cleveref}

\begin{document}

\title{Investigating Anharmonicities in Polarization-Orientation Raman Spectra of Acene Crystals with Machine Learning}

\author{Paolo Lazzaroni}
\author{Shubham Sharma}
\author{Mariana Rossi}
\email{mariana.rossi@mpsd.mpg.de}

\affiliation{MPI for the Structure and Dynamics of Matter, Luruper Chaussee 149, 22761 Hamburg, Germany}

\date{\today}

\begin{abstract}
We present a first-principles machine-learning computational framework to investigate anharmonic effects in polarization-orientation (PO) Raman spectra of molecular crystals, focusing on anthracene and naphthalene. By combining machine learning models for interatomic potentials and polarizability tensors, we enable efficient, large-scale simulations that capture temperature-dependent vibrational dynamics beyond the harmonic approximation. Our approach reproduces key qualitative features observed experimentally. We show, systematically, what are the fingerprints of anharmonic lattice dynamics, thermal expansion, and Raman tensor symmetries on PO-Raman intensities. However, we find that the simulated polarization dependence of Raman intensities shows only subtle deviations from quasi-harmonic predictions, failing to capture the pronounced temperature-dependent changes that have been reported experimentally in anthracene. 
We propose that part of these inconsistencies stem from the impossibility to deconvolute certain vibrational peaks when only experimental data is available. This work therefore provides a foundation to improve the interpretation of PO-Raman experiments in complex molecular crystals with the aid of theoretical simulations.
\end{abstract}

\maketitle

\section{Introduction}

Molecular crystals are a class of solids where the molecular units, which constitute the basis of the crystal lattice, are kept together by intermolecular (non-covalent) interactions. The weak nature of these forces, when compared to those responsible in covalent bonds, 
allows one to identify two distinct energy scales in the nuclear lattice dynamics. At higher energies, intramolecular vibrations usually give rise to phonon branches with flat dispersion, as a result of the weakly perturbed internal motion of the individual molecules. In contrast, at low energies, the collective translations and hindered rotations of the molecular units are responsible for the unique spectral fingerprints of a given crystal structure. 

These low-frequency modes, typically below 200 cm$^{-1}$, dictate numerous properties of organic molecular crystals. It has been discussed, for example, how this region determines polymorph ordering at finite temperatures, by strongly contributing to the vibrational free energy \cite{raimbault_anharmonic_2019,rossi_anharmonic_2016,krynski_efficient_2021}, or how it dominates heat \cite{konstantinov_heat_2003,legenstein_heat_2025} and charge \cite{fratini_map_2017,coropceanu_charge_2007} carrier mobility in organic semiconductors. Because the  weak intermolecular interactions lead to large-amplitude motion, these collective lattice vibrations can exhibit pronounced anharmonic character. Taking into account deviations from the ideal harmonic crystal picture 
is especially relevant when modeling these systems.

Recently, it was shown \cite{asher_anharmonic_2020,asher_chemical_2022} that polarization-orientation (PO) vibrational Raman scattering experiments conducted on several molecular crystals, including some linear oligoacenes, reveal an effect that has been attributed to anharmonic coupling between low-frequency intermolecular phonon modes. This effect consists of a deviation of the polarization dependence of the Raman intensity of some phonon modes from the prediction of the harmonic crystal approximation. To rationalize this finding, a two-mode model, allowing phonon-phonon coupling in the description of inelastic light scattering, was presented in Ref. \cite{benshalom_phononphonon_2023}. Parameters that were fitted on this model, based on the experimental data, could successfully explain the pattern that was measured and led to the conclusion that anharmonic mode-coupling terms are necessary for a complete description of the Raman scattering process in molecular crystals.

In this paper, we provide an \textit{ab initio} quality framework to reproduce, explain and provide quantitative insights into anharmonic effects directly probed by PO-Raman spectroscopy. As a case study, we focus on linear oligoacene molecular crystals, in particular naphthalene and anthracene. The standard approach of combining \textit{ab initio} molecular dynamics (AIMD) with density functional perturbation theory (DFPT) to obtain Raman spectra guarantees that all orders of anharmonicity of the potential energy surface (PES) are included, but is prohibitively expensive for this task. This is due, for example, to the vast range of experimental conditions one needs to explore and the large cell sizes required to accurately describe the phonon manifold. Furthermore, to capture subtle variations in the angle-resolved Raman intensity, the spectra have to be thoroughly converged, and finite-size artifacts due to periodic boundary conditions need to be mitigated. While ML-based frameworks for computing  Raman spectra without explicitly accounting for the crystal orientation have been demonstrated~\cite{raimbault_using_2019,berger_polarizability_2024,gurlek_accurate_2025,grumet_delta_2024}, this work instead presents a ML framework to obtain accurate polarization-orientation Raman spectra across multiple temperatures. This extension is non-trivial: PO-Raman intensities provide a significantly more stringent test of ML accuracy than angle-integrated spectra, as subtle errors in polarizability tensor components or normal mode symmetries dramatically alter angular dependence while remaining invisible in unpolarized spectra. 
We thus combine state-of-the-art equivariant ML models for both interatomic potentials and polarizability tensors with the computationally efficient $\Gamma$RGDOS approximation proposed in Refs.~\cite{hashemi_efficient_2019,berger_raman_2023} to enable quantitative predictions of temperature-dependent anharmonic PO-Raman signals.

In this manuscript, we begin by introducing the theoretical background of PO-Raman scattering, emphasizing the limitations of the harmonic approximation and the need for anharmonic treatments. After benchmarking our ML models and validating the $\Gamma$RGDOS approximation against full ML calculations, we simulate the temperature-dependent PO-Raman spectra of anthracene and naphthalene, capturing anharmonic effects. By comparing our results with experimental data, we analyze the polarization dependence of Raman intensities and discuss the subtle deviations from the harmonic predictions. Our findings highlight the interplay between anharmonic lattice dynamics and Raman tensor symmetries, providing insights into the origins of the temperature-dependent behavior of low-frequency Raman modes and suggesting guidelines for the interpretation of the rich amount of information contained in PO-dependent Raman spectra. This work demonstrates the power of combining machine learning with established theoretical frameworks to unravel complex anharmonic phenomena in molecular crystals.

\section{Theory of Polarization-Orientation Raman Scattering\label{sec:theory-po}}

The Raman effect, first described in 1928~\cite{raman_new_1928}, is a process where photons undergo inelastic scattering on matter, resulting in a finite frequency (and polarization) shift between incident and scattered radiation. 
At a certain energy range, the spectrum of frequency differences is a manifestation of vibrational transitions occurring in the scattering medium, providing unique fingerprints of the vibrations of the material. 

Single crystal polarization-orientation Raman (also referred to as angle-resolved polarized Raman) experiments are often conducted in a back- (or forward-) scattering geometry, where the incident ($\bm{e}_1$) and scattered ($\bm{e}_2$) fields are linearly polarized and propagate through the crystal, which is positioned at a certain orientation \cite{asher_anharmonic_2020,kim_anomalous_2015,xu_polarized_2021}. The scattered radiation is then collected with polarization either perpendicular ($\bm{e}_1 \perp \bm{e}_2$) or parallel ($\bm{e}_1 \parallel \bm{e}_2$) to that of the incident beam. Once the configuration is fixed, the experiment is repeated spanning an angle $\theta$ between $\bm{e}_1$ and an arbitrary reference, either by rotating the sample or the direction of $\bm{e}_1$.

Within a semiclassical treatment, where the light is treated classically and the matter quantum-mechanically, the expression for the Raman cross-section in the electric dipole approximation reads~\cite{mcquarrie_statistical_2000,long_raman_2002}
\begin{equation}
I(\omega, \beta, \theta) \propto \sum_{i,f} \rho^{\beta}_i \left\langle i\left|\bm{e}_1\cdot\bm{\alpha}\cdot\bm{e}_2\right| f\right\rangle^2 \delta(\omega-\omega_{fi}),
\label{eq:raman1}
\end{equation}
where $\bm{\alpha}$ is the polarizability tensor of the matter and, within the Born-Oppenheimer (BO) approximation, the summation runs over all vibrational states, with $\rho_i^\beta$ being the thermal equilibrium population associated to the initial state at inverse temperature $\beta$. Here $\bm{e}_1$ and $\bm{e}_2$ are the polarization vectors defined above, either for the parallel or perpendicular setting. The unpolarized Raman spectrum can be obtained by integrating the Raman intensity over all angles in both configurations, and it is defined as
\begin{equation}
  I^{\text{unpolarized}}(\omega, \beta) = \int_0^{2\pi}\frac{d\theta}{2\pi} [ I^{\perp}(\omega, \beta, \theta) + I^{\parallel}(\omega, \beta, \theta) ].
  \label{eq:unpolarized}
\end{equation}

If we restrict our description to non-resonant Raman, i.e. when the energy of the incident radiation is far below any electronic transition of the system, only the static polarizability tensor of the electronic ground state is needed \cite{long_raman_2002}. Within perturbation theory, its components take the form
\begin{equation}
\alpha_{qp} = 2\sum_{\sigma>0}\frac{\left\langle 0\left|\mu_{p}\right| \sigma\right\rangle\left\langle \sigma\left|\mu_{q}\right| 0\right\rangle}{\varepsilon_{\sigma}-\varepsilon_{0}},
\label{eq:alpha}
\end{equation}
 where $\left| 0\right\rangle$ is the electronic ground state,  the summation runs over all the excited electronic states, $\bm{\mu}$ is the electric dipole operator and $q, p$ denote Cartesian components. With this definition of the polarizability tensor, Eq.~\ref{eq:raman1} becomes the well known Placzek approximation \cite{long_raman_2002,walter_ab_2020} to the full expression of the light scattering cross section derived first by Kramers, Heisenberg and Dirac \cite{kramers_uber_1925,dirac_quantum_1927}. Note that, as recently highlighted in Ref.~\cite{knoop_ab_2024}, by treating the quantity in Eq.~\ref{eq:alpha} as the polarizability tensor (or dielectric susceptibility) entering Eq.~\ref{eq:raman1}, we are already excluding the effect of finite scattering wave-vectors, essentially assuming that only vibrations at the $\Gamma$-point of the crystal phonon Brillouin zone are probed in the experiment, i.e. $\bm{q} = \hat{\bm{k}}_1 - \hat{\bm{k}}_2 \approx 0$, where the $\hat{\bm{k}}$ are the propagation directions of incoming and scattered fields. This tends to be a very good assumption for first-order Raman scattering, but breaks down in the context of second-order Raman scattering in polar materials \cite{knoop_ab_2024}. 

 At this stage, the polarizability tensor can be regarded as an operator for the nuclei, and, in the BO picture, it depends parametrically on the nuclear coordinates through the electronic states of Eq.~\ref{eq:alpha}. Thus, it is commonly expanded in a Taylor series around the nuclear equilibrium configuration $\bm{Q}_0$
 \begin{equation}
    \alpha_{qp}(\bm{Q}) = \alpha(\bm{Q}_0) + \sum_k \frac{\partial\alpha_{qp}(\bm{Q})}{\partial Q_k}\Bigg|_{\bm{Q} = \bm{Q}_0} Q_k + \dots
    \label{eq:alpha-expansion}
 \end{equation}
where $k$ runs over all nuclear degrees of freedom.
While the first term in the expansion above is responsible for Rayleigh scattering, the higher-order terms, involving polarizability derivatives, enter the expression for the Raman cross-section. If the coordinates are chosen to be the mass-scaled normal modes of the system $\tilde{Q}_k$, these derivatives are referred to as Raman tensors, $\bm{\mathcal{R}}_k$, i.e., at first order
\begin{equation}
\bm{\mathcal{R}}_k^{(1)} =\frac{\partial\alpha_{qp}(\bm{Q})}{\partial \tilde{Q}_k} \label{eq:raman-tensor}.
\end{equation}
Note that $\tilde{Q}_k$ are Cartesian-space vectors, obtained by removing the mass-scaling that exists in the eigenvectors that come out of the diagonalization of the dynamical matrix. As such, they form a non-orthonormal basis that needs to be handled carefully in common operations such as normal-mode projections of molecular dynamics trajectories, as will be relevant later in the manuscript (see Section S1 of SM~\footnotemark[1]).

Retaining only the first order term in the expansion of Eq.~\ref{eq:alpha-expansion}, the matrix element of Eq.~\ref{eq:raman1} is approximated as
\begin{equation}
     \left\langle i\left|\bm{e}_1\cdot\bm{\alpha}\cdot\bm{e}_2\right| f\right\rangle \approx  \sum_k  \left\langle i\left| \tilde{Q}_k \right| f\right\rangle \bm{e}_1\cdot\bm{\mathcal{R}}_k\cdot\bm{e}_2 .
\label{eq:raman2}
\end{equation}
At this stage, if the vibrational Hamiltonian is assumed to be harmonic in the normal coordinates, the initial and final states can be written as product states of harmonic eigenstates. It is easy to see, within these two approximations, that the matrix elements of the position operators in Eq.~\ref{eq:raman2} connect only states that differ by a single vibrational excitation in the respective mode, giving rise to the common first-order Raman selection rule.

 For the same reason, when the square modulus of Eq.~\ref{eq:raman2} is taken, the cross-products between different eigenmodes vanish, leaving us with the expression for the Raman cross-section 
\begin{equation}
I(\omega, \beta, \theta) \propto \sum_k \rho^{\beta}_k \left| \bm{e}_1\cdot\bm{\mathcal{R}}_k\cdot\bm{e}_2\right|^2 \delta(\omega-\omega_{k}),
\label{eq:harm}
\end{equation}
that will depend trivially on temperature through the population of the initial states. These derivations can also be found in the SM of Ref.~\cite{benshalom_phononphonon_2023}.
We will often refer to the $\theta$ angular dependence, for a given $k$, as the ``PO-pattern'' of the corresponding mode in the following.

The result of Eq.~\ref{eq:harm} is the well known expression for the (harmonic) polarized Raman scattering cross section that is commonly employed to rationalize experimental results in terms of Raman tensors and factor group analysis~\cite{deangelis_factor_1972}. As shown above and already clearly stressed in \cite{benshalom_phononphonon_2023}, this simple expression is the result of what is often termed a ``double harmonic'' approximation, as it relies on both harmonic normal modes and the truncation at first order of the expansion in Eq.~\ref{eq:alpha-expansion}. Clearly, this formula would fall short of accounting for any effect of anharmonic lattice dynamics in the Raman spectrum. Particularly relevant for this work is the fact that in this approximation, the dependence of the (normalized) scattered intensity on the polarization angle $\theta$ is fully determined by the shape, symmetry and relative values of the Raman tensor components of a given mode, and by the geometry of the experiment. Notably, the thermal populations enter only as a multiplicative factor. As the space group of the crystal and the molecular sites do not change with temperature (assuming no phase transitions), the shape and symmetry of $\mathcal{R}_k$ are fixed. A common interpretation of this observation is that besides a change in its absolute values, the angular dependence of the intensity calculated with Eq.~\ref{eq:harm} will be temperature independent in the harmonic picture. 
As we will see in this paper, though, keeping to Eq.~\ref{eq:harm} but including temperature-dependent lattice changes can lead to modifications in sign and relative magnitude of certain components of $\mathcal{R}_k$, which in turn lead to pronounced variations in the PO pattern that are not connected to dynamical anharmonic effects.

Beyond this, any further change with temperature of the PO-patterns must then be investigated taking into account the dynamical anharmonicity of molecular vibrations, which presents an added challenge to simulations. In particular, dynamical methods which are not based on perturbation theory for the nuclear degrees of freedom provide an attractive platform to include these effects in their totality. 
For practical purposes, it is convenient to switch from the energy representation of Eq.~\ref{eq:raman1} to the Heisenberg representation in the time domain, resulting in
\begin{equation}
    I(\omega, \beta, \theta) \propto \int_{-\infty}^{+\infty}dt \,\, e^{i\omega t} \, \left<\alpha(\theta, t)\alpha(\theta, 0)\right>_\beta
    \label{eq:tdraman}
\end{equation}
where we have conveniently defined
\begin{equation}
    \alpha(\theta) := \bm{e}_1\cdot\boldsymbol{\alpha}\cdot\bm{e}_2,
\end{equation}
$\left<\cdot\right>_{\beta}$ denotes a quantum thermal average, $\omega$ is frequency and $t$ is time. Note that, within the BO approximation, the time dependence of the polarizability is given by the time dependence of the nuclear positions. The expression of Eq.~\ref{eq:tdraman} can be rewritten as sum over component-wise cross-sections,  
\begin{equation}
  I(\omega, \beta, \theta) \propto \sum_{rsqp} e_1^r e_1^s I_{rsqp} (\omega, \beta) e_2^q e_2^p
  \label{eq:fourth}
\end{equation}
highlighting the fourth-rank tensorial nature of $I_{rsqp} (\omega, \beta)$, where $r, s, q, p$ run over Cartesian components, as recently stressed in both Refs. \cite{benshalom_phononphonon_2023} and \cite{knoop_ab_2024}.

The time-correlation function appearing in Eq.~\ref{eq:tdraman} can be well approximated in many situations by considering a classical time-evolution for the nuclei, especially at higher temperatures or when focusing on low-energy vibrational modes. A full quantum-mechanical evaluation of these correlation functions for large and complex systems is too cumbersome with available methodologies, which require the use of explicit wave-function methods or real-time path integrals \cite{beck_multiconfiguration_2000,chatterjee_real-time_2019}. However, these expressions are also amenable to approximate methods that join quantum-statistics with classical nuclear dynamics, based on ring-polymer molecular dynamics \cite{craig_quantum_2004,braams_short-time_2006,habershon_ring-polymer_2013,rossi_how_2014,althorpe_path_2024} and centroid molecular dynamics \cite{jang_derivation_1999,trenins_path-integral_2019,lawrence_fast_2023,musil_quantum_2022}. These methods provide a way to incorporate zero-point-energy and nuclear delocalization effects in the time-correlation functions, but disregard dynamical quantum coherence of the nuclear wavepackets. When coherence effects are small, for example at moderate to high temperatures and in the condensed phase \cite{markland_nuclear_2018}, these methods become suitable for including nuclear quantum effects in molecular crystals.

\section{Simulating PO-Raman Signals \label{sec:ml-raman}}

In this work, we focus on capturing, through first-principles simulations, anharmonic effects in the PO-Raman signal coming from low-frequency vibrational modes on polyacene molecular crystals. Going beyond the harmonic approximation without relying on perturbative expansions amounts to evaluating Eq.~\ref{eq:tdraman}, where the time-dependence of the polarizability tensor through the nuclear coordinates is commonly taken into account by means of AIMD simulations.

Achieving first-principles accuracy in this evaluation relies on being able to obtain $O(10^6)$ calculations of forces and polarizability tensors, through the realization of AIMD trajectories and perturbation theory calculations, resulting in a very high computational cost \cite{hoja_first-principles_2017}. Previous studies~\cite{raimbault_anharmonic_2019,putrino_anharmonic_2002,pagliai_anharmonic_2008} have used AIMD to successfully compute anharmonic unpolarized or powder Raman spectra of molecular crystals, albeit computing very few trajectories of the order of 10~ps, resulting on a high uncertainty in peak intensities. In the present study, converged peak intensities are necessary in order to capture subtle variations in polarization-orientation patterns at different temperatures.

A practical and efficient solution is achieved by evaluating the quantities needed in Eq.~\ref{eq:tdraman} with highly accurate ML models. In this paper we use two models: one model targets the prediction of energies and forces, needed for the evaluation of normal modes or nuclear time propagation, and the other model targets the prediction of polarizability tensors, needed for the evaluation of the Raman cross sections. When combining ML models in order to calculate a particular physical observable, one needs to take into account that errors from both models will compound in the calculation. This calls for a  stringent evaluation of the accuracy and performance of the framework.

Because the training and usage of machine-learning interatomic potentials (MLIPs) is nowadays common in the literature, we only cover the basic aspects of the models used in this work in Methods Section~\ref{sec:ml-pots}. Below, we go into more detail on the models used for the polarizability tensors.

\subsection{Machine Learning Models for Polarizability Tensors \label{sec:ml-polarizability}}

The polarizability tensor is a second-rank tensor with an isotropic component that transforms as spherical harmonics of $l=0$ and an anisotropic component that transforms as spherical harmonics of $l=2$. As such, symmetry-adapted equivariant ML methods can deliver a noticeable boost in efficiency of training. All models we discuss in this paper account for the symmetry of the tensor in the learning problem, being thus equivariant. We train two such models here, which are based on an atomic decomposition of this quantity, i.e.
\begin{equation}
\bm{\alpha} = \bar{\alpha}\mathbf{1} + \tilde{\bm{\alpha}} = \sum_I \left[\bm{\alpha}^{(l=0)}(\mathcal{A}_I)+ \bm{\alpha}^{(l=2)}(\mathcal{A}_I) \right]
\end{equation}
where $\bar{\alpha}$ is the isotropic, traceless part of the polarizability tensor, $\tilde{\bm{\alpha}}$ the anisotropic one and $I$ runs over all atoms that define the center of the atomic environments $\mathcal{A}$.

The first method we employ to learn this tensor is a MACE model \cite{batatia_mace_2022,batatia_design_2025}, which is an equivariant message-passing graph neural-network (NN). Due to its equivariant formulation, it can be seamlessly extended to predict vectorial and tensorial properties \cite{kapil_first-principles_2024, martin_general_2025}, allowing one to set the tensorial rank of the prediction and the symmetries of its components. We will refer to this model as MACE-$\alpha$.

The other method is the symmetry-adapted Gaussian process regression (SA-GPR) proposed in Ref. \cite{grisafi_symmetry-adapted_2018}. As the name proposes, this method is not based on a NN architecture, but instead on a Bayesian Gaussian process regression \cite{deringer_gaussian_2021}. This technique casts the problem of learning the target quantity as a problem of finding the coefficients of a linear expansion of the target quantity on a combination of similarity-kernel functions \cite{willatt_atom-density_2019}. It is possible to define different kernels corresponding to the different rotational symmetries of each component of $\bm{\alpha}$, leading to an equivariant ``symmetry-adapted" version of the learning problem. We will refer to this model as SA-GPR-$\alpha$.

Both methods mentioned above have already been employed in the context of predicting polarizability tensors that were then used to calculate standard vibrational Raman spectra \cite{raimbault_using_2019, berger_polarizability_2024, martin_general_2025, jana_learning_2024, kapil_first-principles_2024,grumet_delta_2024}. Here we instead assess them for the more stringent test of predicting the PO-pattern of each Raman peak.

\subsection{The RGDOS Approximation to Polarizability Tensors \label{sec:rgdos}}

The RGDOS approximation to the polarizability tensor, as introduced in Ref. \cite{hashemi_efficient_2019} and extended for molecular dynamics in Ref. \cite{berger_raman_2023}, is an efficient approach to reconstruct such tensors of a large supercell as a weighted sum of polarizabilities of the multiple unit cells that constitute it. In Ref. \cite{berger_polarizability_2024}, \textit{Berger et al.} showed the good performance of this approximation in reproducing DFT-quality polarizabilities for large supercells and yielding good Raman spectra, including second-order and resonant contributions. The method relies on the existence of a unique equilibrium reference structure and its corresponding eigenmodes, being therefore unable to describe different phases of a material within a single model, but performing well when no phase transitions occur.

Following Ref.\cite{berger_polarizability_2024}, one expands the polarizabiliy of the supercell as
\begin{equation}
    \boldsymbol{\alpha}(\bm{u}) \approx \boldsymbol{\alpha}(\bm{u}_0) + \sum_{n,\nu, \bm{q}} \frac{1}{n!} \mathcal{R}_{\nu, \bm{q}}^{(n)}P_{\nu, \bm{q}}^{n}(\bm{u})~\label{eq:rgdos-exp}
\end{equation}
where $\bm{u}$ is an atomic displacement vector in the supercell, $n$ is the order of the expansion term, $\nu$ is the primitive-cell phonon branch index and $\bm{q}$ is the phonon wavevector,  $\mathcal{R}_{\nu, \bm{q}}^{(n)}$ is the corresponding $n$-th order Raman tensor and $P_{\nu, \bm{q}}(\bm{u})$ is the projection coefficient of the supercell displacement onto the primitive-cell eigenmode, as defined in Section S1 of SM~\footnotemark[1]. The most important approximation involved in the expression above is that $n$-th order cross-derivative terms involving different phonon modes are disregarded, while they would be present in an exact expansion of the polarizability tensor for anharmonic systems.

The flexibility of the method stems from the fact that the expansion can be truncated at any desired order. If one is interested only in first-order Raman scattering processes,  the task is considerably simplified. In addition to keeping only the first order in the expansion of Eq.~\ref{eq:rgdos-exp}, only modes at the vicinity of the $\Gamma$-point of the Brillouin zone contribute, due to photon momentum conservation. This is well approximated by keeping only contributions from $\Gamma$-point phonon modes, i.e. 
\begin{equation}
\boldsymbol{\alpha}(t) = \boldsymbol{\alpha}(\bm{u}(t)) \approx \boldsymbol{\alpha}(\bm{u}_0) + \sum_{\nu} \mathcal{R}_{\nu, \Gamma} P_{\nu, \Gamma}(t).
\label{eq:rgdos1}
\end{equation}
 Once a phonon calculation is performed with the primitive cell and the corresponding Raman tensors are obtained, the $\Gamma$-eigenmodes of the supercell are reconstructed by tiling, i.e. concatenating primitive-cell eigenmodes up to the supercell dimensionality. The projection coefficients are then computed at every time-step to yield the polarizability time series for the supercell according to Eq.~\ref{eq:rgdos1}. In the SM~\footnotemark[1] we provide more details about how we obtain these projection coefficients in a non-orthogonal basis. Note that a different calculation (phonons + Raman tensors) has to be performed at every temperature if lattice thermal expansion is taken into account, as the equilibrium reference structure will be affected by the change in lattice vectors. The $\Gamma$RGDOS PO-Raman spectrum is then obtained by plugging Eq.~\ref{eq:rgdos1} into Eq.~\ref{eq:tdraman}, yielding
\begin{multline}
I^{\Gamma\mathrm{RGDOS}}(\theta) \propto 
\sum_{\nu \mu} \left(\bm{e}_1 \cdot \mathcal{R}_{\nu, \Gamma} \cdot \bm{e}_2\right) 
   \left(\bm{e}_1 \cdot \mathcal{R}_{\mu, \Gamma} \cdot \bm{e}_2\right) \\
\times \int_{-\infty}^{\infty} dt\, 
   \langle P_{\nu, \Gamma}(0)  P_{\mu, \Gamma}(t) \rangle_{\beta} \,  e^{i \omega t},
\label{eq:rgdos-corr}
\end{multline}
 where anharmonic effects are included in the correlation function of the projected supercell displacements. A further possible simplification regards the truncation of Eq.~\ref{eq:rgdos1} with respect to the number of $\Gamma$ phonons considered. The evaluation of Eq.~\ref{eq:rgdos-corr} then requires fewer Raman tensors and projection coefficients. We demonstrate in the SM~\footnotemark[1] (Section S2) how retaining only the intermolecular modes in the expansion has no visible effect in the resulting low-frequency spectrum of the polyacene crystals studied here, making this method particularly attractive for the investigation of low THz regions of molecular crystals.

\subsection{Proposed Framework\label{sec:framework}}

We assessed the quality of the machine-learning interatomic potentials (MLIPs) we trained for the frequencies of the crystal vibrations. A particularly challenging region, where we must guarantee that the errors are acceptable, is the low-THz part of the spectrum, which is determined by weak intermolecular interactions often poorly described in MLIPs that are inherently local. As detailed in  Section~\ref{sec:ml-pots}, we compared a second-generation Behler-Parrinello high-dimensional neural network potential~\cite{behler_generalized_2007}, which we labeled HDNNP-MLIP, and an equivariant graph neural-network MACE~\cite{batatia_mace_2022} interatomic potential, which we labeled MACE-MLIP.

Our benchmarks, detailed in the SM Section S3~\footnotemark[1], show that MACE-MLIP outperforms the HDNNP-MLIP in the prediction of $\Gamma$-point harmonic phonon frequencies of the anthracene crystal, resulting in a lower average percentage error of $\approx$0.2\% against the $\approx$1.6\% obtained from HDNNP-MLIP. Moreover, MACE-MLIP exhibits a significantly tighter distribution of errors, with a low median and minimal variability, highlighting the consistency and accuracy of this model. In contrast, HDNNP-MLIP   shows a broader error distribution, indicating less reliable predictions. 
When focusing specifically on the low THz region, the error on the harmonic frequency of Raman active modes is almost negligible for the MACE-MLIP model, while the HDNNP-MLIP largely underestimates the harmonic frequency of the three highest frequency librational modes, with errors up to 20\%. These conclusions are consistent with the broader benchmark on the accuracy of vibrational properties obtained from MLIPs that we performed in Ref.~\cite{gurlek_accurate_2025}, in which the reasons for the superior performance of MACE-MLIPs for vibrational frequencies of molecular crystals are discussed.

Next, we assessed the quality of our ML models for the prediction of the polarizability tensors. Our benchmarks, detailed in the SM Section S3, show that MACE-$\alpha$ predictions are almost a factor two more accurate than the SA-GPR predictions for diagonal and off-diagonal components of the tensor. We conclude that MACE  is the most suitable solution for achieving higher prediction accuracy, especially when applied to larger datasets and system sizes.

\begin{figure}[ht!]
  \centering
  \includegraphics[width=0.49\textwidth]{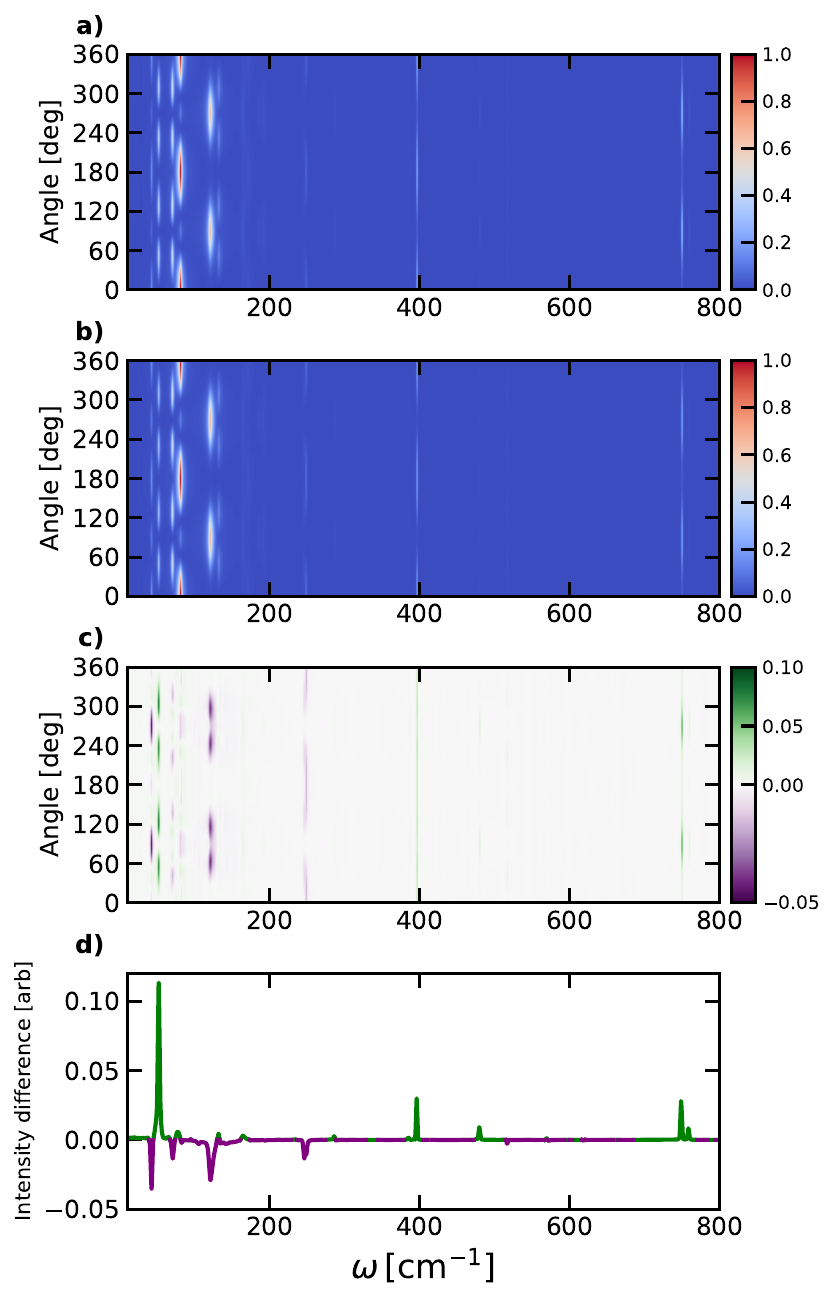}
  \caption{a) PO-Raman spectrum of anthracene at 100K obtained with the ``full-ML" framework (4x4x4 supercell). b) PO-Raman spectrum of anthracene at 100K obtained with the $\Gamma$RGDOS-ML framework (4x4x4 supercell). Intensities are normalized to the highest peak amplitude in both cases. c) Difference between the PO-Raman spectra obtained with the two frameworks. Negative values correspond to features that are more intense in full-ML than in $\Gamma$RGDOS-ML and viceversa. d) Difference between the unpolarized Raman spectra obtained with the two frameworks.}
  \label{fig:methodvs}
 \end{figure}

Lastly, we analyzed the differences between (i) evaluating Eq.~\ref{eq:tdraman} with the MACE-MLIP potential for performing the nuclear time evolution combined with the MACE-$\alpha$ model for the polarizability evaluations and (ii) evaluating the RGDOS approximation in Eq.~\ref{eq:rgdos-corr} with the same MACE-MLIP potential for the nuclear dynamics and fixed Raman tensors computed directly from DFPT. In the following, we name the first procedure ``full-ML'' and the second ``$\Gamma$RGDOS-ML''.

In all the calculations performed in this work, the incident and scattered light are polarized in the $xy$ plane and propagate along the $z$ direction in the lab-frame. This means that only the $x$ and $y$ components of the response are relevant. The PO-Raman spectra we show in the following figures are computed in the parallel scattering configuration, as defined in Section~\ref{sec:theory-po}.

In Fig.~\ref{fig:methodvs}a and b, we show the PO-Raman spectra of the anthracene crystal at 100~K in the range of 0 to 800~cm$^{-1}$, obtained with full-ML and $\Gamma$RGDOS-ML, respectively. All the spectra were computed from $4 \times 4 \times 4$ supercell trajectories, which amounts to more than 3000 atoms in the case of anthracene. More details are given in Section \ref{sec:methods}. The overall visual comparison makes it clear that the differences between these two methods are almost imperceptible in this frequency range, with both yielding very similar polarization dependence of all peaks, and following a very similar relative-intensity pattern when regarding different peaks. 

The small discrepancies that arise are shown in more detail in Fig.~\ref{fig:methodvs}c, where the difference between the PO-Raman spectra of anthracene at 100~K obtained with each procedure is shown. Negative values correspond to features that are more intense in full-ML than in $\Gamma$RGDOS-ML and vice-versa. 
In Fig.~\ref{fig:methodvs}d we report the same difference spectrum, this time for the unpolarized Raman spectrum, as defined in Eq.~\ref{eq:unpolarized}.
In general, $\Gamma$RGDOS-ML demonstrates remarkable accuracy for PO-Raman analysis in the low-THz region of interest, with discrepancies in the order of $\sim 5-10\%$ of the maximum amplitude. The larger discrepancy is found in the $B_g^{(1)}$ mode of anthracene located below 100~cm$^{-1}$, where $\Gamma$RGDOS-ML overestimates the unpolarized peak amplitude $\sim$ 12\%. Consequently, as our focus is limited to first-order Raman -- which contains contributions almost solely from the vicinity of $\Gamma$-point phonons --  we opted to proceed with the $\Gamma$RGDOS-ML framework due to its efficiency and additional advantage of removing uncertainties in the prediction of polarizability tensors.

\section{Results and discussion \label{sec:results}}
\subsection{Quasi-Harmonic Approximation to PO-Raman\label{sec:quasiharm}}

 \begin{figure*}[t]
  \centering
  \subfloat[Anthracene]{
    \includegraphics[width=0.48\linewidth]{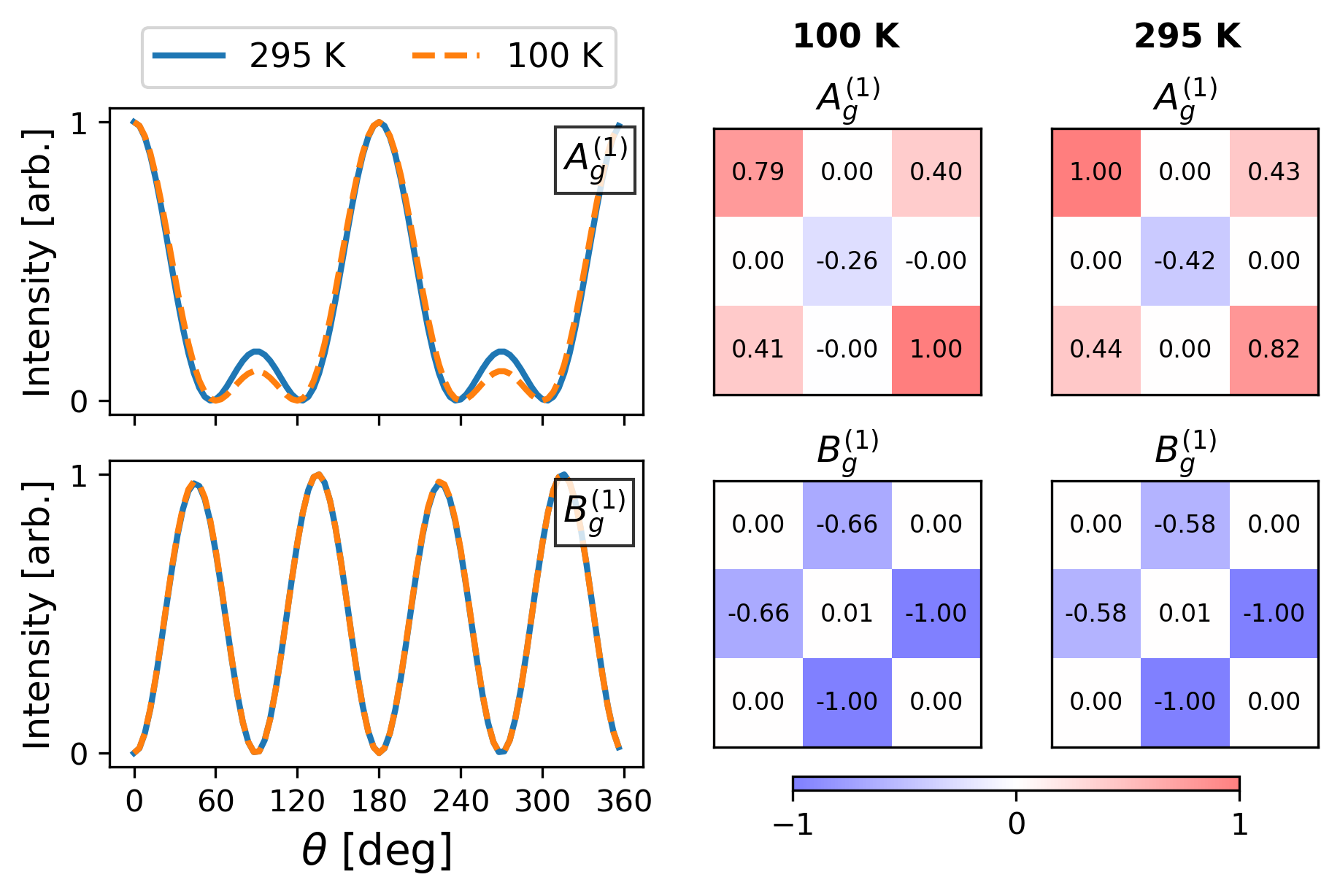}
    \label{fig:harmf2}
  }
  \hfill
  \subfloat[Naphthalene]{
    \includegraphics[width=0.48\linewidth]{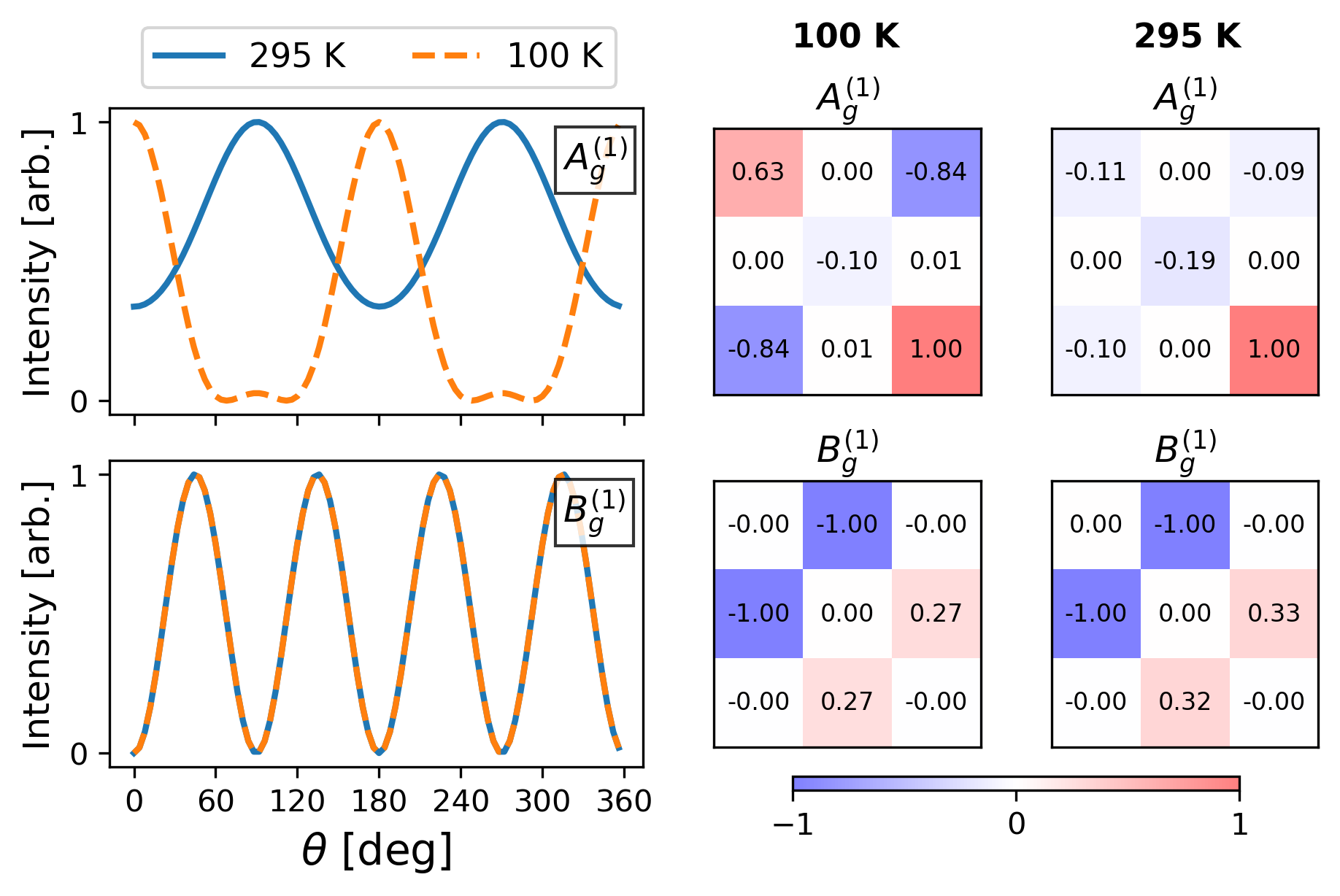}
    \label{fig:harmf2-naph}
  }
 \caption{Harmonic polarization dependence of the Raman intensity of modes $A_g^{(1)}$ and $B_g^{(1)}$ in anthracene (a) and naphthalene (b), obtained with the harmonic approximation of Eq.~\ref{eq:harm} in the parallel configuration. On the right, the corresponding Raman tensors for every mode computed with finite differences from DFPT polarizabilites along normal mode displacements of the MACE-MLIP at the 100 K and 295 K lattice vectors. Tensor components are normalized to their highest value for each mode. In the parallel configuration, only $A_g$ modes can show a change in the polarization pattern with temperature (lattice expansion) within the quasi-harmonic approximation. This effect is particularly pronounced in the case of naphthalene, where structural changes impact the $A_g^{(1)}$ mode. The full set of harmonic polarization patterns and a more detailed analysis can be found in Section S4 of SM.}
  \label{fig:harmonic}
\end{figure*}

As a first step towards the analysis of the impact of temperature on the polarization dependence of the Raman intensity, we compute the quasi-harmonic PO-dependence (parallel configuration) of the Raman intensity in anthracene and naphthalene using only the harmonic approximation of Eq.~\ref{eq:harm}. This calculation is purely harmonic and non-dynamical: the Raman tensors are obtained directly through DFPT at the optimized geometries corresponding to different lattice parameters, and no MD simulations are involved. Thermal lattice expansion is taken into account at 100~K and 295~K by using experimentally-determined lattice constants. In this subsection, when we refer to different temperatures, we are simply referring to the corresponding different lattice constants. This approach is commonly referred to as quasi-harmonic approximation. 

Both materials studied here crystallize in the same space group, $P2_1/a$, with two molecules per primitive cell occupying the same sites in both crystals. Factor group analysis thus leads to the same symmetries for their Raman active modes. Namely, the $A_g$ and $B_g$ modes are respectively the three in-phase and the three out-of-phase librational modes of the two molecular units in the primitive cell, and are Raman active \cite{suzuki_polarized_1968,zhizhin_ii_1995}. From the three out-of-phase translational modes, two bear $A_u$ and one bears $B_u$ symmetry, being instead IR active. As already mentioned, the crystal orientation in our simulations is along the crystallographic $ab$ plane (001), meaning only the $xy$ components of the Raman cross-section will be measured. 

Within the quasi-harmonic picture, the polarization pattern, i.e the angular dependence of the Raman intensity for a given mode $k$ in Eq. \ref{eq:harm}, is fundamentally determined by the shape, symmetry, and the relative magnitude and sign of the components of the corresponding $\mathcal{R}_k$. In Section S4 of the SM~\footnotemark[1] we provide explicit expressions for the general angular dependencies of the Raman intensity patterns generated by $A_g$ and $B_g$ tensors. In the case of naphthalene and anthracene, with the crystals oriented in the crystallographic $ab$ plane (001), we show how, once the intensity is normalized at each temperature, no variations with lattice expansion can occur in the polarization patterns of $B_g$ modes, both in the parallel and perpendicular configuration, within the harmonic approximation of Eq.~\ref{eq:harm}. The same is true for $A_g$ modes in the perpendicular configuration. In contrast, we show how $A_g$ modes in the parallel configuration are affected by changes in the relative value of their Raman tensor components, which can be induced by thermal lattice expansion.

We report in Fig.~\ref{fig:harmonic} the $A_g^{(1)}$ and $B_g^{(1)}$ modes of both crystals as interesting example cases. For completeness, we show the patterns for all modes in parallel and perpendicular configurations in the SM Section~S4~\footnotemark[1]. We start by discussing the PO-patterns of the anthracene crystal. All $B_g$ modes show a four-fold oscillating pattern across polarization angles, as exemplified in Fig.~\ref{fig:harmf2}, and the $A_g$ modes show more variability (see also Fig. S4 of SM~\footnotemark[1]). Specifically, $A_g^{(1)}$ and $A_g^{(2)}$ have $xx$ and $yy$ tensor components of opposite sign, resulting in a four-fold pattern with two maxima of different intensities, whereas $A_g^{(3)}$ displays a two-fold oscillation due to having these diagonal elements of the same sign. This is an important observation, as despite the fact that a square modulus appears in Eq.~\ref{eq:harm}, the relative sign of the Raman tensor components is qualitatively relevant in the resulting PO-dependence of $A_g$ modes, and can therefore be extracted from a fit of the experimental data. 
Variations of the PO-patterns at different temperatures are minimal.  As a matter of fact, we only observe a change in the relative values of the diagonal components of the $A_g^{(1)}$ mode, resulting in an intensity variation of the relative maximum at $\theta =$ 270$^\circ$ in Fig. \ref{fig:harmf2}. 

In the case of naphthalene, the same analysis holds true for the $B_g$ modes, which show a four-fold oscillating pattern, as expected from the fact that these crystals belong to the same space group. The $A_g$ modes again show either a two-fold or a four-fold pattern with maxima of different intensities, depending on the sign of the diagonal components of their Raman tensors (see Fig. S5 of SM~\footnotemark[1]). For example, the $A_g^{(2)}$ mode of naphthalene has a weak PO dependence in the parallel configuration, while having zero intensity in the perpendicular one. An inspection of its Raman tensor at 100~K explains this, as the $xx$ and $yy$ components are almost equal in magnitude and sign, acting like an identity matrix in Eq. \ref{eq:harm}.

Quite surprisingly, though, in naphthalene we observe pronounced changes in the overall polarization patterns of $A_g$ modes with thermal lattice expansion. This is particularly evident in the case of mode $A_g^{(1)}$, where a significant change in the relative values of the diagonal components of the Raman tensor leads to a transition from a four-fold pattern with two maxima of different intensities at 100 K to a two-fold pattern at 295 K (see Fig.~\ref{fig:harmf2-naph}). This observation highlights that even within the quasi-harmonic approximation, temperature-induced lattice expansion can lead to substantial modifications in the polarization dependence of certain Raman active modes. In Section~S4 of SM~\footnotemark[1], by comparing the normal modes displacement vectors at the two different temperatures, we show that the $A_g^{(1)}$ and $A_g^{(2)}$ modes of naphthalene present a measurable change in direction with the different lattice parameters, leading to significant variations in their Raman tensor components and consequently in their polarization patterns. This is not the case for anthracene, where the normal modes obtained for the structures at different temperatures overlap almost perfectly with each other.

As a further note, we demonstrate in Section~S5 of SM~\footnotemark[1], that even small inaccuracies in the normal modes can lead to Raman tensors that yield incorrect polarization dependencies. These changes may be overlooked when the $\theta$ variable is integrated out, underlynig again how PO-patterns are a much stricter test for the quality of the theoretical framework, compared to unpolarized Raman spectra.

In summary, once the experimental configuration is fixed (either $\bm{e}_1 \perp \bm{e}_2$ or $\bm{e}_1 \parallel \bm{e}_2$), only the PO-dependence of modes of a certain symmetry can be affected by thermal lattice expansion in the quasi-harmonic approximation. In the case of anthracene and naphthalene, these are the $A_g$ modes in the parallel configuration. The extent to which the PO-pattern of these modes is affected varies significantly between different materials, as exemplified by the comparison between anthracene and naphthalene. While we prove that not all temperature-dependent changes in PO-Raman patterns will stem from dynamical coupling between different vibrational modes, we also provide a clear baseline for the following analysis of temperature-dependent spectra. It is important to note that the changes we capture within the quasi-harmonic approximation in Fig.~\ref{fig:harmonic}, although sometimes pronounced, are still encoded in the form of second-rank Raman tensors of fixed symmetry. The quasi-harmonic results thus establish a baseline for interpreting temperature effects that arise purely from lattice expansion. Any further changes in the polarization patterns beyond this baseline, as observed in experiment or in the dynamical simulations we present next, must originate from anharmonic dynamical effects involving the vibrational degrees of freedom, where the full, fourth-rank nature of the Raman response function (Eq. \ref{eq:fourth}) is revealed.

\subsection{Temperature Dependence of PO-Raman Signals}

\begin{figure*}[tb]
  \centering
  \includegraphics[width=0.8\textwidth]{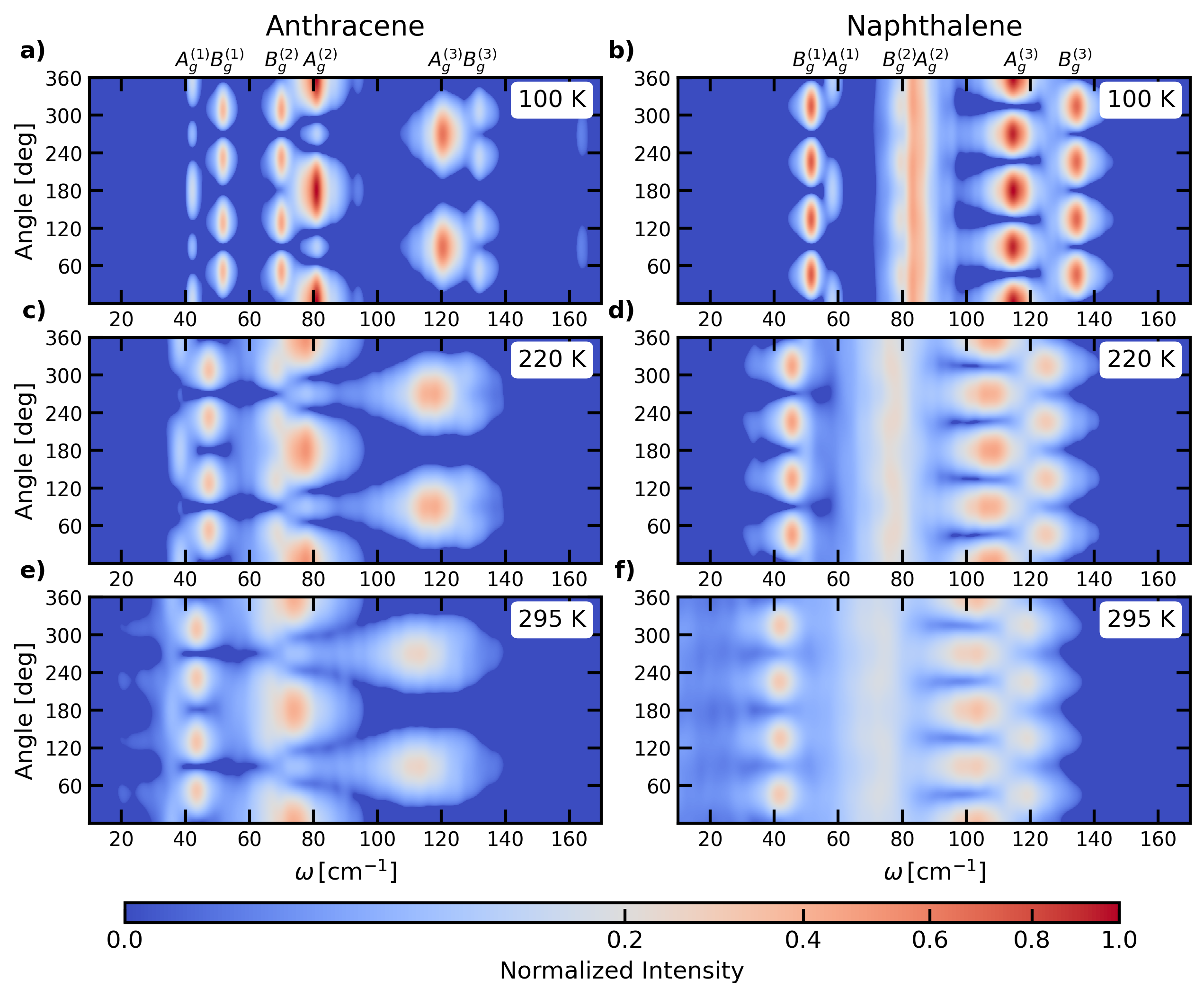}
  \caption{Temperature dependence of the parallel configuration PO-Raman spectra of anthracene (a,c,e) and naphthalene (c,d,f) in the intermolecular motion range of 10 to 170~cm$^{-1}$, obtained with the $\Gamma$RGDOS-ML method. The intensity is normalized to the maximum intensity of the 100K spectrum.}
  \label{fig:tscan-RGDOS}
 \end{figure*}

We now turn our attention to the investigation of the temperature dependence of the PO-Raman spectra of anthracene and naphthalene crystals. In Fig. \ref{fig:tscan-RGDOS} we report the temperature evolution of the parallel configuration anharmonic PO-Raman spectra in the intermolecular motion range of 10 to 170~cm$^{-1}$, obtained with the $\Gamma$RGDOS-ML method. The perpendicular configuration PO-spectra can be found in the SM~\footnotemark[1] (Section S6) for completeness. Computational details on these production simulations are given in Section \ref{sec:dynmethod}.

In panels c–f of Fig. \ref{fig:tscan-RGDOS}, a clear redshift and broadening of all peaks is observed as the temperature increases in both crystals. The extracted peak positions are listed in Table \ref{tab:temp}. The shift is similar across different modes, spanning a range of $\sim 3-9$ cm$^{-1}$. This effect is a direct result of the incorporation of anharmonicity, in particular the temperature-dependent lattice expansion. As shown in Section S7 of SM~\footnotemark[1] for anthracene, the redshift is almost negligible when lattice expansion is not taken into account, despite the anharmonic dynamics. This  highlights once more the strong influence of lattice parameter variations on the dynamics of low-frequency intermolecular modes. Notably, some modes even exhibit a blue shift in the absence of lattice expansion. Beyond these classical anharmonic effects, we also assessed the role of nuclear quantum effects (NQEs) on the PO-patterns. As shown in Section S8 of SM~\footnotemark[1] (see additional Refs. \cite{seiler_nuclear_2021,kowalski_theory_2023,litman_temperature_2020,rossi_progress_2021} therein), based on path-integral molecular dynamics simulations of the PO-Raman patterns of anthracene at 100 K and 295 K~\cite{wang_machine_2019,musil_quantum_2022,castro2025}, NQEs are found to have a negligible impact on the polarization patterns of these modes below 400~cm$^{-1}$, confirming that classical-nuclei simulations are very accurate at the frequency range considered here.

In the following subsections, we will compare our simulated spectra with available experimental data and analyze in detail the signatures of anharmonic mode mixing and peak broadening that emerge at higher temperatures.

\begin{table}[b]
\centering
\begin{ruledtabular}  
\renewcommand{\arraystretch}{1.3} 

\begin{tabular}{lccc@{\hspace{1cm}}lccc} 
\multicolumn{4}{c}{\textbf{Anthracene} [cm$^{-1}$]} & 
\multicolumn{4}{c}{\textbf{Naphthalene} [cm$^{-1}$]} \\
\hline
Mode & 100K & 220K & 295K & Mode & 100K & 220K & 295K \\
\hline
$A_g^{(1)}$ & 42 & 38 & 36 & $B_g^{(1)}$ & 51 & 45 & 41 \\
$B_g^{(1)}$ & 52 & 47 & 44 & $A_g^{(1)}$ & 58 & 50 & 48 \\
$B_g^{(2)}$ & 70 & 68 & 67 & $B_g^{(2)}$ & 79 & 75 & 69 \\
$A_g^{(2)}$ & 81 & 77 & 74 & $A_g^{(2)}$ & 84 & 79 & 72 \\
$A_g^{(3)}$ & 121 & 117 & 112 & $A_g^{(3)}$ & 115 & 108 & 103 \\
$B_g^{(3)}$ & 132 & 128 & 124 & $B_g^{(3)}$ & 134 & 125 & 119 \\
\end{tabular}

\end{ruledtabular}
\caption{Evolution of the peak frequencies with temperature in anthracene and naphthalene from the PO-maps of Fig.~\ref{fig:tscan-RGDOS}. Peak positions are determined from the fitting procedure detailed in Section~\ref{sec:methods}. For naphthalene, $B_g^{(2)}$ and $A_g^{(2)}$ peak positions were extracted at 220 K and 295 K from the VDOS decomposition of Fig. \ref{fig:cvdos} due to their broadness and overlap.}
\label{tab:temp}
\end{table}

\subsubsection{Contrasting Simulations with Experimental Data \label{sec:exp-theory}}

\begin{figure*}[tb]
  \centering
  \subfloat[Parallel configuration]{
    \includegraphics[width=0.48\linewidth]{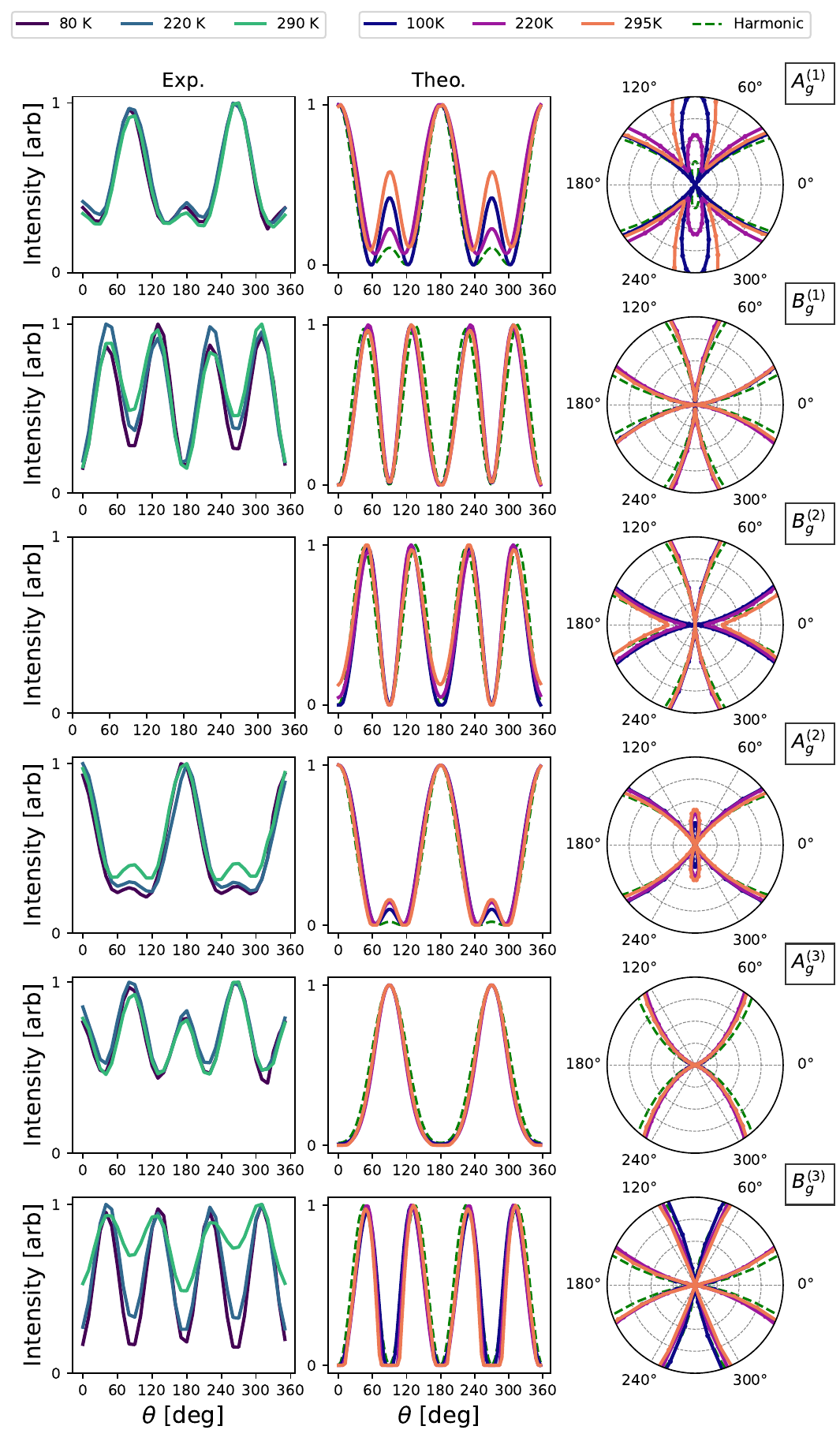}
    \label{fig:fitting_parallel}
  }
  \hfill
  \subfloat[Perpendicular configuration]{
    \includegraphics[width=0.48\linewidth]{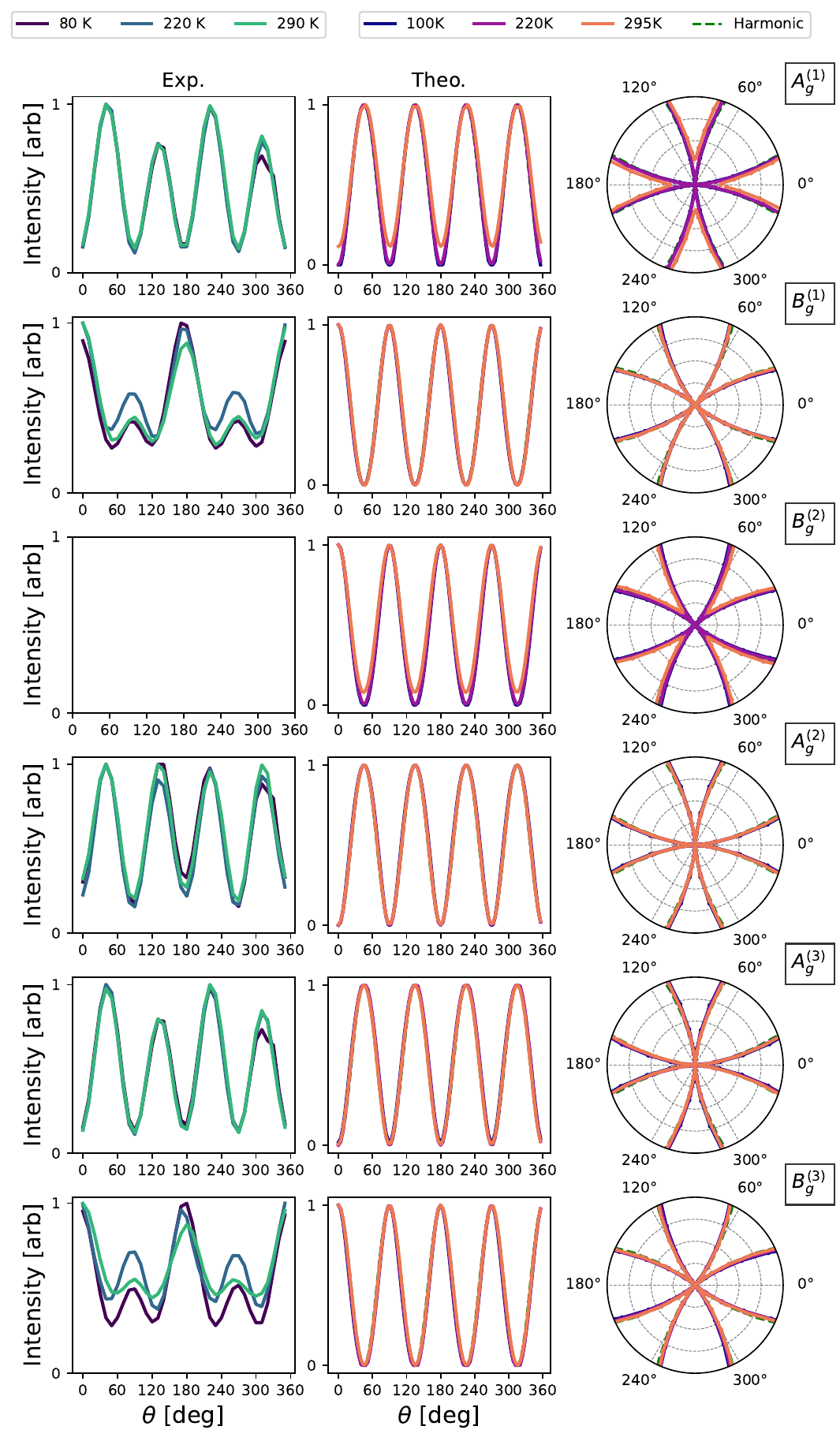}
    \label{fig:fitting_perpendicular}
  }
  \caption{Polarization dependence of the integrated intensity of low THz Raman peaks in anthracene for (a) Parallel configuration and (b) perpendicular configuration. Left columns are experimental results reproduced with permission from Ref. \cite{asher_anharmonic_2020} at 80 K, 220 K and 290 K. Center and right columns fitted with Eq.~\ref{eq:multilorentz} from the simulated PO-maps of Fig.~\ref{fig:tscan-RGDOS} (parallel) and Fig. S8 of SM (perpendicular) obtained with the $\Gamma$RGDOS-ML framework at 100 K, 220 K, and 295 K. Intensity is normalized with respect to the highest intensity of every peak at the corresponding temperature. For comparison, the harmonic PO dependence computed with Eq.~\ref{eq:harm} is also shown. Polar plots zoomed in the region of intensities below 0.4, to highlight small differences arising with varying temperature.}
  \label{fig:fitting_combined}
\end{figure*}

The first observations of an anomaly in the polarization dependence of the Raman peaks in anthracene and pentacene were made by \textit{Asher et al.} in 2020 \cite{asher_anharmonic_2020}. For example, the polarization dependence of some of the low-frequency Raman peaks of anthracene was found to gradually become less pronounced with increasing temperature. These effects were later reported for other classes of organic molecular crystals by the same authors, and extensive work has been performed in altering (quenching) this behavior through chemical substitutions \cite{asher_chemical_2022}. 

This phenomenon, which was described as a dynamical symmetry breaking or mode-specific ``liquid-like" transition, has been recently attributed by \textit{Benshalom et al.}~\cite{benshalom_phononphonon_2023} to the manifestation of temperature activated, anharmonic coupling between the intermolecular modes. In their work, the authors tackled the problem from an analytical point of view. An expression for the Raman intensity similar to Eq.~\ref{eq:tdraman} was evaluated with fixed Raman tensors in a two-mode model, allowing for anharmonic coupling between modes of same and different symmetries. By relying on fourth-rank tensors for the Raman cross-sections like the ones in Eq.~\ref{eq:fourth}, this approach was successful in fitting the experimental PO-dependence of the Raman intensity when the harmonic expression of Eq.~\ref{eq:harm} failed. These results led the authors to conclude that the observed variation in the PO-dependence of the Raman peaks is a consequence of the anharmonic coupling between intermolecular modes, activated by temperature. Note that the harmonic approximation leading to Eq.~\ref{eq:harm}, as explained in Section \ref{sec:theory-po}, can only account for deviations due to thermal lattice expansion that conserve the symmetry of the Raman tensors. Wherever this approach fails, this can be regarded as a sign of a  dynamical anharmonic effect.

From our point of view, it is instructive to understand whether a first-principles procedure without fitting parameters can capture and provide deeper insight into the observed changes in PO-Raman intensities with increasing temperature. In this study, we are considering the full expression of the anharmonic intensity (i.e Eq.~\ref{eq:tdraman}), thus retaining the generalization to the fourth-rank tensor of the scattering cross section and the complete mode coupling given by the first-principles full-dimensional anharmonic PES. The expression Eq.~\ref{eq:rgdos-corr} has the same form as the starting point of the analysis of \textit{Benshalom et al.}~\cite{benshalom_phononphonon_2023}, containing all the cross-correlations terms between phonon modes, i.e. the off-diagonal components of the phonon spectral function, albeit only at the $\Gamma$-point. 
Moreover, our simulations give us the chance to directly compare harmonic and anharmonic results, with and without thermal lattice expansion, providing a tool to decouple and unravel the different effects that temperature has on the resulting Raman spectra.

Before focusing on the angle dependence of the integrated intensity, we highlight some general discrepancies between our simulated and the experimental spectra of anthracene. We observe good overall agreement with the experimental peak positions as reported in Fig. S7 of the SM of Ref.~\cite{asher_anharmonic_2020}, with redshifts in the order of $\approx$ 5-10 cm$^{-1}$ in the simulated spectra. Notably, the $B_g^{(2)}$ mode has negligible Raman intensity in the experiments, where no polarization dependence could be extracted. In contrast, the mode shows appreciable intensity in our results. 
This discrepancy is attributed to the level of theory employed, and is not affected by the subsequent dynamics. In fact, it is known to be already present when calculating a harmonic unpolarized Raman spectrum with the same level of theory (PBE+MBD), as reported in Ref.~\cite{asher_anharmonic_2020}. In the SM, Section S13, we further show that using different functionals to calculate the crystal vibrations shift the position of the peaks in a non-uniform fashion in this frequency range, but do not alter the relative peak intensities.

In this work, we attempt to extract the PO-dependence of the integrated intensity of every peak from the anharmonic PO-maps of Fig.~\ref{fig:tscan-RGDOS} by fitting the Raman spectra of anthracene at 100 K, 220 K and 295 K to a multi-Lorentzian function at each angle. The fitting procedure mimics the one employed for the experiments, as detailed in the SM of Ref.~\cite{asher_anharmonic_2020}, and reported in Section~\ref{sec:methods}.

In Fig. \ref{fig:fitting_combined} we show the fitted PO-dependence of the integrated intensity of the low-frequency Raman peaks of anthracene in both the parallel and perpendicular experimental configurations. The fitted data is normalized to the maximum intensity of each peak at the corresponding temperature. For comparison, we also include the harmonic PO-dependence 
and the experimental results from Ref. \cite{asher_anharmonic_2020}, at the closest available temperatures of 80 K, 220 K and 290 K. We remind the reader that for anthracene, the harmonic PO-pattern is insensitive to lattice expansion. In Fig. S13 of the SM~\footnotemark[1], we show the result of attempting the same fit to the simulated spectra of naphthalene, which fails at higher temperatures due to excessive broadening of the peaks.

The fitted intensity oscillation patterns from the simulated data in the parallel configuration shown in Fig.~\ref{fig:fitting_combined}a are in general agreement with the experimental results, with the notable exception of the $A_g^{(3)}$ mode. While both theory and experiment agree that the temperature variation has no impact on the PO-dependence of this mode, its polarization pattern differs. In experiment, it shows a four-fold pattern, with two relative maxima. In the simulations, only a two-fold pattern is present. This discrepancy is entirely determined by the form of the Raman tensor predicted by the specific level of theory employed, and cannot attributed to the subsequent dynamics. Specifically, our PBE $A_g^{(3)}$ Raman tensor has $xx$ and $yy$ components of the same sign, which cannot give rise to a four-fold pattern. One must note that experimental artifacts, such as birefringence, can alter the polarization dependence of the Raman intensity in experiment, making the direct comparison even more challenging. Artificially including birefringence in the form of Jones matrices \cite{kranert_raman_2016} can alter the simulated PO-patterns of $A_g$ modes, but not to a sufficient extent to explain the observed discrepancies (see Section S9 of SM~\footnotemark[1]).  In the perpendicular configuration, all modes present a four-fold pattern in our simulations. This is partially consistent with experiments, where indeed every mode appears as four-fold, but with $B_g$ having two relative maxima of different intensities. 

In our simulations, the $A_g^{(1)}$ mode shows variations in the PO pattern with temperature, although we attribute a portion of this variation to the change in the relative magnitudes of the diagonal components of the Raman tensor, discussed in Section~\ref{sec:quasiharm} and shown in Fig.~\ref{fig:harmonic}.

The experimental results in Fig. \ref{fig:fitting_combined} show a clear change of the PO-dependence with temperature of the $B_g^{(1)}$ and $B_g^{(3)}$ modes in the parallel configuration. The largest change is seen in going from 220 K to 290 K. A similar effect is present for the $A_g^{(2)}$ mode in experiment, but less pronounced. In all cases, the effect can be described as a gradual loss of the oscillation pattern with increasing temperature. When comparing  to our simulated, anharmonic PO-maps, we do not find a one-to-one, quantitative correspondence of this phenomenon. In some cases the trend is reproduced correctly, such as the $A_g^{(2)}$ mode, where the oscillation pattern is indeed less pronounced at higher temperatures in both experiment and theory in the parallel configuration. In the case of the $B_g^{(1)}$ and $B_g^{(3)}$ modes, which show a strong loss of polarization dependence in experiment, our simulated pattern is not affected by temperature. Nonetheless, we observe variations of the PO pattern in the $B_g^{(2)}$ mode, with a consistent trend of gradual attenuation of the oscillation pattern with increasing temperature. In the perpendicular configuration of Fig.~\ref{fig:fitting_combined}b, $B_g^{(2)}$ still shows subtle deviations in the simulated patterns. In contrast, the experimental data is substantially less clear regarding the temperature dependence of the PO patterns when the perpendicular geometry is employed. In this geometry, changes are still observed in experiment for $B_g^{(1)}$ and $B_g^{(3)}$, but no monotonic trend can be identified with the increasing temperature. We proceed with a deeper theoretical analysis of these observations.

\subsubsection{The Impact of Mode Mixing and Mode Broadening on Anharmonic Spectral Weights \label{sec:mix-broad}}

 \begin{figure}[tb]
  \centering
\includegraphics[width=0.49\textwidth]{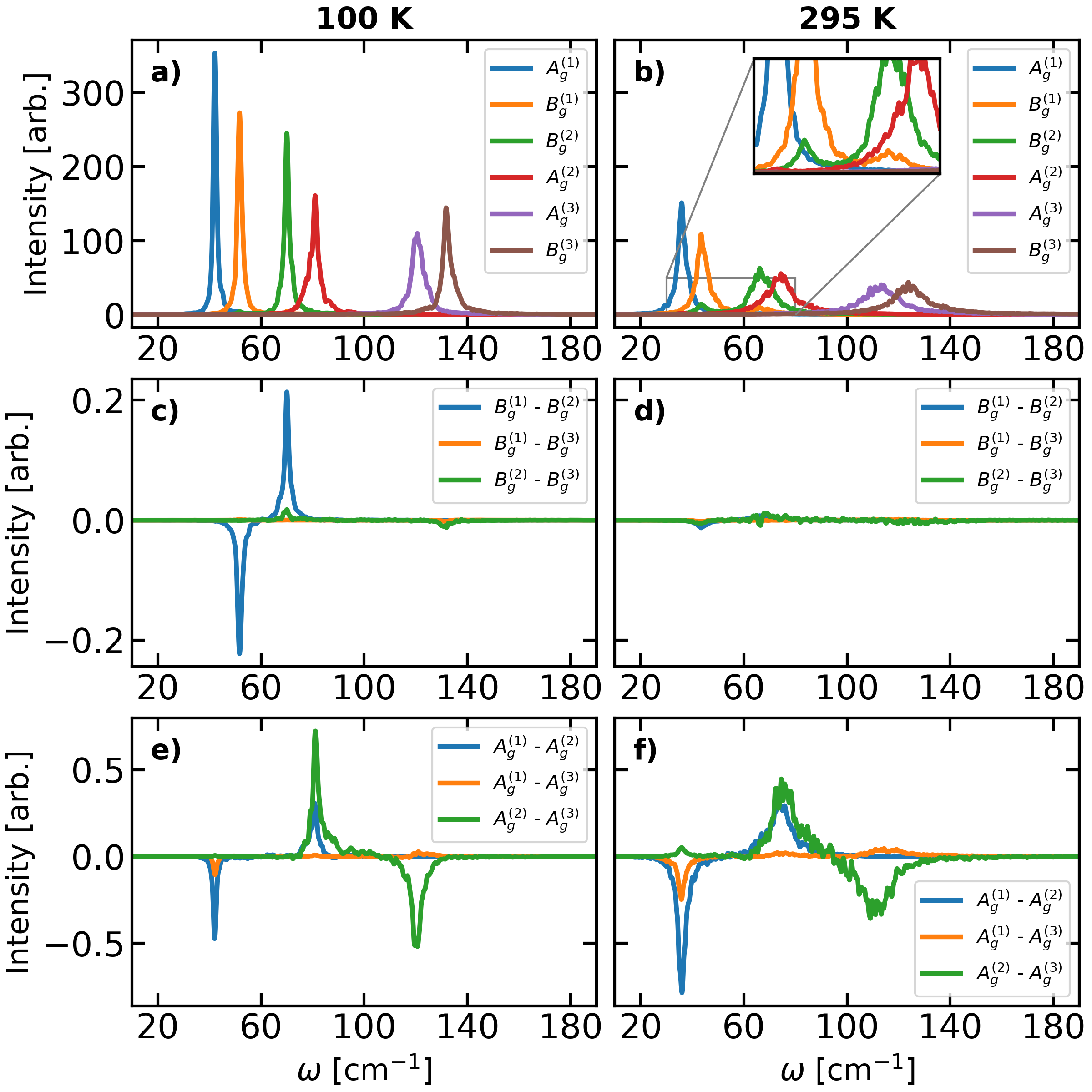}
  \caption{Vibrational density of states (VDOS) and cross-VDOS (CVDOS) computed in the normal mode basis for the intermolecular Raman active modes of anthracene ($\Gamma$ point only) at 100 K (left column) and 295 K (right column). Panels a) and b) show the VDOS projected on each mode, while panels c)-f) show the CVDOS between all pairs of $B_g$ and $A_g$ modes. Intensities are not normalized to give a better idea of the relative strength of the cross-correlations. The inset in panel b) zooms in on the spectral mixing between the $B_g^{(1)}$ and $B_g^{(2)}$ modes, as discussed in the text.}
  \label{fig:cvdos}
 \end{figure}

In order to address the role of anharmonicity in the vibrational dynamics of anthracene, we analyze the vibrational density of states (VDOS) and the cross-VDOS (CVDOS) computed in the normal mode basis for the $\Gamma$-point intermolecular modes. 
These quantities are defined in Eq.~\ref{eq:cvdos} of Section \ref{sec:methods}. The VDOS provides a measure of the distribution of vibrational frequencies, while the CVDOS captures correlations between different modes, revealing signatures of mode coupling. 

In Fig.~\ref{fig:cvdos} we show the VDOS and CVDOS at the $\Gamma$ point of anthracene at 100~K and 295~K. For completeness, we include the same figure for naphthalene in the SM~\footnotemark[1] (Section S10). Panels a-b display the mode-projected VDOS, whereas panels c–f report the CVDOS between all pairs of modes with the same symmetry. 

At low temperatures, the VDOS of anthracene shows clear Lorentzian-like peaks for each mode. The corresponding CVDOS shows hints of mode coupling between modes $B_g^{(1)}$-$B_g^{(2)}$, $A_g^{(1)}$-$A_g^{(2)}$ and $A_g^{(2)}$-$A_g^{(3)}$. The CVDOS of modes with different symmetry is zero and not shown. As the temperature increases, we observe the emergence of spectral weight in the VDOS of a given mode stemming from other modes. This is particularly evident for the $B_g^{(1)}$-$B_g^{(2)}$ case at 295 K (inset of Fig. \ref{fig:cvdos}b) and provides a clear indication of anharmonic mode mixing. Note how in the CVDOS of Fig. \ref{fig:cvdos}d, the cross-correlation intensity between these two modes vanishes going from 100 K to 295 K. Instead, the VDOS gains spectral weight for the $B_g^{(1)}$ and $B_g^{(2)}$ modes, signaling that the phonon basis is a less accurate representation of the vibrational dynamics at higher temperature. This observation can explain the attenuation of the PO-pattern of the $B_g^{(2)}$ mode observed in the simulated data of Fig.~\ref{fig:fitting_combined}.
The phenomenon of mode mixing has also been addressed in the analysis of \textit{Benshalom et al.}~\cite{benshalom_phononphonon_2023} mentioned at the beginning of the previous section. They showed how, instead of isolated Lorentzian peaks, the spectral functions acquire contributions from multiple modes, resulting in nontrivial intensity redistribution and the possibility of temperature-dependent changes in the PO-patterns at a given frequency.  

In addition, the broadening of the peaks in VDOS and CVDOS with increasing temperature, evident in Fig.~\ref{fig:cvdos}, makes it increasingly difficult to assign spectral features to individual modes when one only has access to the PO-Raman maps such as those shown in  Fig.~\ref{fig:tscan-RGDOS}.
This delicate interplay of overlap and mixing between different modes stresses the need for robust deconvolution procedures when analyzing temperature-dependent PO-Raman spectra, as both true anharmonic coupling and spectral overlap can affect the PO-patterns extracted from the fitting procedure of the global maps of Fig.~\ref{fig:tscan-RGDOS}.

This analysis suggests that, while genuine mode coupling like the one that emerges in Fig.~\ref{fig:cvdos} can induce real changes in the polarization dependence, the additional overlap of non-Lorentzian, broadened peaks can also lead to apparent intensity sharing between modes, complicating the interpretation of PO-pattern evolution with temperature. Modes like $B_g^{(1)}$ and $B_g^{(3)}$, which show a pronounced change in the PO-pattern in experiment (Fig.~\ref{fig:fitting_combined}b), both overlap substantially with an intense peak of a different symmetry: $A_g^{(2)}$ and $A_g^{(3)}$, respectively. Especially at polarization angles where one of the two overlapping features is in principle not present, any deviation of the other mode spectral feature from a perfect Lorentzian will lead to intensity sharing, thus potentially hiding real anharmonic coupling effects or giving rise to artificial contributions. 

Finally, we note how the $\Gamma$RGDOS method explained in Sec.~\ref{sec:rgdos} amounts to a weighting of the $\Gamma$-point VDOS and CVDOS with Raman tensors. This operation adds an intensity modulation that is angle-resolved, and can highlight or suppress mode coupling at different polarization angles. Therefore, one can also perform a mode decomposition of the PO-maps of Fig. \ref{fig:tscan-RGDOS} in terms of self- and cross-mode contributions at each polarization angle. We present in Section S12 of SM~\footnotemark[1] examples of this decomposition for anthracene at selected angles, at both temperatures. While such analysis provides a powerful tool to aid the deconvolution of experimental PO maps, the conclusions drawn from the analysis of the VDOS and CVDOS remain unaltered.

\section{Conclusions}

In this work, we have established a computational framework to investigate anharmonic effects in polarization-orientation Raman spectra of molecular crystals, applying it to naphthalene and anthracene. By combining machine learning models for interatomic potentials with the $\Gamma$RGDOS approximation, we achieve
\textit{ab initio} accuracy for very large supercells and systematically explore temperature-dependent effects. In particular, we showcased how focusing on the polarization dependence of Raman intensities provides a stringent test of the underlying theoretical framework.

Already at the quasi-harmonic level, we observe that when lattice expansion affects the nature of the normal modes, as in naphthalene, significant changes in the polarization patterns can occur, which are still rationalized within a simplified, second-rank Raman tensor formalism. 
On top of this, our simulations capture clear anharmonic signatures, including temperature-driven redshifts and peak broadening, which are strongly influenced by lattice thermal expansion. We also confirm that we capture anharmonic mode mixing and that nuclear quantum effects have negligible impact on PO patterns of low-frequency modes. While the polarization dependence of Raman intensities remains largely consistent with harmonic predictions, subtle deviations emerge, reflecting the interplay between anharmonic lattice dynamics and Raman tensor symmetry, complementing the phenomenological approach of Ref. \cite{benshalom_phononphonon_2023}. 

Notably, even though our approach captures the anharmonic effects on peak positions and intensities, the simulated PO-patterns show only modest temperature dependence compared to the pronounced changes observed experimentally in Ref. \cite{asher_anharmonic_2020}. A careful assessment shows that the extraction of PO intensity variations from the full 2D maps that can be gathered from experiments or simulations can be tricky because of the fitting procedure involved. In particular, the simple multi-Lorentzian fitting becomes increasingly ambiguous as temperature-induced peak broadening leads to intensity overlaps between different spectral features. These factors can obscure genuine dynamical effects and complicate a direct comparison between theory and experiment.

Based on the analysis we have presented in this paper, we can propose a way forward. The ability of the $\Gamma$RGDOS-ML framework to decompose the PO-Raman patterns into self- and cross-contributions from different vibrational modes could be used to direclty assign intensities to a given spectral feature in the simulated PO-Raman maps, avoiding the need to fit the entire pattern at once. While in this manuscript we chose to compare on equal grounds the global fit of both the simulation and experiment, this alternative first-principles decomposition could be used in the future to guide the experiments towards a well-informed fitting procedure that takes mode mixing and non-trivial spectral overlap into account \cite{benshalom_phononphonon_2023}.

It is important to emphasize that a perfect match between simulated and experimental PO-Raman patterns is not expected, as discrepancies already arise at the harmonic level due to the specific level of theory employed for the electronic structure and Raman tensor calculations.
Other limitations include the use of finite supercell sizes which impose a finite sampling of the Brillouin zone (here a $4 \times 4 \times 4$ $q$-grid), and exclude $q$-points in the close vicinity of the Brillouin zone center. While the large supercell still mitigates finite-size effects and indirectly affects the dynamics at the $\Gamma$-point, our analysis of the Raman signals is fundamentally restricted to the $\Gamma$-point itself, as we stressed from the beginning. We note that allowing for contributions from the vicinity of the $\Gamma$-point in Raman spectroscopy could strongly affect the mode-coupling picture, by breaking fundamental symmetry restrictions and allowing modes of different irreducible representations to couple~\cite{benshalom_phononphonon_2023}. These contributions are very challenging to assess by molecular dynamics alone, requiring very large system sizes.  

Nevertheless, our framework demonstrates that, starting from a given theoretical baseline, we can systematically include temperature and dynamical (anharmonic) effects on top of it, providing a predictive route to assess their impact on the PO-Raman response.
Overall, we showed that ML-driven simulations combined with established theoretical frameworks can unravel complex anharmonic phenomena in molecular crystals. The insights gained here advance the interpretation of PO-Raman spectra and provide a route for future improvements, including more robust deconvolution methods for analysis of 2-dimensional polarization orientation Raman patterns and larger-scale simulations to further bridge theory and experiment.

\section{Methods and computational Details}
\label{sec:methods}

\subsection{Training Machine Learning Models for Interatomic Interactions~\label{sec:ml-pots}}

 In this paper we investigate two types of MLIPs based on neural-network (NN) architectures. One architecture is that of high-dimensional neural-network potentials (HDNNP) \cite{behler_generalized_2007,behler_four_2021}, which uses two-body and three-body local atomic descriptors in the form of atom-centered symmetry functions. In this representation, there are only a few hidden layers and a few nodes on the NN attached to each atomic species are needed, and the construction of the potential is successful for quite complex systems \cite{omranpour_insights_2025, artrith_high_2011, artrith_high_2012, sosso_neural_2012, natarajan_neural_2016}. This architecture does not explicitly encode the equivariance of the transformation of each component of the symmetry functions. The target of the optimization of the NN parameters is the total energy of the system, constructed from the sum of the individual contributions of the atomic energies predicted by the NN, and forces are readily obtained by back-propagation (i.e., the model is conservative). We have referred to this model as HDNNP-MLIP in the main text.

The other architecture is the equivariant message-passing graph NN employed in the MACE model \cite{batatia_mace_2022,batatia_design_2025}. In this architecture, atoms are represented as nodes of a graph and the interactions between pairs of atoms are the edges. The messages are constructed as many body expansions where each term is approximated by a linear combination of a complete basis of equivariant features \cite{batzner_e3-equivariant_2022}. These messages are thus of high-body order and can be understood as an efficient evaluation of a many-body expansion on ACE descriptors \cite{drautz_atomic_2019}. Forces are obtained by automatic differentiation and the model is also conservative. The use of effective high body-order messages delivers highly accurate predictions. The imposition of equivariance reduces the complexity of the problem by incorporating the rotational properties of the representation, resulting in the need of less training data points and removing the need of explicit data augmentation. We have referred to this model as MACE-MLIP in the main text.

In order to construct the training dataset for anthracene, we used the experimentally available lattice parameters at temperatures of 100~K and 295~K from Ref. \cite{asher_anharmonic_2020} and performed a 2 ns NVT-MD run with the foundational model MACE-OFF \cite{kovacs_mace-off_2025} to generate trial structures. A total of 1200 configurations were selected through farthest point sampling and the corresponding total energies and forces were recomputed at the PBE \cite{perdew_generalized_1996} level of theory with many-body dispersion correction (MBD) \cite{tkatchenko_accurate_2009} using the FHI-aims \cite{blum_ab_2009} electronic structure package, with the recommended \textit{tight} basis-set.  Dispersion corrections are essential to capture the cohesive forces in molecular crystals and the MBD scheme has proven to be very accurate for structural properties of these systems \cite{hunnisett_seventh_2024}. Both of our models for anthracene were subsequently trained on this dataset, with 10\% of the structures left as test set. The accuracy of the models on training and test set is reported in Table \ref{tab:antml}.

\begin{table}[t]
\centering
\begin{ruledtabular}
\begin{tabular}{lcccc}
 & \multicolumn{2}{c}{Energy (meV/atom)} & \multicolumn{2}{c}{Force (meV/\AA)} \\
\hline
 & Training & Test & Training & Test \\
\hline
HDNNP-MLIP & 0.3 & 2.5 & 45.6 & 47.7 \\
MACE-MLIP  & 0.2 & 0.2 & 5.3  & 7.1  \\
\end{tabular}
\end{ruledtabular}
\caption{Training and validation errors as RMSE for energies and forces of MACE-MLIP and HDNNP-MLIP for anthracene.}
\label{tab:antml}
\end{table}

\begin{table}[t]
\centering
\begin{ruledtabular}
\begin{tabular}{lcccc}
 & \multicolumn{2}{c}{Energy (meV/atom)} & \multicolumn{2}{c}{Force (meV/\AA)} \\
\hline
 & Training & Test & Training & Test \\
\hline
MACE-MLIP & 0.1 & 0.1 & 2.6 & 3.4 \\
\end{tabular}
\end{ruledtabular}
\caption{Training and validation errors as RMSE for energies and forces of MACE-MLIP for naphthalene, reproduced from Ref.~\cite{gurlek_accurate_2025}.}
\label{tab:napml}
\end{table}

For naphthalene, we used the MACE potential decribed in Ref.\cite{gurlek_accurate_2025}. The underlying electronic-structure is also PBE+MBD with \textit{tight} basis-set. For any further detail on the training procedure, the reader is referred to Ref.\cite{gurlek_accurate_2025}. The errors on energies and forces are reported here (Table \ref{tab:napml}) for completeness.

\subsection{Training Procedure of the Machine Learning Polarizability Models}

The dataset for both polarizability models consisted in 1000 structures sampled from NVT MD simulations generated by our MACE-MLIP. The structures were sampled at 100~K, 150~K and 293~K for anthracene and 30~K, 80~K and~295 K for naphthalene, employing available experimental lattice constants from Refs.~\cite{asher_anharmonic_2020,capelli_molecular_2006}. The polarizability tensors were computed from DFPT as implemented in the FHI-aims code using the LDA functional, which has been shown in Ref.~\cite{Shang2018kf} to be accurate in capturing the polarizability variations that enter the Raman intensity calculation as long as the underlying dynamics is generated by a high quality potential. We also performed an explicit comparison of LDA vs PBE for Raman tensors in Section S5 of SM~\footnotemark[1].

The training set was split into 85\% training and 15\% validation sets. The MACE-$\alpha$ model was trained with the \textit{mu\_alpha} branch of the MACE code \cite{macerepo}, while the SA-GPR model was trained with a modified version of the TENSOAP code \cite{grisafi_symmetry-adapted_2018} available at \cite{alanTensoap}.

\subsection{Phonons calculations and geometry optimizations}

For the comparison of Section S3 of SM~\footnotemark[1] we computed the harmonic phonon frequencies for anthracene on the minimum energy atomic configuration with lattice vectors fixed to the experimental structure at 295K from Ref. \cite{asher_anharmonic_2020} with both our ML models and DFT. The geometry optimizations were separately performed with FHI-aims (PBE+MBD, \textit{tight} basis set) for the DFT case and with our two ML potentials. Phonon calculations were subsequently performed using the phonopy code \cite{togo_first-principles_2023,togo_implementation_2023}.

For the production simulations, energy minimization and phonon calculations needed to obtain the normal modes for the RGDOS projection coefficients of Section \ref{sec:rgdos} were performed with the i-PI \cite{litman_i-pi_2024} code, using the same MLIP as used in the dynamics as the force driver.

\subsection{Raman tensors for $\Gamma$-RGDOS}
The Raman tensors needed for the RGDOS expansion were obtained from finite differentiation of DFPT polarizabilities of displaced minimum energy structures along the normal coordinates. The procedure is repeated for every material at every different lattice parameters considered. The LDA functional was used for the benchmarks of Section \ref{sec:framework} for a direct comparison with the LDA-trained SA-GPR model, while the PBE functional was employed in the production calculations of the $\Gamma$-RGDOS spectra. All DFPT calculations employed \textit{tight} basis-sets. While our comparison of LDA vs PBE Raman tensors in Section S5 of SM~\footnotemark[1] shows nearly identical results, we opted for PBE in production calculations due to reduced computational cost of the RGDOS methodology, and to ensure full consistency with the PBE level of theory of the MLIPs employed. Section S5 of SM~\footnotemark[1] (see also additional Ref. \cite{togo_spglib_2024} therein) also includes a brief discussion on \textit{a posteriori} symmetryzation of numercially noisy Raman tensors.

A further benchmark unpolarized harmonic Raman spectra calculated at different levels of theory is provided for completeness in Section S13 of SM~\footnotemark[1]. We show how different DFT functionals (including hybrid-GGA) and dispersion corrections do not significantly alter Raman intensities in anthracene at the harmonic level.

\subsection{Production simulations for Raman spectra \label{sec:dynmethod}}
For the production runs, we performed 48 independent NVE-MD simulations of 100~ps each with a time step of 1~fs for both anthracene and naphthalene at temperatures of 100~K, 220~K and 295~K. For both crystals, we employed $4\times4\times4$ supercells to mitigate finite-size effects. This amounts to a total of 3072 atoms for anthracene and 2304 atoms for the naphthalene crystal. The convergence of the PO-Raman spectra with the supercell dimension is reported in Section S11 of SM~\footnotemark[1]. The NVE trajectories were generated from a 2~ns parent NVT trajectory where temperature was controlled with a stochastic velocity-rescaling thermostat \cite{bussi_canonical_2007} with a time constant of 100~fs. The polarizability tensors were computed on-the-fly every 10~fs from the RGDOS expansion Eq.~\ref{eq:rgdos1}, with the procedure to obtain projections coefficients described in Section S1 of SM~\footnotemark[1]. The Raman spectra were subsequently computed through numerical Fourier transformation with a Hanning windowing function applied to the time series. No additional broadening was introduced by the windowing process. 

\subsection{Fitting of PO-dependence}
Following Ref. \cite{asher_anharmonic_2020} we fitted the PO-dependence of the integrated intensity of every peak in the PO-maps, like the ones in Figs.~\ref{fig:tscan-RGDOS}, to a multi-Lorentzian function of the form
\begin{equation}
I(\omega) = \sum_i \frac{A_i \Gamma^2_i\omega_{0,i}\omega}{\omega^2 \Gamma_i^2 + \left(\omega^2-\omega^2_{0,i}\right)^2},
\label{eq:multilorentz}
\end{equation}
where $A_i$ is the amplitude, $\omega_{0,i}$ the position and 
$\Gamma_i$ the width of the $i$-th peak. The position and the width for each peak were determined from a first round of fitting at fixed polarization angles at which peaks were as clearly separated as possible. Once determined, under the well-founded assumption that the polarization angle does not influence the peak position and width but only the amplitude, we fitted the oscillating intensities.

\subsection{VDOS and CVDOS calculations}

The VDOS and CVDOS of Fig. \ref{fig:cvdos} were computed from Fourier transformation of the velocity autocorrelation function of the $\Gamma$-point projections $\tilde{v}_k$ with $k$ being a $\Gamma$-point normal mode, following the same procedure as Ref. \cite{carreras_dynaphopy_2017}. The final expression reads as 
\begin{equation}
C_{ij}(\omega) = \mathcal{F}\left\{\langle \tilde{v}_i(t)\tilde{v}_j(0) \rangle\right\},
\label{eq:cvdos}
\end{equation}
where $i=j$ gives the VDOS of mode $i$ and $i \neq j$ gives the CVDOS between modes $i$ and $j$. When defined like this, the VDOS is normalized to twice the vibrational kinetic energy of the system, $2\langle K \rangle$. The projection velocities $\tilde{v}_k$ are obtained through finite differentiation of normal mode projection coefficients $\tilde{q}_k$ along the MD trajectory. These are defined as 
\begin{equation}
\bm{u}(t) = \sum_k \tilde{Q}_k \tilde{q}_k (t),
\end{equation}
where $\bm{u}(t)$ are the atomic displacements at time $t$ and $\tilde{Q}_k$ is the cartesian normal mode $k$. In order to extract only the $\Gamma$-point projection coefficients, we follow the procedure we described in Section S1 of SM~\footnotemark[1]. No windowing was applied to the time series before Fourier transformation.

\section*{Author contributions}
P.L. and M.R. designed and conceptualized the research. S.S. and P.L. trained machine-learning models. P.L. ran simulations, conducted theoretical derivations, analyzed data and produced all figures in the manuscript. M.R. contributed to the data analysis the discussion. P.L. and M.R. wrote the first draft of the paper. All authors edited the draft.

\section*{Conflicts of interest}

The authors declare no conflicts of interest.

\section*{Data availability}

Data pertaining to this paper will be made available at \url{https://github.com/sabia-group/Paper-PO-Raman} upon publication.

\section*{Acknowledgements}

M.R. acknowledges funding by the European Union (ERC, QUADYMM, 101169761). P.L. and S.S. acknowledge support from the UFAST International Max Planck Research School. The authors are grateful to Nimrod Benshalom, Noam Pinsk, and Omer Yaffe for insightful discussions, valuable input, and for inspiring this work through their earlier experiments. We further acknowledge Nimrod Benshalom for his critical reading of the manuscript. We also thank Burak Gurlek, Elia Stocco, and George Trenins for helpful discussions.


 \footnotetext[1]{See Supplemental Material at [URL will be inserted by publisher] 
 for details on normal modes projections, supercell convergence, perpendicular configuration PO-maps, Raman tensors benchmarking and full harmonic PO-Raman results.}

\bibliography{main,extra} 

\end{document}


    \title{\textbf{Investigating Anharmonicities in Polarization-Orientation Raman Spectra of Acene Crystals with Machine Learning}\\SUPPLEMENTARY INFORMATION}

    \author{Paolo Lazzaroni}
    \affiliation{MPI for the Structure and Dynamics of Matter, Luruper Chaussee 149, 22761 Hamburg, Germany}
    
    \author{Shubham Sharma}
    \affiliation{MPI for the Structure and Dynamics of Matter, Luruper Chaussee 149, 22761 Hamburg, Germany}

    \author{Mariana Rossi}
    \affiliation{MPI for the Structure and Dynamics of Matter, Luruper Chaussee 149, 22761 Hamburg, Germany}

    \maketitle

   \section{RGDOS Normal Modes Projection and the $\Gamma$RGDOS Approximation}
        \label{sec:nmproj}

The mass-scaled Cartesian normal modes $\tilde{\mathbf{Q}}_i$ of the
supercell do not form an orthogonal basis. Therefore, care must be taken
when obtaining the projection coefficients $P_{k,\Gamma}$ for the RGDOS
approximation (Section~\ref{sec:rgdos} of the main text). Following
Ref.~\cite{berger_raman_2023}, these coefficients can be obtained from the
linear system
\begin{equation}
    \sum_i \tilde{\mathbf{Q}}_i \cdot \tilde{\mathbf{Q}}_j \, P_i
    = \tilde{\mathbf{Q}}_j \cdot \mathbf{u},
\end{equation}
with $\mathbf{u}$ the displacement vector of the $N_1 \times N_2 \times N_3 = N$ supercell.
Here, both $\mathbf{u}$ and the $\tilde{\mathbf{Q}}_i$ are vectors of
dimension $3MN$, where $M$ is the number of atoms in the unit (primitive) cell, and
$i$ runs over all supercell normal modes. Introducing the matrix
$\mathbf{A}$ whose columns are the $\tilde{\mathbf{Q}}_i$ (which we assume to be normalized), this can be
written in matrix form as
\begin{equation}
    \mathbf{A}^T \mathbf{A}\,\mathbf{P} = \mathbf{A}^T \mathbf{u}.
\end{equation}
Since the full set of supercell modes spans the $3MN$-dimensional
displacement space, $\mathbf{A}$ is square and (up to numerical
subtleties associated with the acoustic $\Gamma$-modes) invertible.
The projection coefficients may therefore be written explicitly as
\begin{equation}
    \mathbf{P} = (\mathbf{A}^T \mathbf{A})^{-1} \mathbf{A}^T \mathbf{u}.
\end{equation}
This expression, however, requires the knowledge of all supercell normal
modes and is therefore not the most practical.

For first-order Raman scattering, only the phonon modes pertaining to the $\Gamma$-point of the crystal primitive cell are relevant. The corresponding $N_1 \times N_2 \times N_3 = N$ supercell normal modes can thus be obtained
obtained by concatenating the primitive cell $\Gamma$-point eigenvectors $N$ times. Defining
$\mathbf{a}$ as the $3M \times 3M$ matrix of primitive-cell Cartesian normal modes, we
construct
\[
\tilde{\mathbf{A}} =
\begin{bmatrix}
\mathbf{a} \\
\mathbf{a} \\
\vdots \\
\mathbf{a}
\end{bmatrix},
\]
which has dimension $3MN \times 3M$. The projection onto the 
$\Gamma$-point modes is then
\begin{equation}
    \mathbf{P}_\Gamma
    = (\tilde{\mathbf{A}}^T \tilde{\mathbf{A}})^{-1}\tilde{\mathbf{A}}^T \mathbf{u}
    = N^{-1} (\mathbf{a}^T \mathbf{a})^{-1} \tilde{\mathbf{A}}^T \mathbf{u}.
\end{equation}
Since $\mathbf{a}$ is square and invertible,
$(\mathbf{a}^T \mathbf{a})^{-1} = \mathbf{a}^{-1}\mathbf{a}^{-T}$,
such that
\begin{equation}
    \mathbf{P}_\Gamma
    = N^{-1} \mathbf{a}^{-1} (\tilde{\mathbf{A}}\mathbf{a}^{-1})^T \mathbf{u}.
\end{equation}
Because $\tilde{\mathbf{A}}$ consists of $N$ copies of $\mathbf{a}$,
the product $\tilde{\mathbf{A}}\mathbf{a}^{-1}$ reduces to $N$ stacked
copies of the $3M \times 3M$ identity. The product of its transpose with $\bm{a}^{-1}$ is a horizontal concatenation of $N$ copies of the inverse of the primitive-cell $\Gamma$-point normal modes matrix. When multiplying the superecell displacement vector $\mathbf{u}$, this yields
\begin{equation}
    \mathbf{P}_\Gamma
    = N^{-1}
      \begin{bmatrix}
      \mathbf{a}^{-1} & \mathbf{a}^{-1} & \cdots & \mathbf{a}^{-1}
      \end{bmatrix}
      \begin{bmatrix}
      \mathbf{u}_1 \\ \mathbf{u}_2 \\ \vdots \\ \mathbf{u}_{N}
      \end{bmatrix}
    = N^{-1} \sum_{n=1}^{N} \mathbf{a}^{-1} \mathbf{u}_n,
\end{equation}
where $\mathbf{u}_n$ denotes the displacement vector of the $n$-th unit
cell in the supercell.

Thus, the projection of the supercell normal modes onto $\Gamma$-point modes of primitive cell can be built by separately projecting every constituent unit-cell. This requires only the inversion
of the unit-cell matrix $\mathbf{a}$. We name this approximation $\Gamma$RGDOS.

    \section{$\Gamma$RGDOS with different cutoffs of gamma modes}
        \label{sec:cutoffrg}

    \begin{figure*}[h!]
        \centering
        \includegraphics[width=0.72\textwidth]{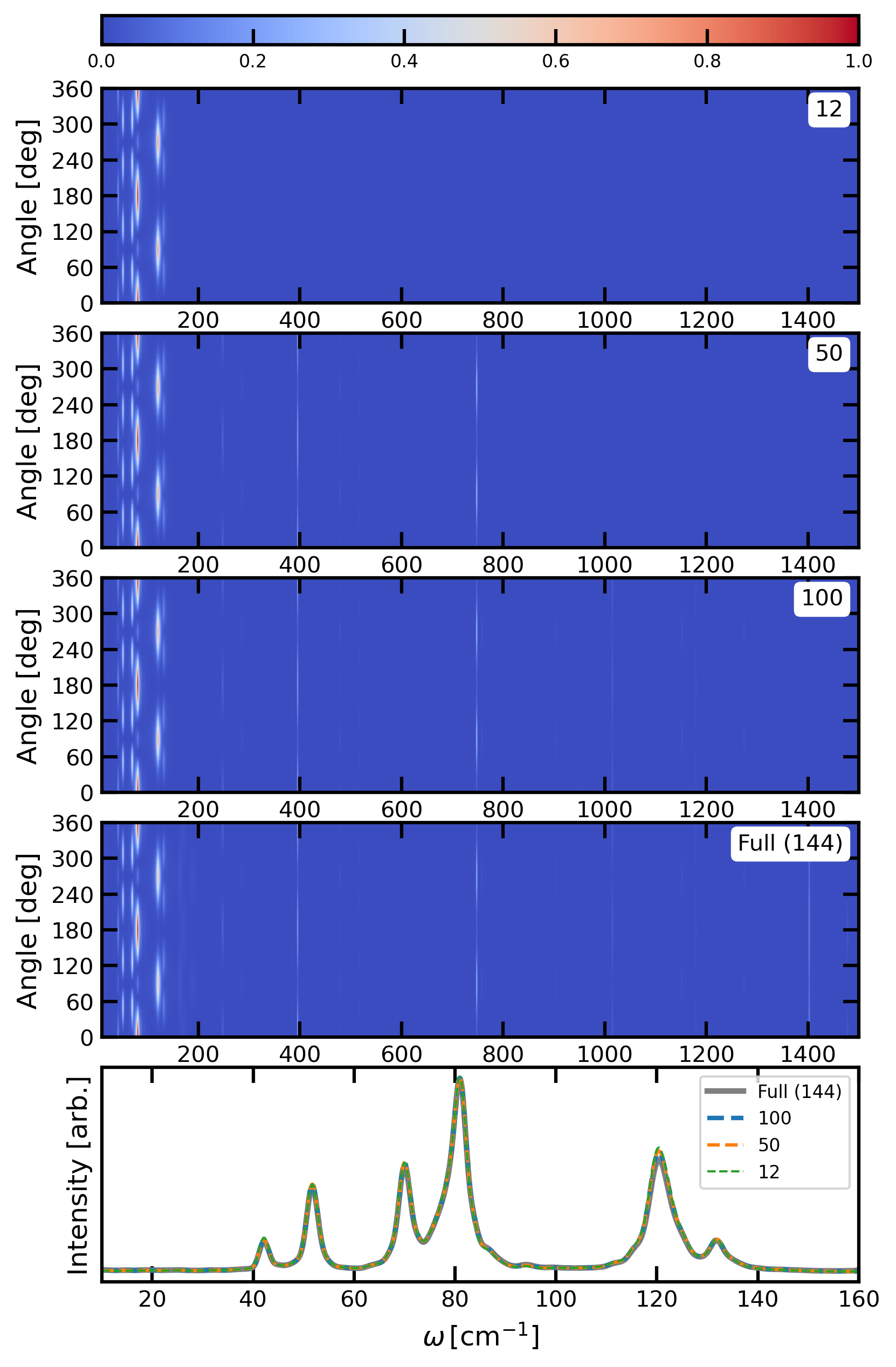}
        \caption{PO-Raman spectrum of Anthracene at 100~K in the parallel configuration considering different amounts of $\Gamma$-point modes for the $\Gamma$RGDOS-ML method (Eq. \ref{eq:rgdos1} main text). The number of modes considered, starting from the one corresponding to the lowest frequency, is shown in the panels. The bottom panel is the average parallel spectrum in the low-THz range, i.e., Eq. \ref{eq:unpolarized} (main text) with only the parallel component. The inclusion of more modes leads to the appearance of more peaks at higher and higher frequencies. However, the low-frequency region is unaltered by the cutoff choice, as long as the intermolecular modes are included.}
        \label{fgr:modescutoff}
       \end{figure*}

    \newpage

\section{Assessment of the Framework}
\label{sec:quality}

In the following, we will compare different models for MLIPs and polarizability tensors of polyacene molecular crystals. \textit{Berger et al.} have performed benchmarking of similar frameworks for different classes of inorganic materials\cite{berger_polarizability_2024}. Here, we apply this approach to the realm of molecular crystals and include a comparison with a state-of-the-art equivariant neural network model for polarizabilities, with the ultimate goal of obtaining \textit{ab initio} quality PO-Raman spectra.

\label{sec:ml}
\begin{figure}[ht!]
    \centering
    \includegraphics[width=0.5\linewidth]{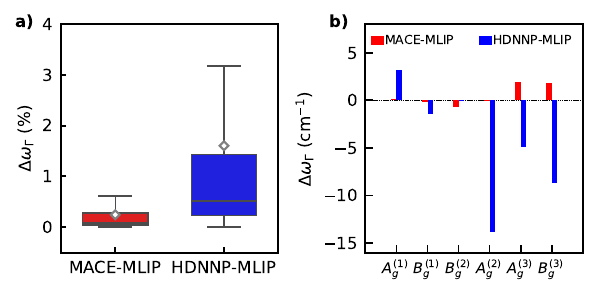}
    \caption{a) Comparison of percentage (\%) error on $\Gamma$-point harmonic phonon frequencies of anthracene for MACE-MLIP and HDNNP-MLIP with respect to DFT (PBE+MBD) reference values. b) Bar plot showing the absolute errors on $\Gamma$-point harmonic phonon frequencies for Raman active intermolecular modes of anthracene (labeled by  their symmetry) with MACE-MLIP and HDNNP-MLIP.}
    \label{fig:phonon-errors}
\end{figure}

\begin{figure*}[h]
    \centering
    \includegraphics[width=\linewidth]{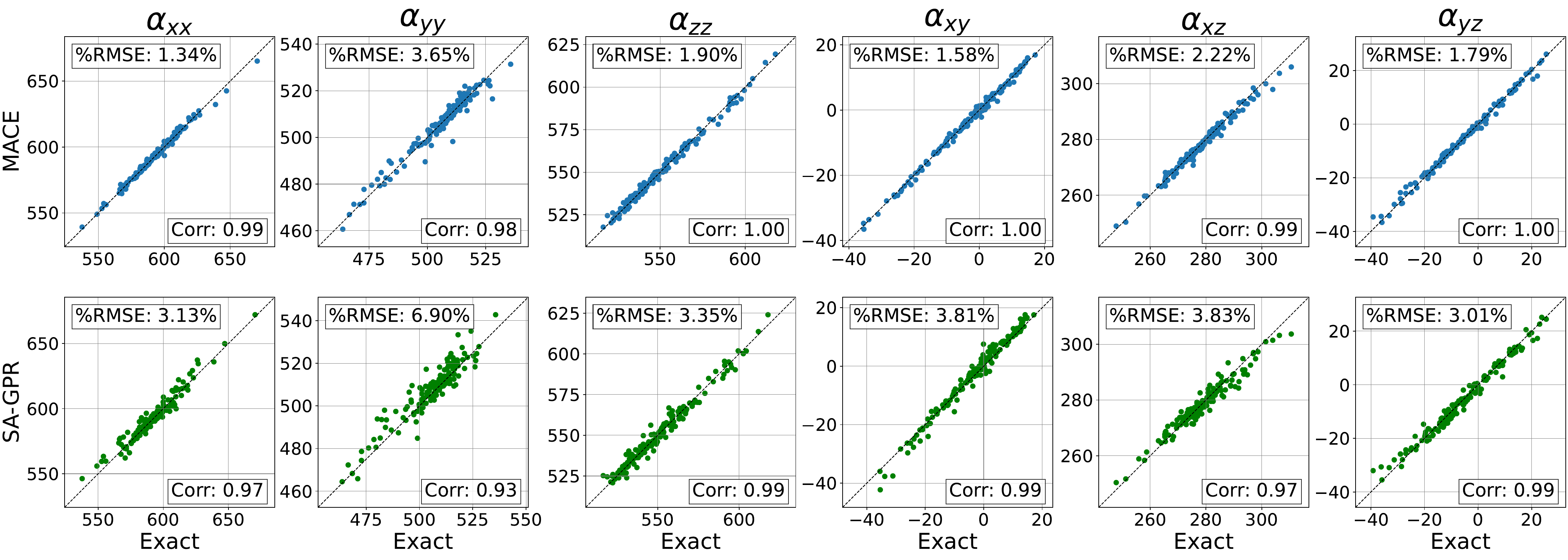}
    \caption{Correlation plots of the different components of the polarizability tensor as predicted by MACE-$\alpha$ (top row) and SA-GPR (bottom row) with respect to the DFPT values. Predictions are obtained from a test set of 150 independent structures sampled from configurations at three different temperatures (100 K, 150 K, 295 K) and different lattice parameters.}
    \label{fig:alpha-corr}
\end{figure*}

All the details regarding the generation of the training datasets, training procedures and level-of-theory employed throughout the manuscript are given in Methods Section in the main text.

Here, we start by assessing the quality of the MLIPs we trained for a crucial quantity in the simulation of Raman spectra, namely the frequencies of the crystal vibrations. Even though we will employ these potentials in dynamical simulations, we choose the harmonic phonons at the $\Gamma$-point of the Brillouin zone as a deterministic measure of the quality of our potentials, since they are free of any statistical uncertainties. As mentioned in the main text, we compare a second-generation Behler-Parrinello high-dimensional neural network potential~\cite{behler_generalized_2007}, which we label HDNNP-MLIP and a equivariant graph neural-network MACE~\cite{batatia_mace_2022} interatomic potential, which we label MACE-MLIP.  

We take the anthracene molecular crystal as a benchmark case for evaluating the quality of the machine-learned models for the PES. Details of the training dataset and procedure are given in Section \ref{sec:methods} (main text). As shown in Fig.~\ref{fig:phonon-errors}a MACE-MLIP outperforms the HDNNP-MLIP in the prediction of $\Gamma$-point harmonic phonon frequencies resulting in a lower average percentage error of $\approx$0.2\% against the $\approx$1.6\% obtained from HDNNP-MLIP. As mentioned in the main text, MACE-MLIP exhibits a significantly tighter distribution of errors, with a low median and minimal variability, highlighting the consistency and accuracy of this model. In contrast, HDNNP-MLIP   shows a broader error distribution, indicating less reliable predictions. 
The larger interquartile range and higher mean error for HDNNP-MLIP suggests the presence of higher error values, which could negatively impact dynamical simulations of vibrational properties, where broader regions of the PES are sampled. 
Note that the models have been trained on an identical dataset as already specified in the previous section. 

In Fig.~\ref{fig:alpha-corr} we show correlation plots for each component of the polarizability tensor of the anthracene crystal, comparing the predicted quantities with the calculated DFPT values over a separate test set of 150 structures (not used for training).

The data shown in Fig.~\ref{fig:alpha-corr} confirms that both methods are very accurate in predicting these tensors, as indeed previous work in the literature has shown \cite{raimbault_using_2019, berger_polarizability_2024, martin_general_2025, jana_learning_2024, kapil_first-principles_2024,grumet_delta_2024}.
However, we do  observe that for the same training dataset size, the MACE-$\alpha$ predictions are almost a factor two more accurate than the SA-GPR predictions for diagonal and off-diagonal components of the tensor. One could make the SA-GPR model more accurate by adding more training data. However, it is also important to highlight that MACE, as a parametric ML model, does not result in more expensive predictions when increasing the training data set, while SA-GPR, being a non-parametric model, does. This distinction makes MACE the most suitable solution for achieving higher prediction accuracy, especially when applied to larger datasets and system sizes.


  \newpage  
    \section{Harmonic PO-patterns of anthracene and naphthalene}
           \label{sec:harmonicrest}
    
\begin{figure*}[h!]
        \centering
        \includegraphics[width=0.62\textwidth]{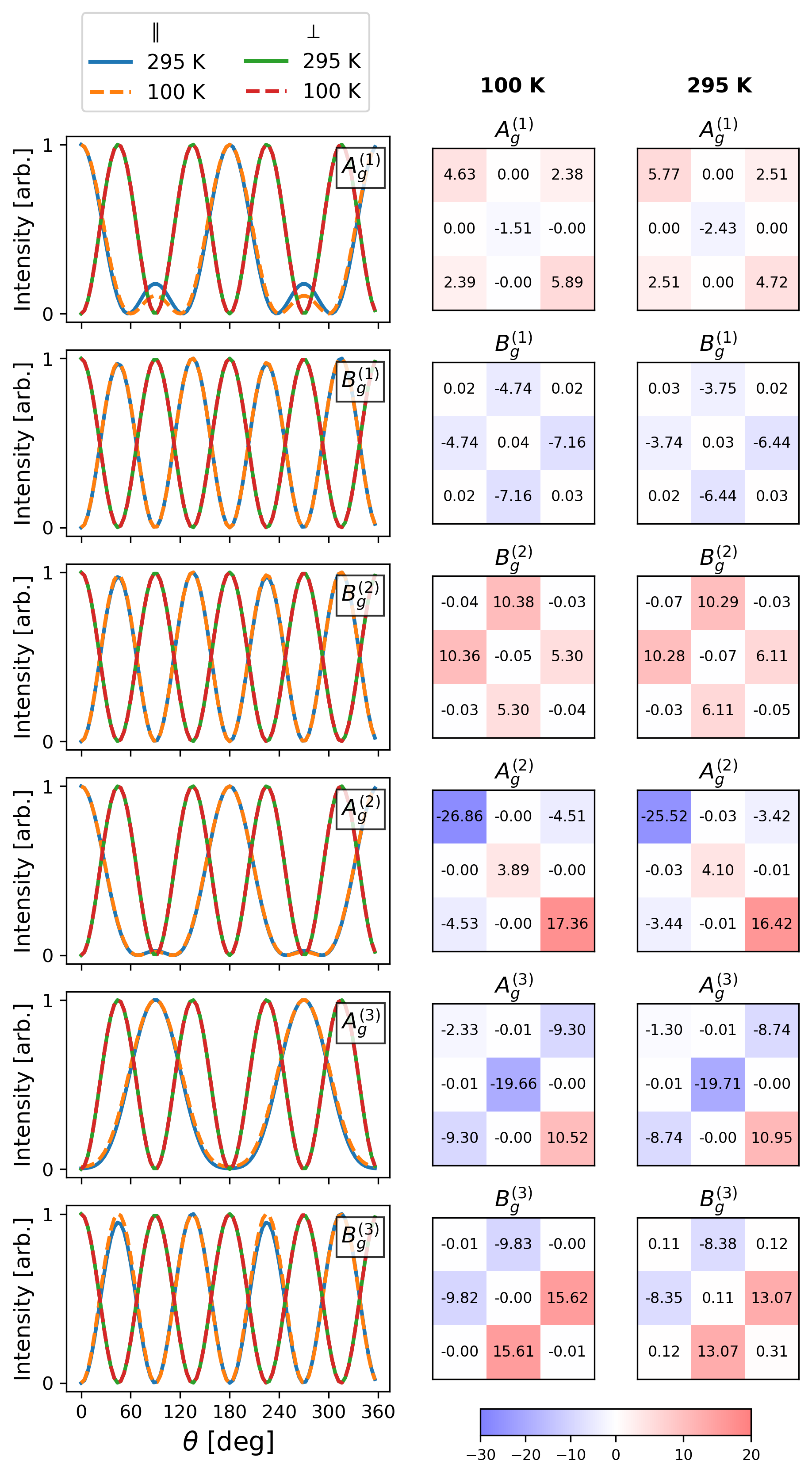}
        \caption{Harmonic polarization dependence of the Raman intensity in anthracene, obtained with the harmonic approximation of Eq. \ref{eq:harm} (main text) in both parallel and perpendicular configurations. On the right, the corresponding Raman tensors for every mode computed with finite differences from DFPT (PBE) polarizabilites along normal mode displacements of the MACE-MLIP at the 100~K and 295~K lattice vectors.}
        \label{fgr:harmperp}
       \end{figure*}

\begin{figure*}[h!]
        \centering
        \includegraphics[width=0.62\textwidth]{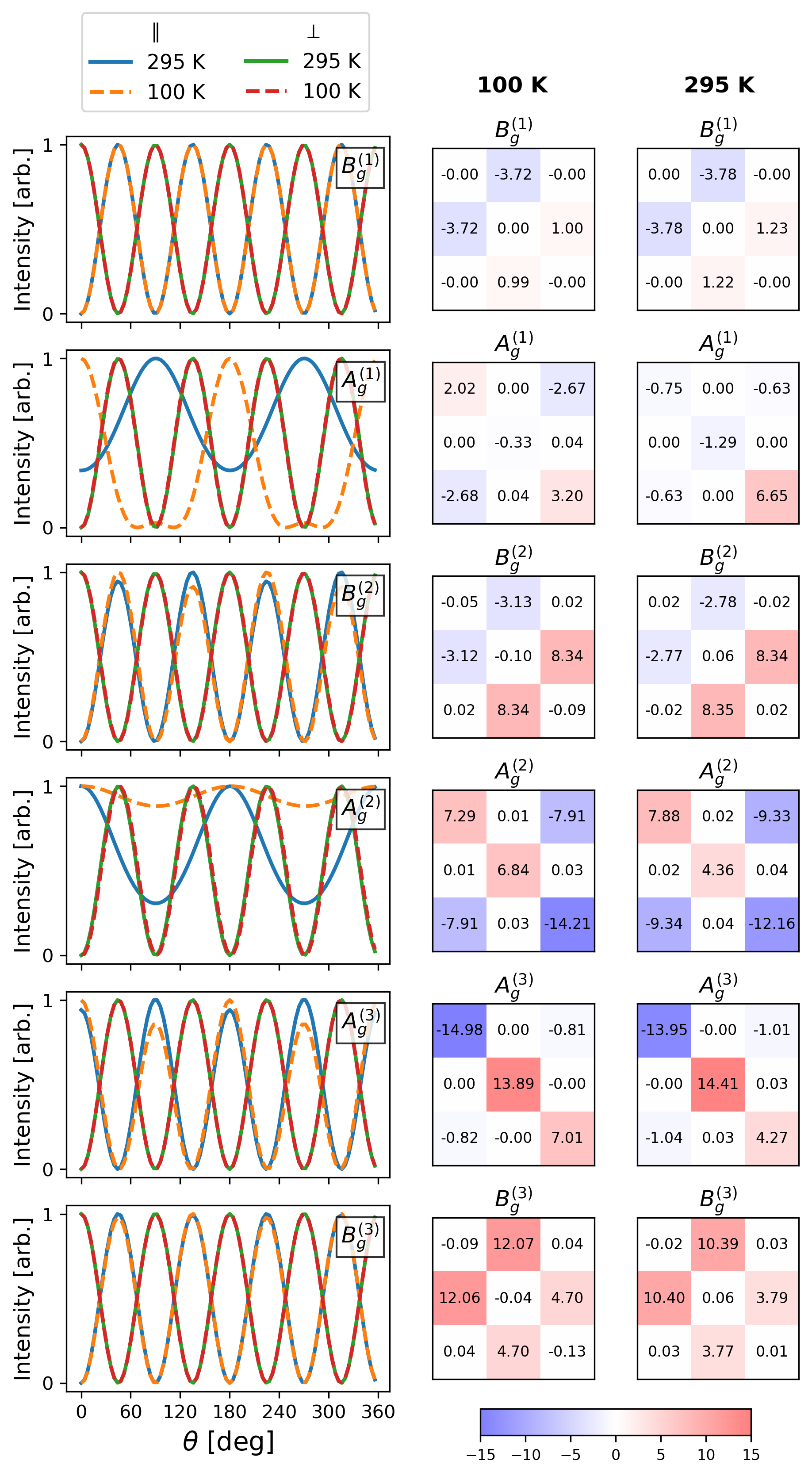}
        \caption{Harmonic polarization dependence of the Raman intensity in naphthalene, obtained with the harmonic approximation of Eq. \ref{eq:harm} (main text) in both parallel and perpendicular configuration. On the right, the corresponding Raman tensors for every mode computed with finite differences from DFPT (PBE) polarizabilites along normal mode displacements of the MACE-MLIP at the 100~K and 295~K lattice vectors. }
        \label{fgr:harmnaph}
       \end{figure*}

    In the following, we report the harmonic PO-Raman patterns of anthracene (Fig. \ref{fgr:harmperp}) and naphthalene (Fig. \ref{fgr:harmnaph}) in both parallel and perpendicular configurations at 100~K and 295~K, obtained for all six Raman active intermolecular modes with the harmonic approximation 
    \begin{equation}
I_k(\theta) \propto \left| \bm{e}_1\cdot\bm{\mathcal{R}}_k\cdot\bm{e}_2\right|^2
\label{eq:harmsi}.
\end{equation}
    Some simple algebra can be used to gauge the angular dependence of every mode in every configuration. In our configuration, the (001) face of the crystal is exposed, and only the $xy$ components of the response are relevant. The polarization vectors $\bm{e}_1$ and $\bm{e}_2$ are defined as
\[\bm{e}_1 =
\begin{bmatrix} \cos(\theta) \\ \sin(\theta) \\ 0 \end{bmatrix};
\quad       
\bm{e}_2 = \bm{e}_1 \quad \text{(parallel configuration)};
\quad
\bm{e}_2 =
\begin{bmatrix} -\sin(\theta) \\ \cos(\theta) \\ 0 \end{bmatrix}
\quad \text{(perpendicular configuration)} 
\]
where $\theta$ is the angle between the incident polarization and an arbitrary reference in the $xy$ plane. With these definitions and using the convention for the tensor components presented in Section~\ref{sec:RamanTensors}, one can easily obtain the angular dependence of every mode in both configurations:
\begin{equation}
I_{A_g}^{\parallel}(\theta) \propto |a \cos^2(\theta) + b \sin^2(\theta)|^2
\quad
I_{A_g}^{\perp}(\theta) \propto \left|\frac{b-a}{2}\sin(2\theta)\right|^2
\label{eq:nodep}
\end{equation}
\[
I_{B_g}^{\parallel}(\theta) \propto |e \sin(2\theta)|^2
\quad
I_{B_g}^{\perp}(\theta) \propto |e \cos(2\theta)|^2
\]
From these expressions, one can see that, with the exception of $I_{A_g}^{\parallel}$, the tensor components (or their difference) enter only as an overall multiplicative factor. Once the intensity patterns are normalized, varying the tensor components has no effect on their angular dependence. This already means that, if a different structure is considered (e.g. at a different temperature accounting for thermal lattice expansion), the PO-patterns of all $B_g$ modes in both configurations and of $A_g$ modes in the perpendicular configuration will remain unchanged within the harmonic approach. The only exception is $I_{A_g}^{\parallel}$, which can show different patterns depending on the relative values of $a$ and $b$. We have reported one example of this in the main text for the $A_g^{(1)}$ mode of both crystals. 

Here, we expand this analysis to all modes of both crystals and both configurations. In the case of anthracene (Fig.~\ref{fgr:harmperp}) we find that all modes have the same PO-pattern at 100~K and 295~K in both configurations, with small variations of $A_g^{(1)}$ in the parallel configuration, as already discussed in the main text. In naphthalene (Fig.~\ref{fgr:harmnaph}), we find that $A_g^{(1)}$ and $A_g^{(2)}$ change their PO-pattern significantly in the parallel configuration when going from 100~K to 295~K. For this analysis, all normal modes were dully symmetrized as described in Section~\ref{sec:RamanTensors}. A small change is observed also in the $A_g^{(3)}$ mode. As discussed above, this is a result of the fact that, for these modes, the relative values of $a$ and $b$ change significantly with temperature. In particular, for $A_g^{(1)}$ the $xx$ and $yy$ tensor components flip sign at 295~K, giving rise to a qualitatively different pattern. In order to give better insight on this effect, we performed a projection analysis of the normal modes at 100~K and 295~K, which is reported in Table~\ref{tab:modes}. These normal modes are the ones obtained from the MACE-MLIP potential as described in the main text, for both crystals.

To quantify the similarity between vibrational modes at different temperatures, we computed the normalized overlap between two displacement vectors as  
\[
O_{ij} = \left| \frac{\mathbf{u}_i^{(T_1)} \cdot \mathbf{u}_j^{(T_2)}}{\|\mathbf{u}_i^{(T_1)}\| \, \|\mathbf{u}_j^{(T_2)}\|} \right| ,
\]
where $\mathbf{u}_i^{(T_1)}$ is the displacement vector of mode $i$ at temperature $T_1$ and $\mathbf{u}_j^{(T_2)}$ is the displacement vector of mode $j$ at temperature $T_2$.  
From the overlap, we define an effective angle  
\[
\theta_{ij} = \arccos \!\left( O_{ij} \right),
\]
which gives a geometric measure of the deviation between the two mode vectors.  
In addition, we considered the residual vector  
\[
\mathbf{r}_{ij} = \mathbf{u}_i^{(T_1)} - 
\frac{\mathbf{u}_i^{(T_1)} \cdot \mathbf{u}_j^{(T_2)}}{\|\mathbf{u}_j^{(T_2)}\|^2}\,\mathbf{u}_j^{(T_2)},
\]
which captures the part of $\mathbf{u}_i^{(T_1)}$ that does not overlap with $\mathbf{u}_j^{(T_2)}$.  
The Cartesian contributions $x_{\mathrm{frac}}, y_{\mathrm{frac}}, z_{\mathrm{frac}}$ reported in the tables correspond to the normalized squared components of $\mathbf{r}_{ij}$ along each Cartesian axis. 

From Table \ref{tab:modes}, we see that all modes of anthracene have a very high overlap between 100~K and 295~K (the different temperatures here only regard the different lattice constants), with values above 0.997. The effective angles are all below 5 degrees. This confirms that the normal modes of anthracene are essentially unchanged when going from 100~K to 295~K, in agreement with the fact that the harmonic PO-patterns remain unchanged for all modes in both configurations. In the case of naphthalene, we see that the $B_g$ modes also have a very high overlap above 0.998 and small effective angles within 1-5 degrees. The $A_g$ modes, instead, show a lower overlap, especially $A_g^{(1)}$ and $A_g^{(2)}$, which have an overlap of 0.972 and 0.964 respectively, and larger effective angles of 13.57 and 15.35 degrees. This indicates that these two modes change more significantly when going from 100~K to 295~K, in agreement with the fact that their harmonic PO-patterns change significantly in the parallel configuration. The residual vector analysis shows that the changes occur mostly along the $x$ direction for $A_g^{(1)}$ and along the $y$ direction for $A_g^{(2)}$. This observation correlates with the fact that we see the largest change in the $xx$ (for $A_g^{(1)}$) and $yy$ (for $A_g^{(2)}$) components of their respective Raman tensors (see Fig.~\ref{fgr:harmnaph}).

\begin{table}[h]
\centering
\caption{Mode overlaps between 100~K and 295~K for anthracene and naphthalene. 
The overlap $O_{ij}$ quantifies the similarity between phonon eigenvectors, 
$\theta_{ij}$ gives the generalized angle in degrees, and 
$x_{\mathrm{frac}}, y_{\mathrm{frac}}, z_{\mathrm{frac}}$ are the normalized Cartesian components 
of the residual vector between the 100~K mode and its projection onto the corresponding 295~K mode.}
\renewcommand{\arraystretch}{1.2} 
\begin{tabular}{l c c c c c}
\hline
\multicolumn{6}{c}{\textbf{Anthracene}} \\
\hline
Mode & Overlap & Angle (deg) & $x_{\mathrm{frac}}$ & $y_{\mathrm{frac}}$ & $z_{\mathrm{frac}}$ \\
\hline
$A_{g}^{(1)}$ & 0.998 & 3.19 & 0.367 & 0.071 & 0.561 \\
$B_{g}^{(1)}$ & 0.998 & 4.01 & 0.252 & 0.317 & 0.431 \\
$B_{g}^{(2)}$ & 0.997 & 4.42 & 0.274 & 0.536 & 0.190 \\
$A_{g}^{(2)}$ & 0.998 & 3.48 & 0.023 & 0.808 & 0.169 \\
$A_{g}^{(3)}$ & 1.000 & 1.35 & 0.220 & 0.480 & 0.300 \\
$B_{g}^{(3)}$ & 1.000 & 1.39 & 0.217 & 0.498 & 0.285 \\
\hline\hline
\multicolumn{6}{c}{\textbf{Naphthalene}} \\
\hline
Mode & Overlap & Angle (deg) & $x_{\mathrm{frac}}$ & $y_{\mathrm{frac}}$ & $z_{\mathrm{frac}}$ \\
\hline
$B_{g}^{(1)}$ & 0.999 & 3.10 & 0.521 & 0.190 & 0.288 \\
$A_{g}^{(1)}$ & 0.972 & 13.57 & 0.784 & 0.069 & 0.147 \\
$B_{g}^{(2)}$ & 0.998 & 3.59 & 0.619 & 0.096 & 0.284 \\
$A_{g}^{(2)}$ & 0.964 & 15.35 & 0.294 & 0.644 & 0.062 \\
$A_{g}^{(3)}$ & 0.989 & 8.66 & 0.592 & 0.084 & 0.324 \\
$B_{g}^{(3)}$ & 0.998 & 3.83 & 0.095 & 0.216 & 0.689 \\
\hline
\end{tabular}
\label{tab:modes}
\end{table}

\newpage

\newpage
    \section{Benchmarking Raman tensors}
           \label{sec:RamanTensors}

                       \begin{figure*}[h!]
        \centering
        \includegraphics[width=0.62\textwidth]{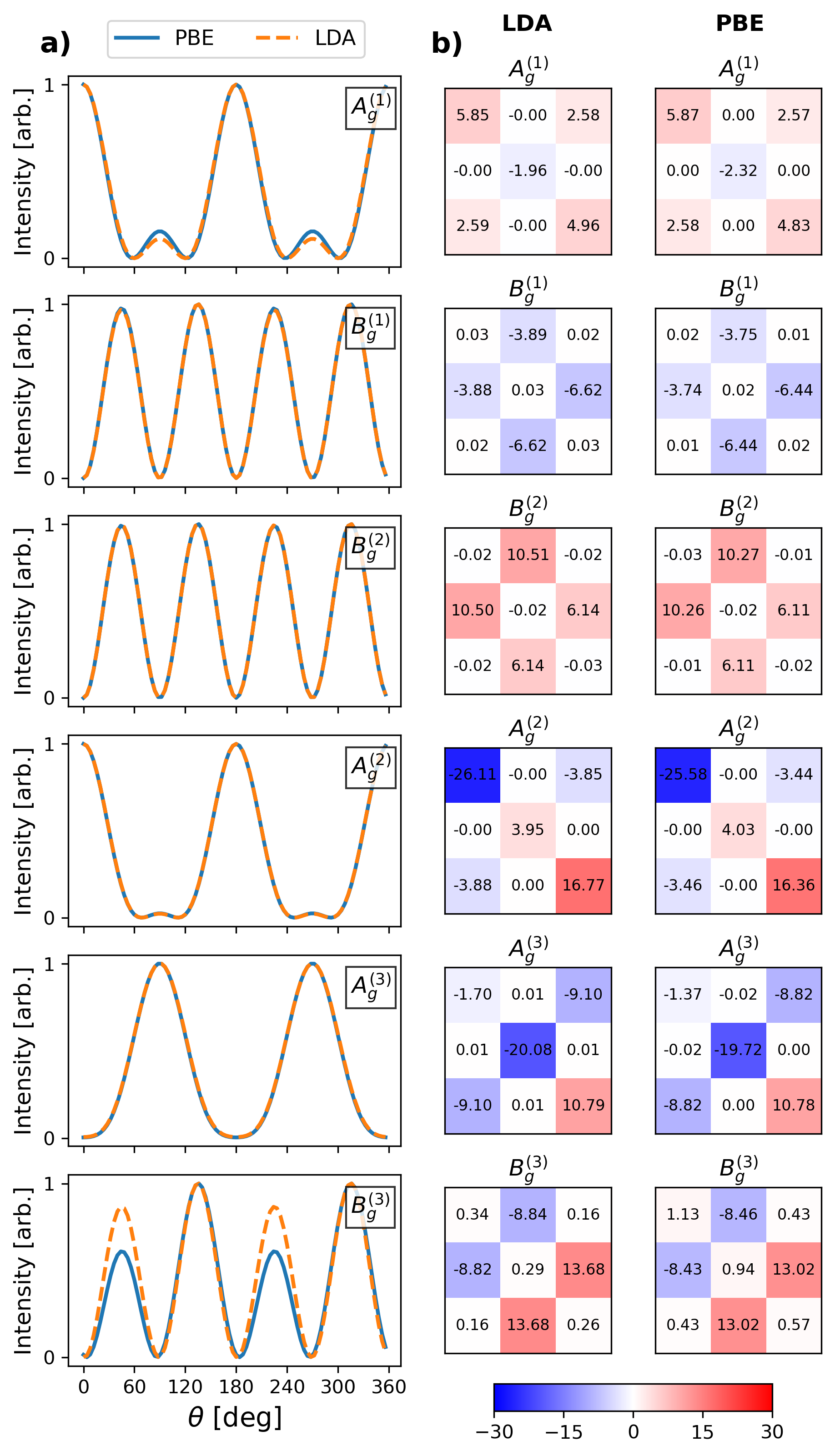}
        \caption{Harmonic polarization dependence of the Raman intensity in anthracene (295~K lattice vectors), obtained with Eq.~\ref{eq:harm} (main text) in the parallel configuration. Panel (a) shows a comparison of the polarization patterns for the LDA and PBE functionals. Panel (b) shows the corresponding Raman tensors obtained directly from finite differences of polarizability tensors calculated from density-functional perturbation theory (DFPT) along normal mode displacements of the MACE-MLIP at the 295~K lattice vectors, using LDA (left) and PBE (right) functionals. Both functionals yield very similar tensors and patterns. 
        }
        \label{fgr:ldavspbe}
       \end{figure*}

    We benchmark and analyze the Raman tensors of the intermolecular modes of anthracene obtained with finite differences from DFPT polarizabilities along normal mode displacements of the MACE-MLIP at the 295~K lattice vectors. In order to obtain the Raman tensors, a fully converged geometry optimization was performed with the MACE-MLIP. The same potential was used to perform a phonon calculation with the i-PI code~\cite{litman_i-pi_2024}. 

    Naphthalene and anthracene both crystallize in the $P2_1 / a$ space group with the two molecules occupying the same sites in both crystals. The Raman tensors of the Raman active modes are of $A_g$ and $B_g$ irreducible representations and have the general shape 
\[
A_g =
\begin{bmatrix}
a & 0 & d \\
0 & b & 0 \\    
d & 0 & c
\end{bmatrix}
\quad
B_g = 
\begin{bmatrix}
0 & e & 0 \\
e & 0 & f \\
0 & f & 0
\end{bmatrix}
\]
where $a,b,c,d,e,f$ are the independent tensor components.
    
    First, we show the Raman tensors obtained with this procedure for both LDA and PBE functionals (Fig.~\ref{fgr:ldavspbe}). The two functionals yield very similar Raman tensors, with only slight differences in the relative intensities of the tensor components, confirming that LDA is still a robust choice for the description of the polarizability in these systems, as previously shown in Ref. \cite{raimbault_anharmonic_2019}. 
    Notice how the $B_g^{(3)}$ Raman tensor has, in both cases, some non zero components along the diagonal, which should be strictly zero if the corresponding normal mode had a numerically exact $B_g$ symmetry. This shows how, even when thoroughly converged, normal modes obtained by ML potentials are not perfect, and can still present some small mixing between different irreps. This discrepancies were quantitatively addressed in Ref.~\cite{gurlek_accurate_2025}. 
    
    In particular, Raman tensors are a very sensitive probe of the correct symmetry of the atomic displacements. The effect of what looks like a small mixing between different irreps is amplified by the resulting harmonic PO-Raman intensity and its oscillation pattern. In this specific case, the $B_g^{(3)}$ mode symmetry is correctly predicted at the 100~K lattice vectors, but some diagonal components appear for the 295~K lattice vectors. Even if small, these inaccuracies completely alter the PO-Raman pattern. If left unnoticed, such an effect could appear as a modification of the PO-dependence due to the higher temperature, while in reality this is just an artifact due to the numerical noise that distorts the symmetry of the $B_g^{(3)}$ mode at 295~K.

               \begin{figure*}[h!]
        \centering
        \includegraphics[width=0.78\textwidth]{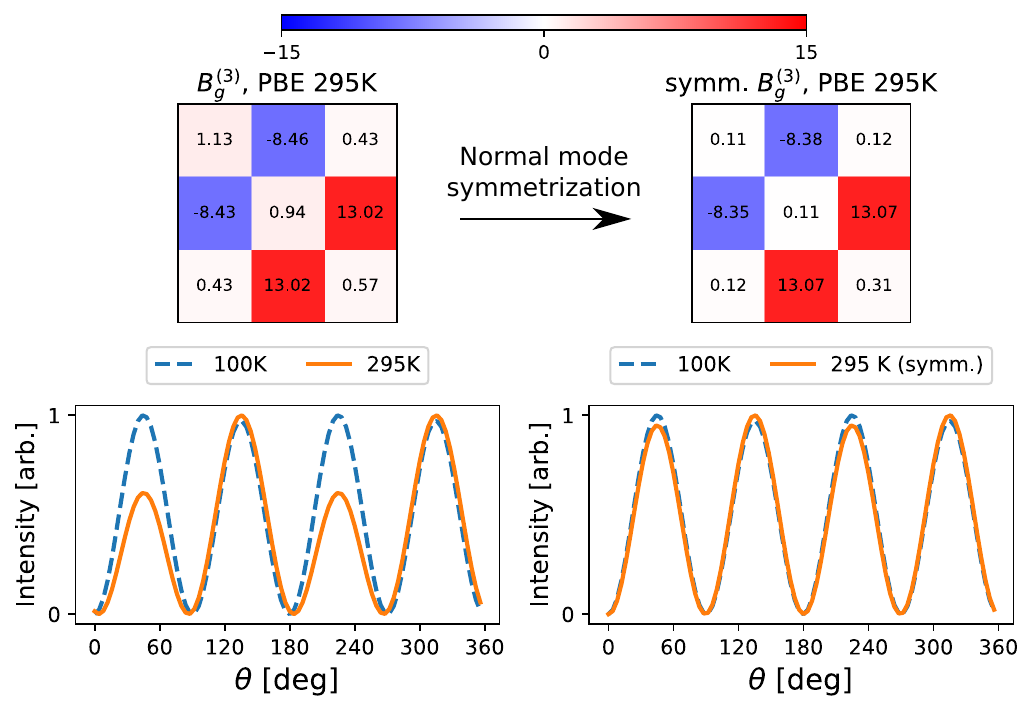}
        \caption{ On the left, Raman tensor of the $B_g^{(3)}$ mode of anthracene at 295~K lattice vectors obtained with PBE (top) and its corresponding PO-pattern (bottom). On the right, the same Raman tensor recomputed after symmetrization of the $B_g^{(3)}$ normal mode and its corresponding PO-pattern (bottom). The symmetrization procedure significantly reduces the small diagonal components of the Raman tensor, which should be zero by symmetry. The resulting PO-pattern is in much better agreement with the one obtained at 100~K.}
        \label{fgr:symmpbe}
       \end{figure*}

              The knowledge of factor group analysis of the correct irrep of each Raman active mode together with the point-group of the crystal allows one to enforce the correct symmetries on the normal modes by projecting them with the correct irrep, thereby symmetrizing the Raman tensors. This can be done with a library such as \texttt{spglib}~\cite{togo_spglib_2024}. In Fig.~\ref{fgr:symmpbe}, we show the effect of this procedure on the $B_g^{(3)}$ of anthracene. Once correctly projected on the $B_g$ irrep, the mode shows almost zero diagonal components. The resulting harmonic PO-Raman pattern is shown in the bottom right panel, and it is now in much closer agreement with the one obtained at 100~K. This is a clear example of how sensitive PO-Raman spectra are to the quality of the PES, the resulting normal modes and their Raman tensors.

\newpage
    \section{PO-Raman of anthracene and naphthalene: perpendicular configuration}
        \label{sec:perpconf}

       \begin{figure*}[h!]
        \centering
        \includegraphics[width=\textwidth]{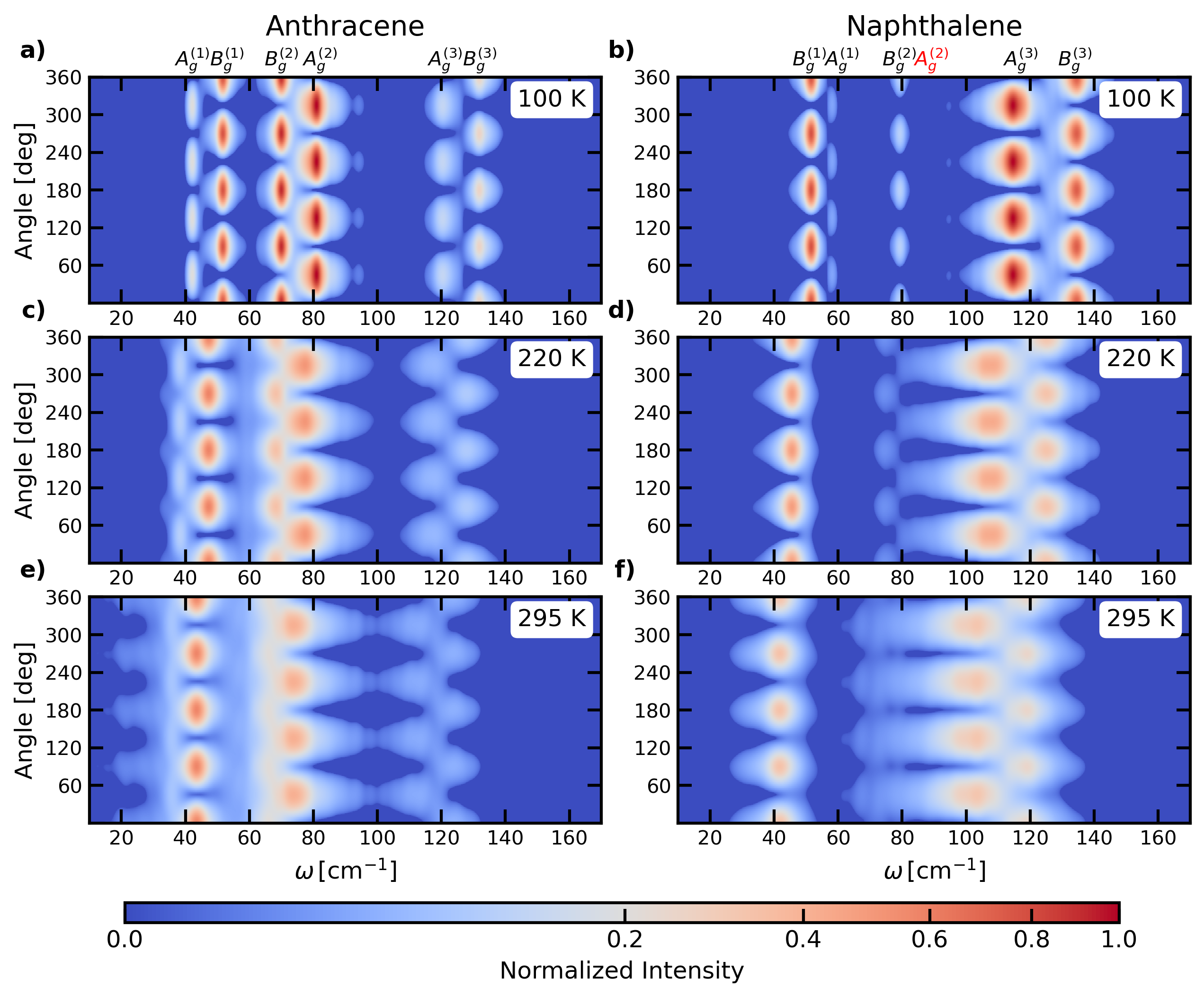}
        \caption{Temperature dependence of the  PO-Raman spectra of Anthracene (a,c,e) and Naphthalene (c,d,f) at several temperatures, in the perpendicular configuration and as obtained by the $\Gamma$RGDOS-ML method. The intensity is normalized to the maximum intensity of the 100~K spectrum. Peak position of mode $A_g^{(2)}$ of naphthalene is marked in red, as it shows no intensity in the perpendicular configuration.}
        \label{fgr:tscan_perp}
       \end{figure*}

    \newpage
    \section{PO-Raman of anthracene: temperature evolution without thermal lattice expansion}
           \label{sec:sameTdiffC}

    \begin{figure*}[h!]
        \centering
        \includegraphics[width=0.8\textwidth]{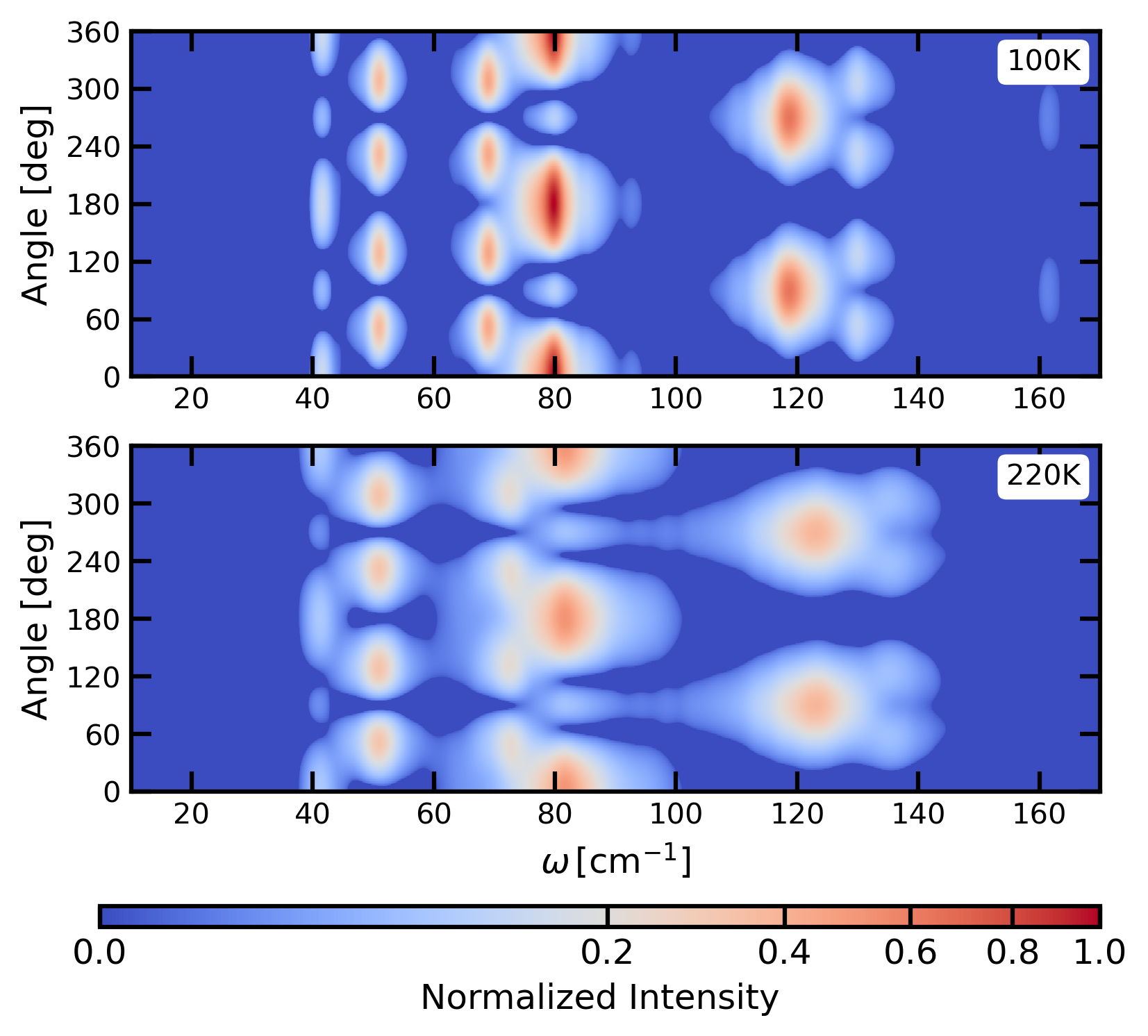}
        \caption{PO-Raman spectrum of Anthracene obtained with the $\Gamma$RGDOS-ML procedure (molecular dynamics) at 100~K and 220~K in the parallel configuration, without considering the thermal expansion of the lattice vectors. The lattice parameters corresponding to 100~K are used for both temperatures. The intensity is normalized to the highest value of the 100~K spectrum. While the peak broadening is still present, the peak positions do not shift as much as in the case where thermal expansion is considered (see Fig.~\ref{fig:tscan-RGDOS} in the main text). Notably, apart from the two lowest frequency modes, all the other peaks even show a slight blue shift at higher temperatures. The anharmonicity of the potential energy surface alone is not sufficient to explain the experimental red-shift of the peaks with increasing temperature, when thermal expansion is not included.}
        \label{fgr:nothermexp}
       \end{figure*}
       \newpage

\newpage
\section{Accounting for nuclear quantum effects in PO-Raman spectra}\label{sisec:nqe}

\begin{figure*}[h!]
  \centering
\includegraphics[width=0.98\textwidth]{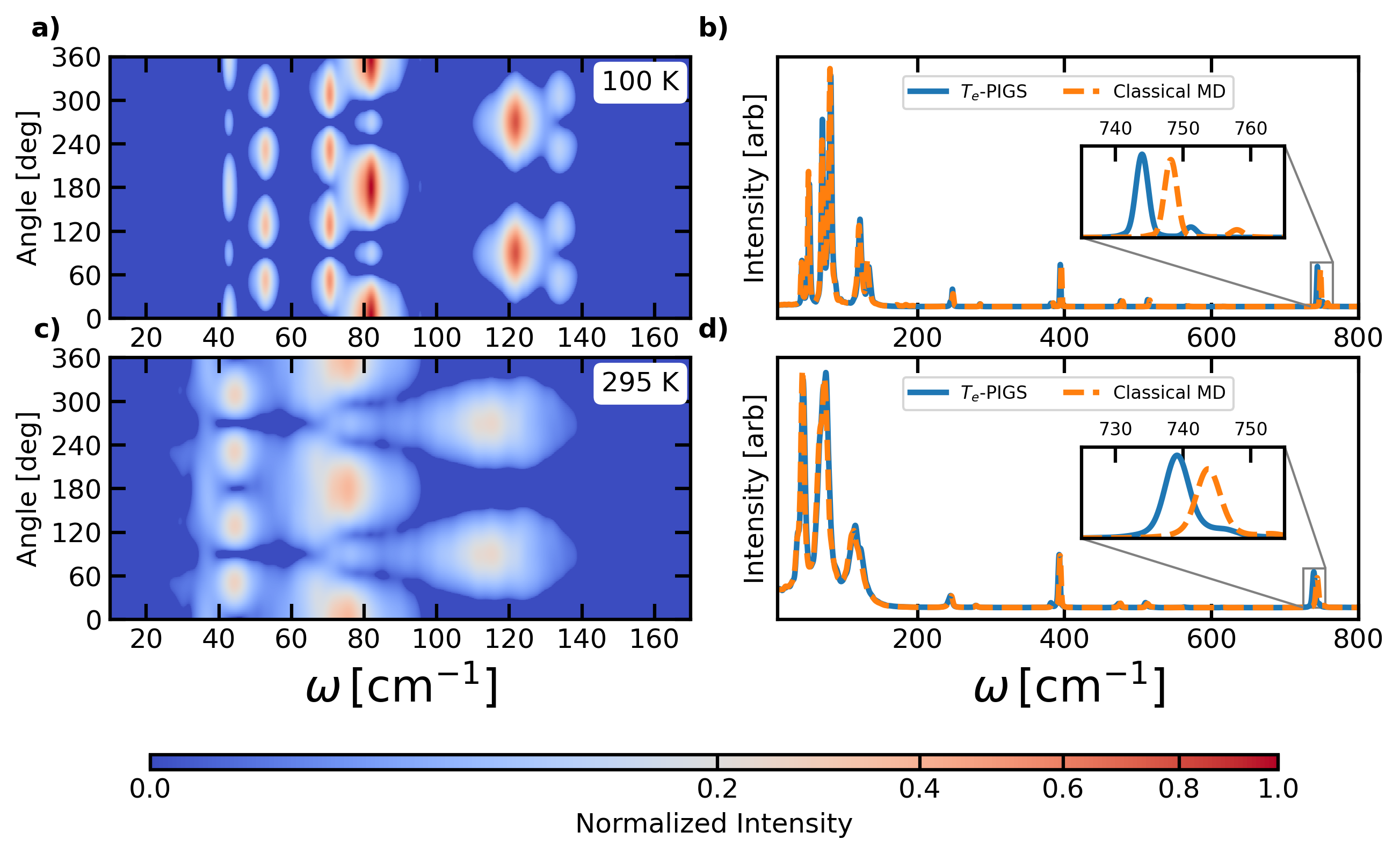}
  \caption{a-c) PO-Raman spectra of anthracene at 100~K and 295~K obtained with the $T_e$-PIGS method~\cite{musil_quantum_2022} to include nuclear quantum effects (NQEs). Intensity is normalized to the highest value at 100 K. b-d) Unpolarized Raman spectra of anthracene at 100 K and 295 K obtained with classical MD and $T_e$-PIGS. Zoom-in on the region around around 750 cm$^{-1}$ of the unpolarized Raman spectra, showing the redshift of the peaks due to NQEs. Minor intensity differences visible between classical and $T_e$-PIGS spectra at low frequencies arise from statistical sampling variations between independent simulations and are not attributable to systematic NQE contributions.}
  \label{fig:nqes}
 \end{figure*}

Low-frequency vibrations such as the ones we discuss in the main text are typically well-modeled by classical nuclear motion. The reason is that thermal energy is larger than the energy-spacing between any two vibrational levels of these modes, even at low temperatures ($100$~K $\approx 69.5$~cm$^{-1}$), making classical Boltzmann statistics a good approximation in these cases. In addition, the effective mass of the modes is also large, making classical dynamics equally appropriate at the temperatures (100 to 295~K) and frequency-ranges (20 to 150 cm$^{-1}$) we are concerned with. 

However, mode coupling between these low-frequency intermolecular vibrations and higher frequency modes at the region of $\approx 300$~cm$^{-1}$ are known to be present in these crystals \cite{seiler_nuclear_2021,kowalski_theory_2023}. Because these higher-frequency modes are more sensitive to nuclear quantum effects, this coupling can lead to non-trivial temperature-dependence that is sensitive to the quantum nature of the nuclei \cite{litman_temperature_2020,rossi_progress_2021}.

To investigate this, we trained a path-integral coarse-grained (PIGS) MACE MLIP, following Refs.\cite{wang_machine_2019,musil_quantum_2022}. The elevated temperature was chosen to be 500~K. We performed ``elevated-temperature'' centroid molecular dynamics ($T_e$-CMD)~\cite{musil_quantum_2022,castro2025} simulations for the PO Raman spectrum of anthracene at 100~K and 295~K using the $T_e$-PIGS method~\cite{musil_quantum_2022}. This method involves performing classical dynamics on a ML potential fitted by force-matching techniques to the centroid potential of mean force at a given elevated temperature. The results are shown in Fig. \ref{fig:nqes}. Comparing the PO-maps in Fig.~\ref{fig:nqes}a-c with the ones presented in Fig. \ref{fig:tscan-RGDOS}a-e (main text), we conclude that the polarization patterns in the intermolecular vibrations region are unaltered by nuclear quantum effects. At higher frequencies, changes are observed. Comparing classical and quantum unpolarized spectra, as shown in Fig. \ref{fig:nqes}b-d, a typical red-shift induced by the inclusion of zero-point motion is observed already at the lower-frequency intramolecular vibrations around 750 cm$^{-1}$ (highlighted in the insets of panels b-d). Importantly, temperature dependent changes of the low-frequency modes are not altered by the inclusion of NQE. We conclude that classical-nuclei simulations are sufficient to obtain anharmonic PO-Raman signals of these molecular crystals at the low-frequency range.

        \newpage
    \section{Birefringence correction for the harmonic PO-Raman}
        \label{sec:biri}

           \begin{figure*}[h!]
        \centering
        \includegraphics[width=0.8\textwidth]{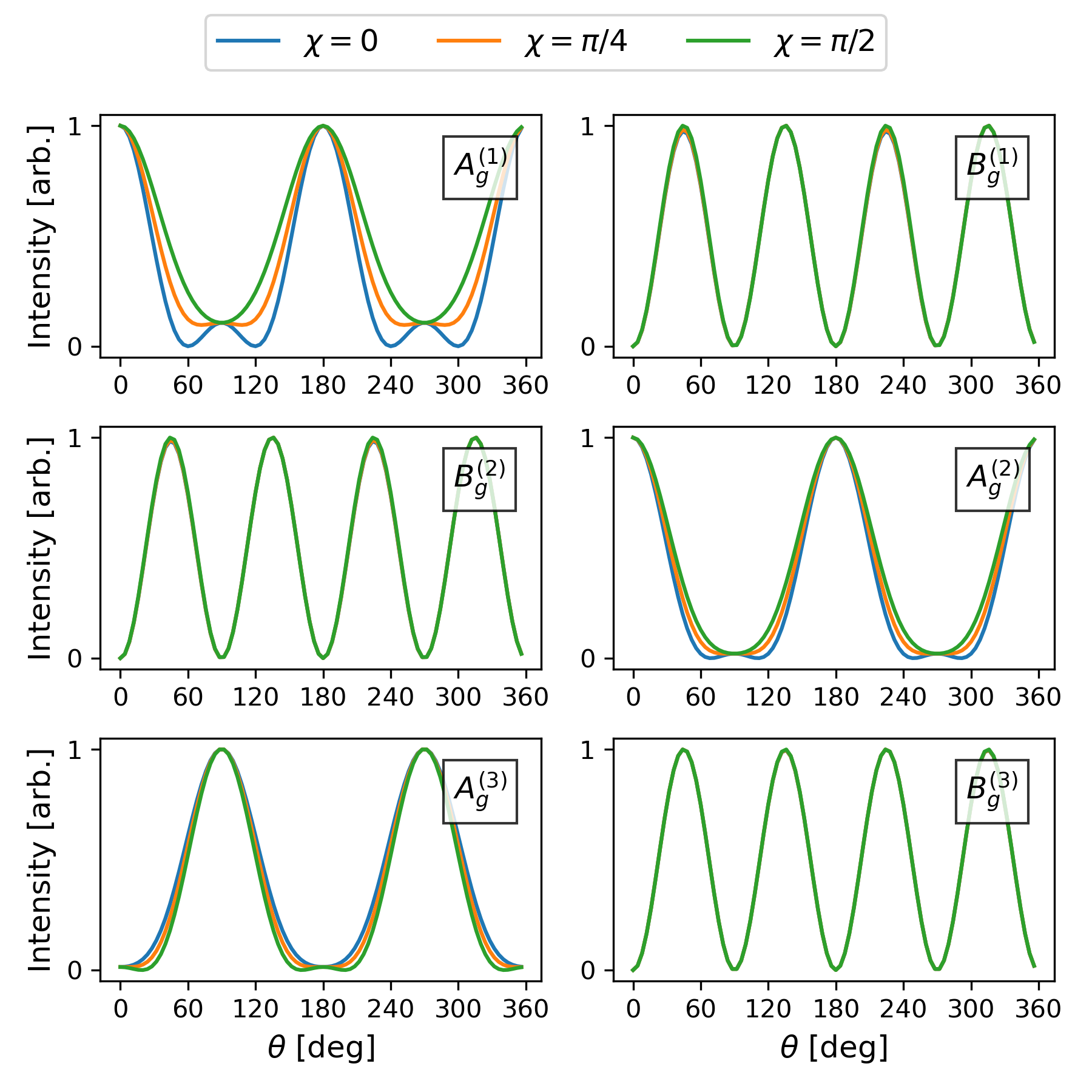}
        \caption{Birifringence correction effect on the harmonic PO-Raman patterns of anthracene in the parallel configuration. The correction is in the form of Jones matrices for different phase shifts. The Jones matrices $\bm{J}(\chi)$ enter the harmonic Eq.~\ref{eq:harm} of the main text as $I \propto \sum_k \rho^{\beta}_k \left| \bm{e}_1\cdot\bm{J}\bm{\mathcal{R}}_k\bm{J}\cdot\bm{e}_2\right|^2$ where $\chi$ is the phase shift. The Jones matrix formalism for polarized Raman scattering in anisotropic materials is thoroughly described in Ref. \cite{kranert_raman_2016}. Reasonable values for the phase shifts were taken from the fits of the experimental data in Ref. \cite{asher_anharmonic_2020}. These plots show that the introduction of a birefringence phase-shift can alter the PO-Raman patterns to some extent. In the present case, only the $A_g$ modes are affected. While this is an important observation, these effects are not sufficient to explain the discrepancies between theory and experiment, e.g. a different number of maxima in the $A_g^{(3)}$ mode in the parallel configuration, as discussed in the main text.}
        \label{fgr:biri}
       \end{figure*}

\clearpage
\section{Analysis of the Fitting Procedure of PO-maps of Naphthalene}
\label{sec:naphtavdos}

 \begin{figure}[h!]
  \centering
\includegraphics[width=0.65\textwidth]{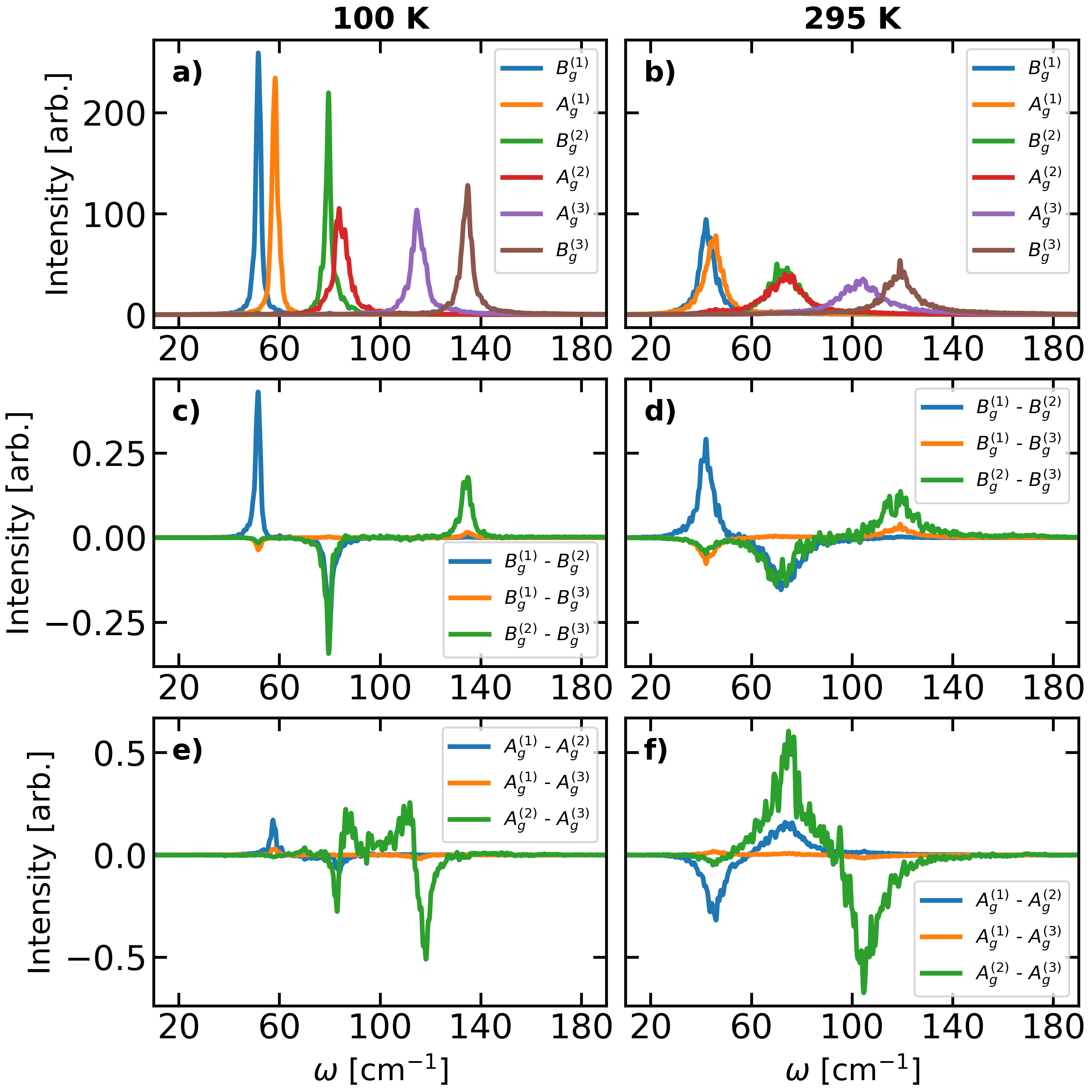}
  \caption{Vibrational density of states (VDOS) and cross-VDOS (CVDOS) computed in the normal mode basis for the intermolecular Raman active modes of naphthalene ($\Gamma$ point only) at 100 K (left column) and 295 K (right column). Panels a) and b) show the VDOS projected on each mode, while panels c)-f) show the CVDOS between all pairs of $B_g$ and $A_g$ modes. Cross correlations between modes of different symmetries are numerically zero and not shown. Intensities are not normalized to give a better idea of the relative strength of the cross-correlations. The strong overlap between the $B_g^{(2)}$ and $A_g^{(2)}$ modes is evident at 295 K, and it is the reason why the fitting procedure of the PO-maps (Fig. \ref{fig:fitnaph}) is not able to extract a reliable polarization pattern for these two modes, even at 220 K. Panels e) and f) shows qualitative differences in the CVDOS at the frequencies of the $A_g^{(1)}$ and $A_g^{(2)}$ modes, highlighting once again the different nature of these modes across the two temperatures.}
  \label{fgr:cvdosnaph}
 \end{figure}

 \begin{figure}[h!]
  \centering
\includegraphics[width=0.65\textwidth]{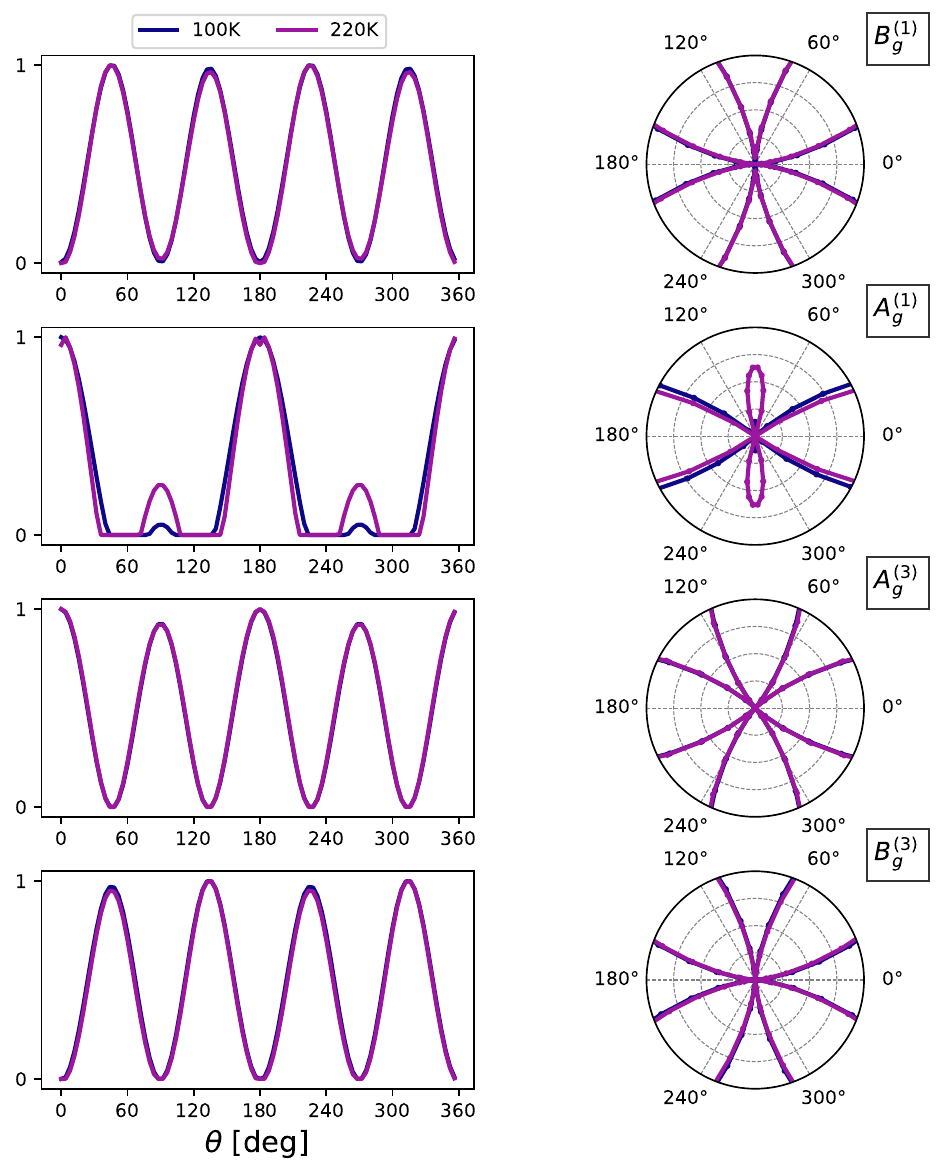}
  \caption{Polarization dependence of the integrated intensity of low THz Raman peaks in naphthalene for the parallel configuration fitted with Eq.~\ref{eq:multilorentz} (main text) from the simulated PO-maps of Fig.~\ref{fig:tscan-RGDOS} (main text) obtained with the $\Gamma$RGDOS-ML framework at 100 K and 220 K. Intensity is normalized with respect to the highest intensity of every peak at the corresponding temperature. Due to the strong overlap of the $B_g^{(2)}$ and $A_g^{(2)}$ modes, as shown in Fig. \ref{fgr:cvdosnaph}, the fitting procedure is not able to reliably extract a physical PO-pattern for those mode even at 220 K. The problem is even more pronounced at 295 K. The variation of the pattern for the $A_g^{(1)}$ with temperature is consistent with the harmonic case analyzed in Sec. \ref{sec:harmonicrest}, no additional dynamical effect is found in this case.}
  \label{fig:fitnaph}
 \end{figure}

 \clearpage
    \section{Finite-size effects and supercell size convergence}
        \label{sec:finitesize}

    \begin{figure*}[h!]
        \centering
        \includegraphics[width=0.8\textwidth]{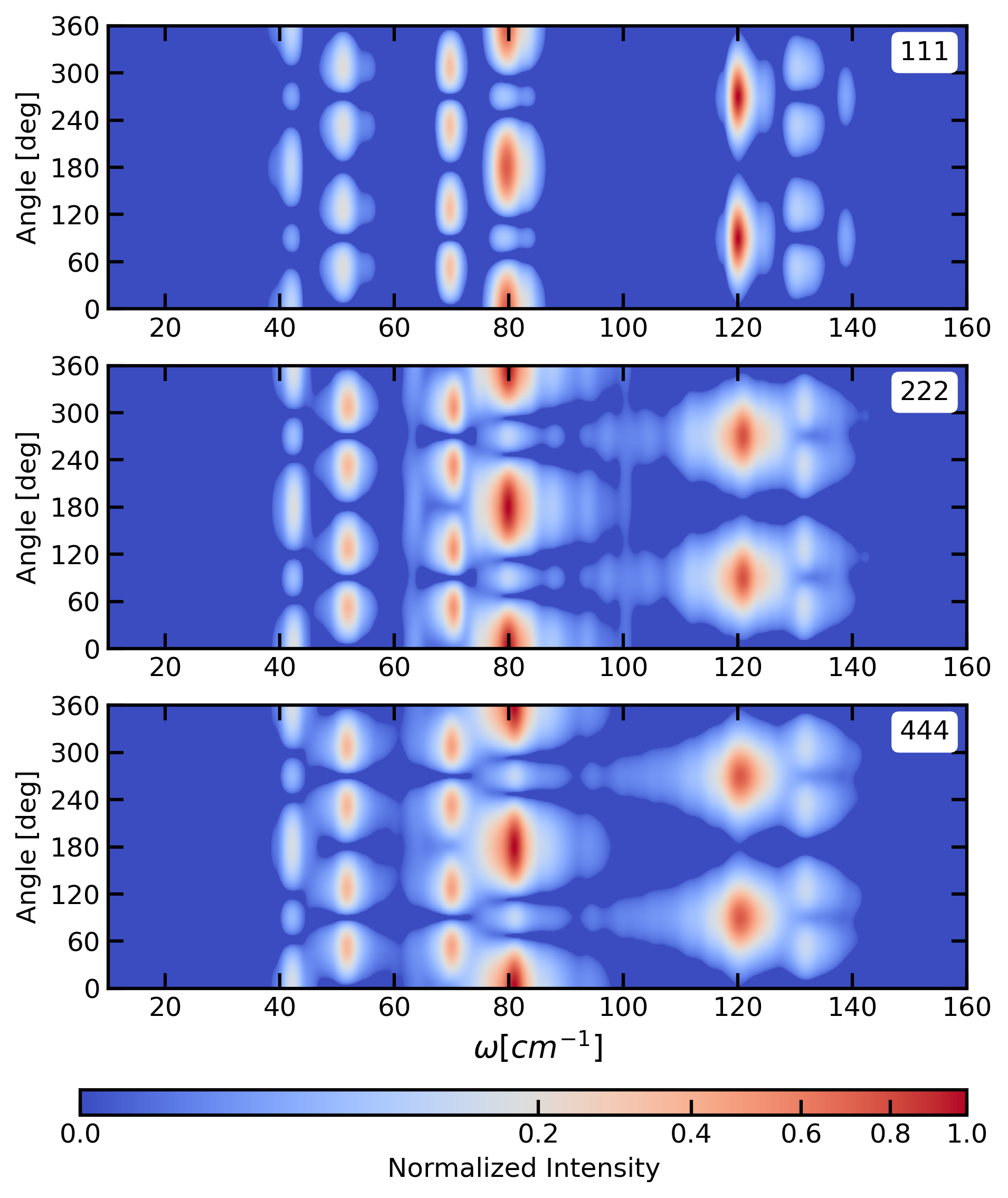}
        \caption{PO-Raman spectra of Anthracene at 100~K in the parallel configuration with different supercell sizes used for the $\Gamma$RGDOS-ML method. The maximum intensity is normalized to 1 for each spectrum. The spectra appear almost visually converged at the 2x2x2 supercell size. The same statistical sampling was used for all the supercell sizes, consisting in 48 NVE trajectories of 100~ps starting from a 2ns parent NVT trajectory. This amount of trajectories led to minimal statistical noise in the intensities even of the smaller supercell.}
        \label{fgr:cellsize}
    \end{figure*}

    \begin{figure*}[h!]
        \centering
        \includegraphics[width=0.84\textwidth]{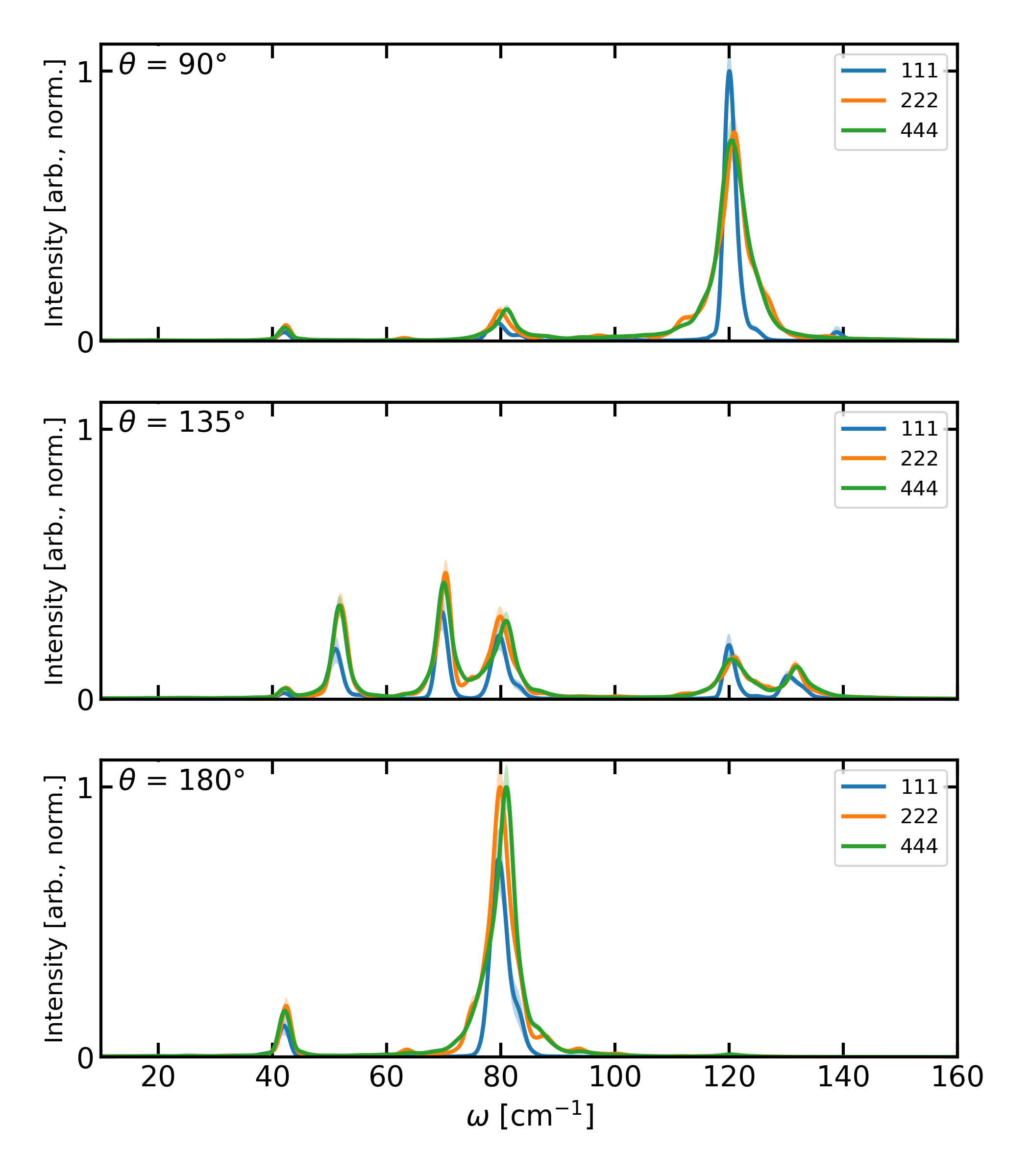}
        \caption{Cuts of the PO-maps of Fig. \ref{fgr:cellsize} at  polarization angles of $90^\circ$, $135^\circ$ and $180^\circ$. The shaded area represents twice the standard error of the mean. A large difference can be seen between the $1\times1\times1$ and the $2\times2\times2$ supercell, especially in terms of peak broadening at all angles. While the spectra are essentially converged at $2\times2\times2$ size, the $4\times4\times4$ peaks appear smoother. In the bottom panel, small oscillations around the 80 cm$^{-1}$ peak can be seen in the $2\times2\times2$ case, which are not present in the larger supercell. Note that the spectra were produced following exactly the same procedure, as outlined in Section \ref{sec:methods} of the main text, with a fixed time windowing cutoff at all supercell sizes.}
        \label{fgr:sccuts}
    \end{figure*}

\clearpage
\section{Mode decomposition of the $\Gamma$RGDOS signals}
\label{sec:decomposedrgdos}

\begin{figure*}[th!]
  \centering
  \subfloat[$\theta = 90^{\circ}, T = 100 K$]{
    \includegraphics[width=0.3\linewidth]{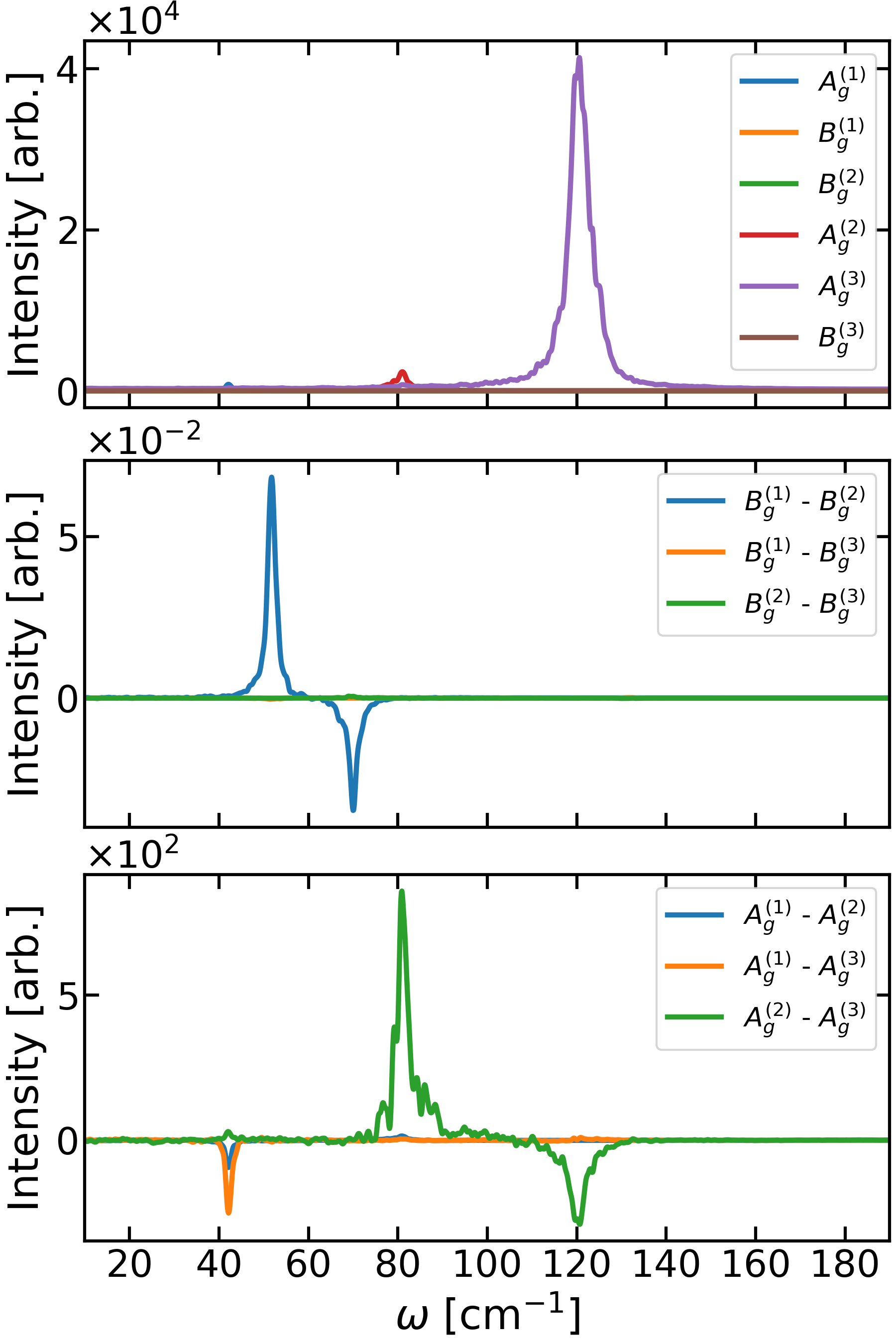}
  }
  \hfill
  \subfloat[$\theta = 120^{\circ}, T = 100 K$]{
    \includegraphics[width=0.3\linewidth]{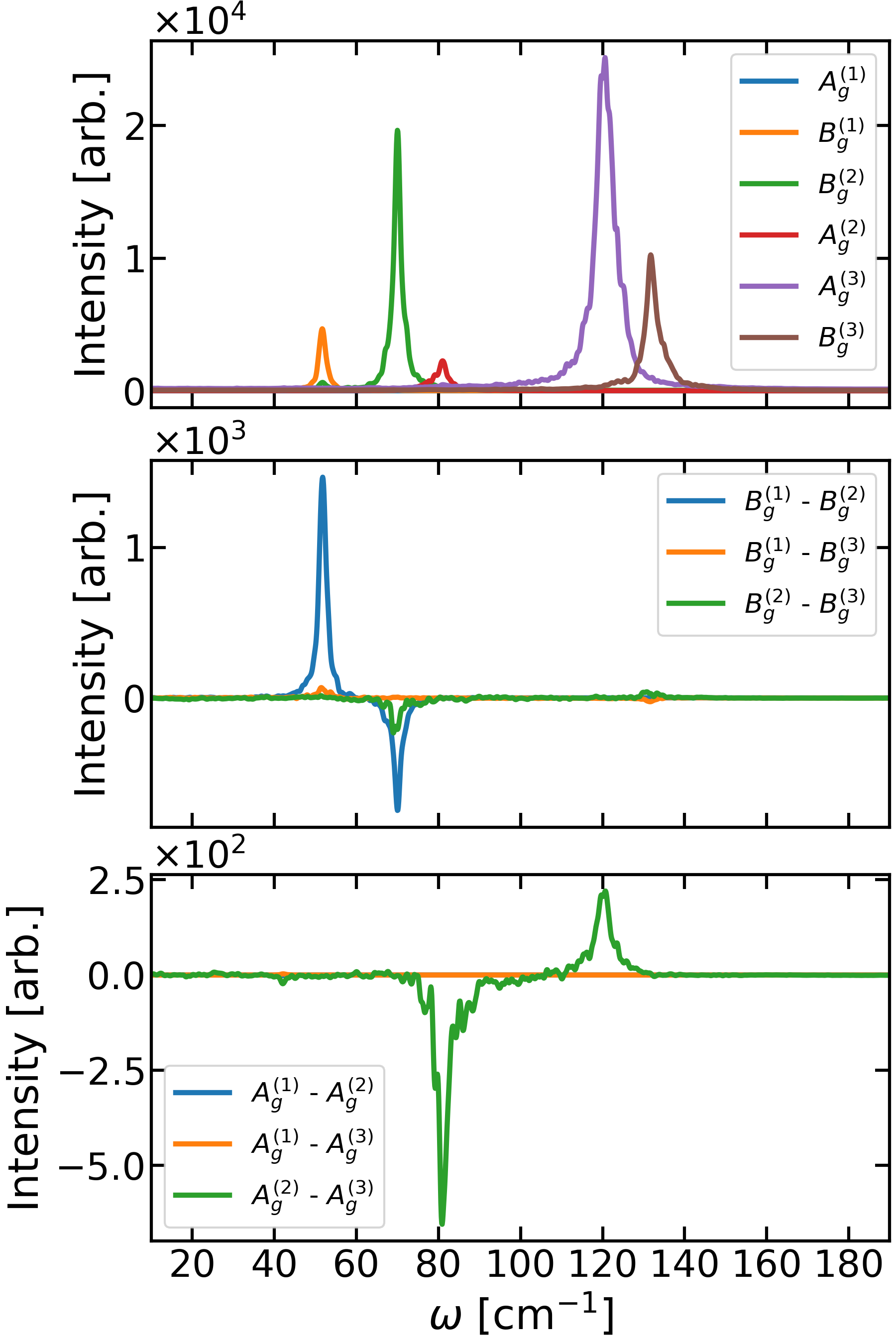}
  }
\hfill
  \subfloat[$\theta = 135^{\circ}, T = 100 K$]{
    \includegraphics[width=0.3\linewidth]{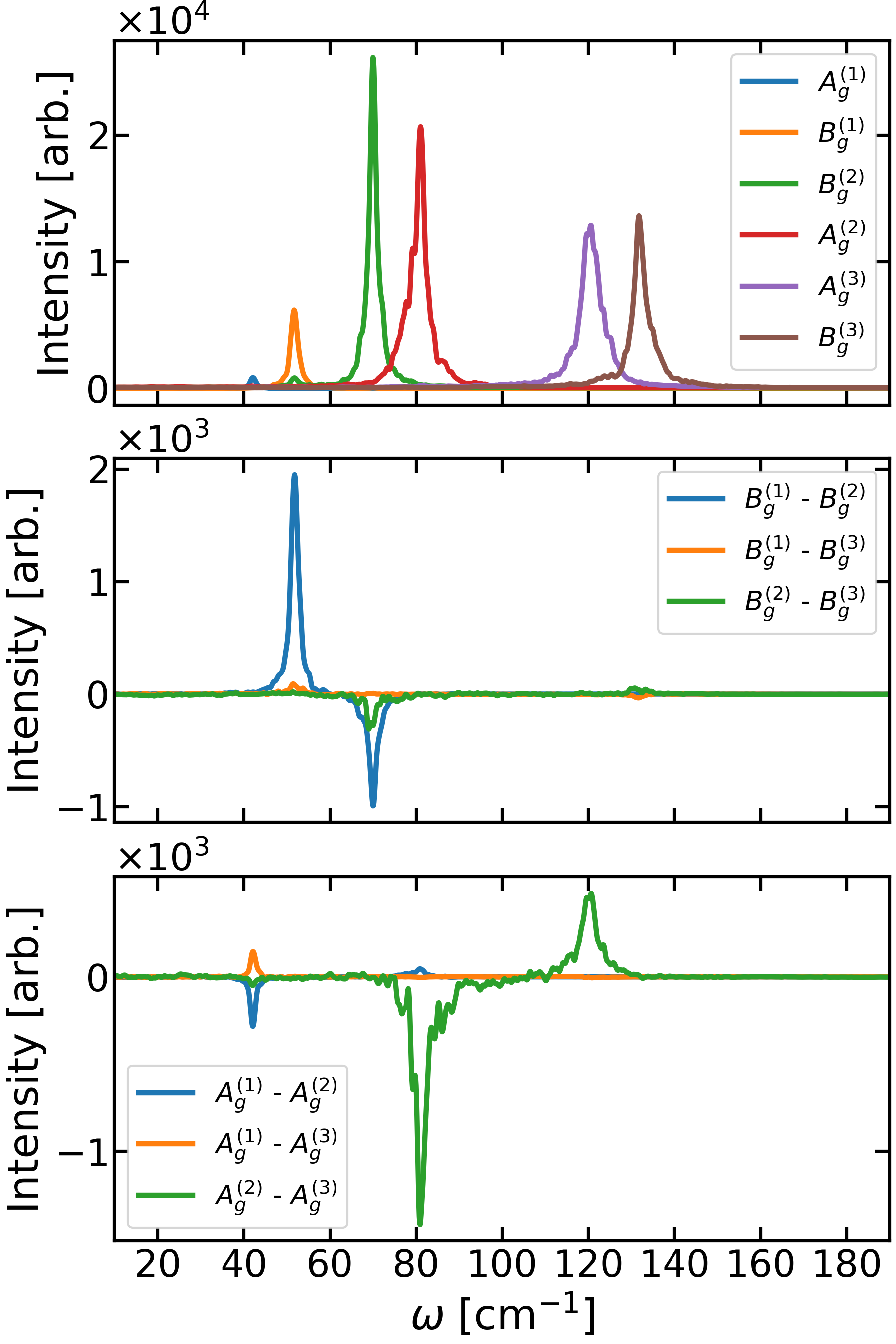}
  }
  \vfill
    \subfloat[$\theta = 90^{\circ}, T = 295 K$]{
    \includegraphics[width=0.3\linewidth]{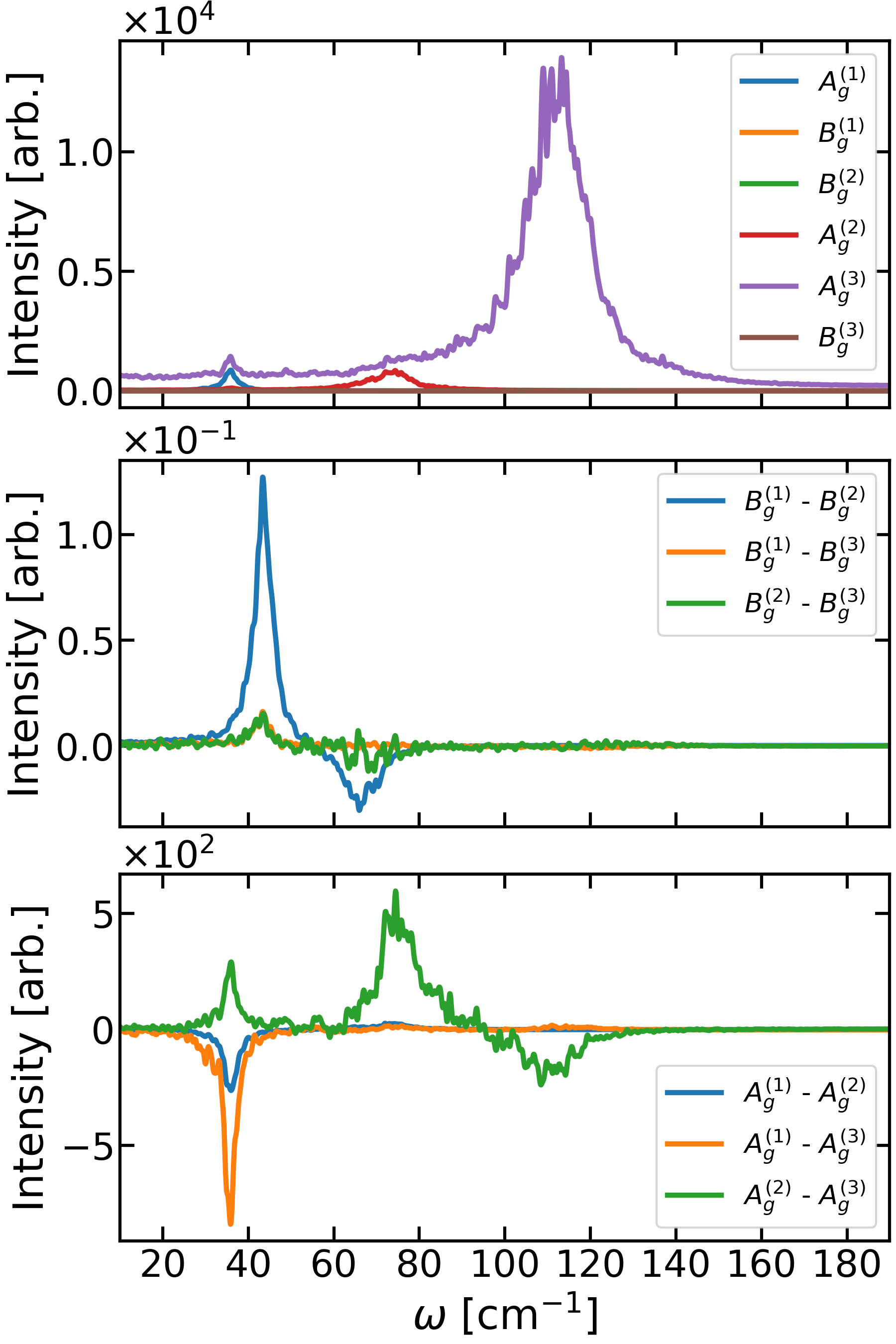}
  }
  \hfill
  \subfloat[$\theta = 120^{\circ}, T = 295 K$]{
    \includegraphics[width=0.3\linewidth]{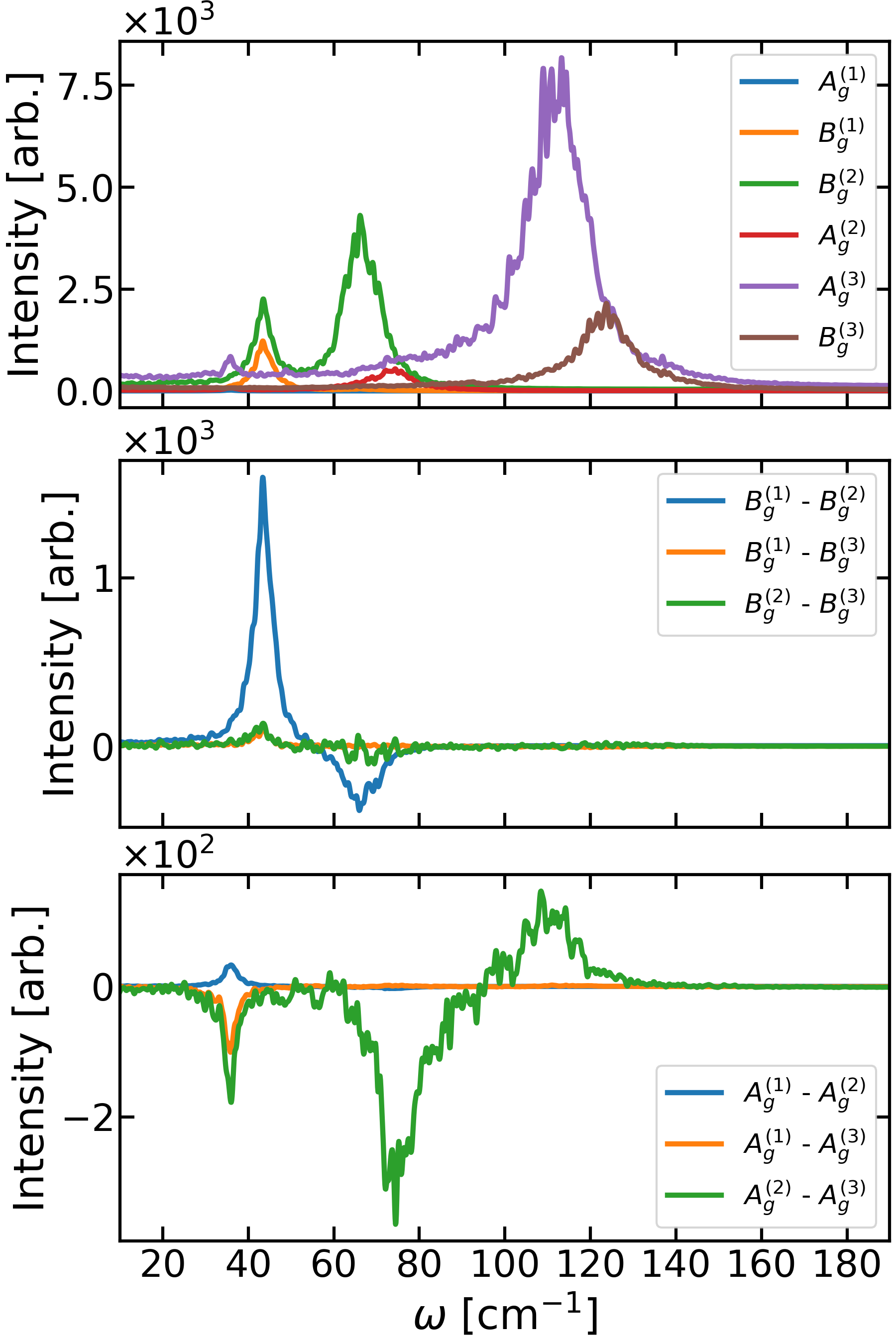}
  }
\hfill
  \subfloat[$\theta = 135^{\circ}, T = 295 K$]{
    \includegraphics[width=0.3\linewidth]{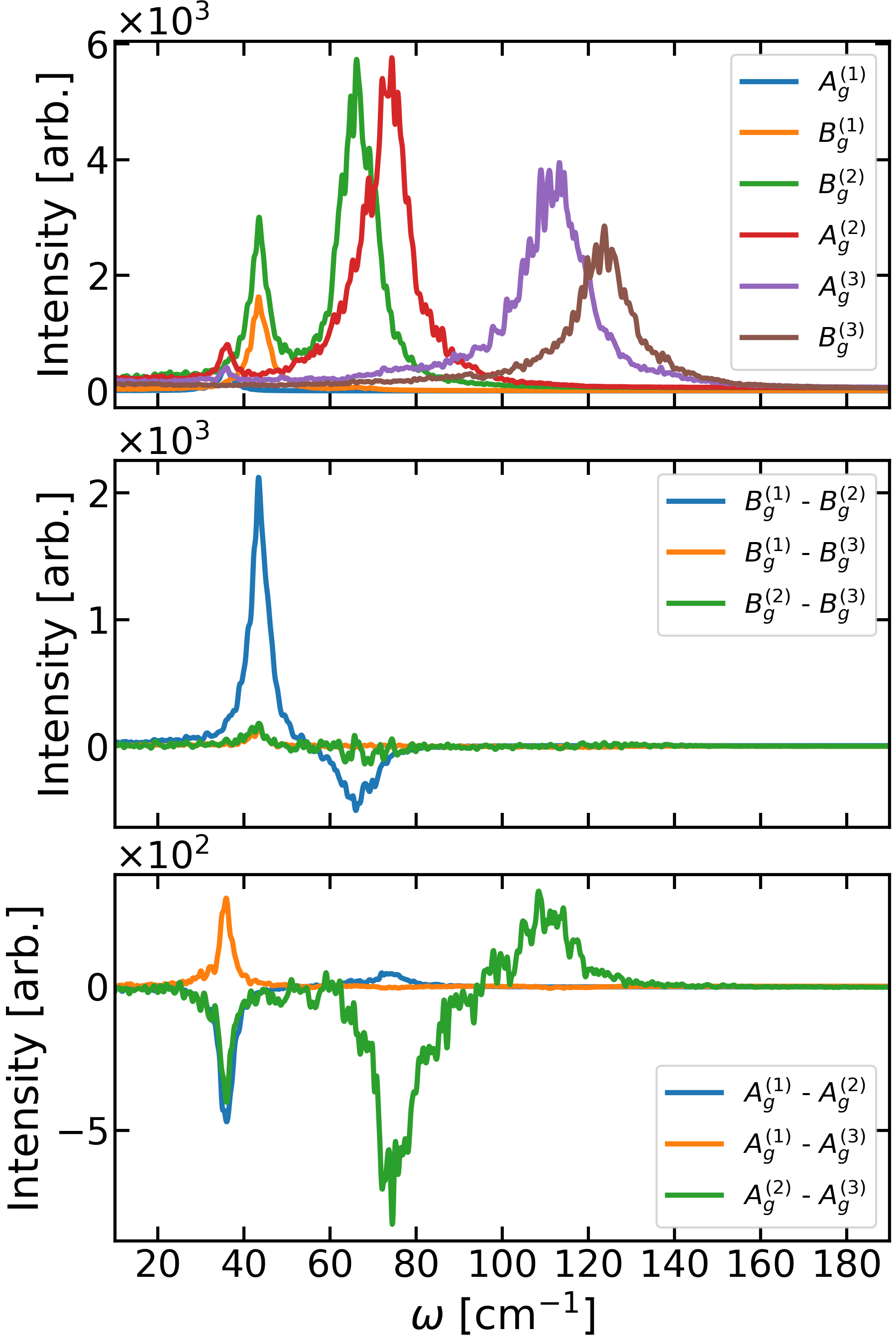}
  }
  \caption{Mode decomposition of the $\Gamma$RGDOS-ML PO-Raman spectra of anthracene at 100 K (top row) and 295 K (bottom row) in the parallel configuration into self- and cross- contributions from individual $\Gamma$-point normal modes. Three different polarization angles are shown as an example: $90^{\circ}$ (left), $120^{\circ}$ (center) and $135^{\circ}$ (right). Note that the spectra are obtained from the raw MD trajectories using the mode-resolved version of Eq. \ref{eq:rgdos-corr} (main text), without any additional windowing or broadening. While the picture that emerges is the same as the one discussed in the main text through simpler VDOS and CVDOS spectra (Section \ref{sec:mix-broad}), this representation allows one to better visualize how the inclusion of the Raman tensors either suppresses or enhances the contribution of self- and cross- terms at different polarization angles, thus shaping the final PO-maps. Cross-terms between modes of different symmetries are not shown, as they are numerically zero.}
  \label{fig:decomposedrgdos}
\end{figure*}

\clearpage
\section{Comparison of DFT functionals and dispersion corrections for the unpolarized Raman spectrum of anthracene}
        \label{sec:dftcomparison}

    \begin{figure*}[h!]
        \centering
        \includegraphics[width=0.8\textwidth]{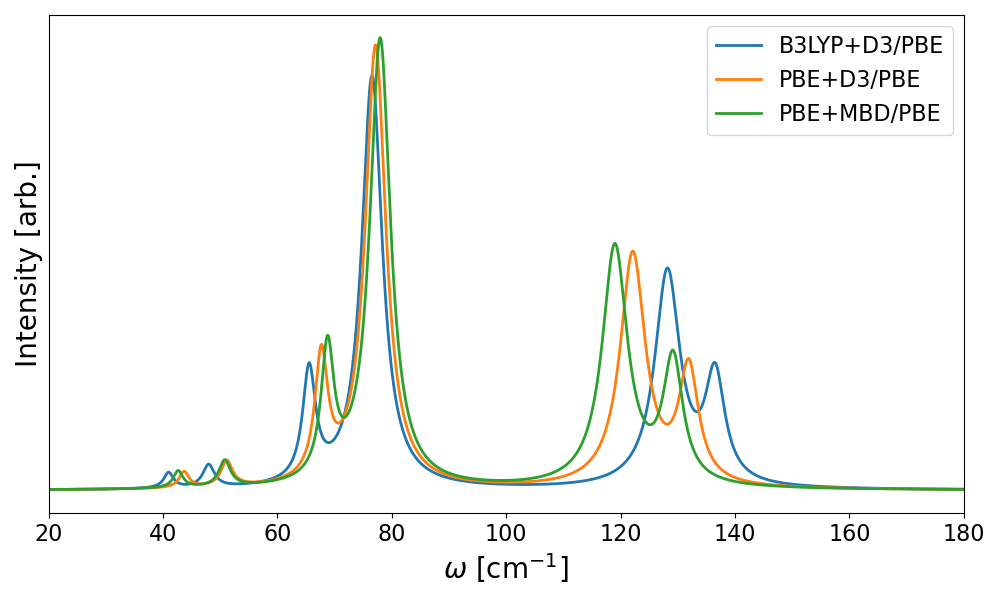}
        \caption{Unpolarized harmonic Raman spectra of anthracene obtained with different combinations of DFT functionals and dispersion corrections. PBE was used in all cases for the DFPT calculations (\textit{tight} basis set). The structure minimization and phonons calculations were done respectively at the PBE+MDB, PBE+D3 and B3LYP+D3 levels of theory, with really tight convergence criteria ($10^{-10} e/a_0^3$ on the density, $10^{-5}$ eV/\AA{} on residual per atom force).  The lattice parameters of the 100 K structure of the main text were used. The unpolarized Raman is defined in Eq. \ref{eq:unpolarized} of the main text. In this case, the intensities come from the harmonic expression of Eq. \ref{eq:harm} (main text) and a Lorentzian broadening was used, employing the fitted linewidths from the anharmonic calculation of the main text.}
        \label{fgr:dftcomparison}
    \end{figure*}

      We note how dispersion corrections like the ones employed here only act through the PES (geometry optimization, harmonic force constants) but do not change the DFPT polarizability directly, as they are effectively post-processing routines on the converged electron density. While they do alter the spectrum through frequency shifts, they can only alter Raman tensors indirectly via changes in the equilibrium structure and the resulting normal-modes. We find in Fig.~\ref{fgr:dftcomparison} some expected frequency shifts when comparing results obtained with PBE+MBD, PBE+D3 and B3LYP+D3 for the geometry optimization and phonon calculations at a fixed PBE level for the DFPT polarizabilities. Notably, the frequency shifts are non-uniform when the hybrid-GGA B3LYP functional is used, with the four  modes below $80$ cm$^{-1}$ undergoing a redshift and the two higher frequency mode showing a blueshift. The absolute and relative Raman intensities remain though essentially unchanged, signaling the the structural effects of these different functionals and dispersion correction methods have minimal impact on the resulting Raman tensors.

\bibliography{main,extra}